\documentclass[12pt]{article}

\let\plural=\relax

\makeatletter
\def\newleaf{\newpage
\newcount\tmp
\tmp=\c@page
\divide\tmp by 2
\multiply\tmp by 2
\ifnum\c@page=\tmp
~\newpage
\fi
}
\makeatother

\expandafter\ifx\csname proofnewpage\endcsname\relax

\fi

\def\color[#1]#2{}

\long\def\nop#1{}

\def\comment{\edef\cps{\the\parskip} \parskip=0.5cm \begingroup \tt}

\def\noproof#1{
\def\proof{{#1}\iffalse}
\let\qed=\fi
}

\begingroup
\makeatletter
\global\def\fakelabel#1#2{
\expandafter\ifx\csname fakelabelsome\endcsname\relax
\let\fakelabelsome\par
\AtEndDocument{\typeout{}}
\fi
\AtEndDocument{\typeout{NOTE: #1 is a fake label, marked #2}\typeout{}}
\@ifundefined {r@#1}
{\global\@namedef{r@#1}{#2}}
{}
}
\endgroup

\long\def\figurearrow#1#2{
\newbox\before
\setbox\before=\hbox{#1}
\newbox\after
\setbox\after=\hbox{#2}
\newdimen\vdim
\vdim=\ht\before
\ifdim\vdim<\the\ht\after
  \vdim=\ht\after
\fi
\hbox{
\vbox to \the\vdim{\vfill\box\before\vfill}
\vbox to \the\vdim{\vfill\hbox to 3cm{\hfill\Huge$\Rightarrow$\hfill}\vfill}
\vbox to \the\vdim{\vfill\box\after\vfill}
}
}

\expandafter\ifx\csname shortcite\endcsname\relax

\fi

\newbox\current

\long\def\plframebox#1{
\setbox\current\vbox{#1}		

\vbox to \ht\current {\hrule\vss
\hbox to \wd\current {%
\vrule \hss\box\current\hss \vrule}
\vss\hrule }
}

\long\def\eatpar#1{%
\ifx#1\par                      
\let\nextmove=\eatpar           
\else
\let\nextmove=#1
\fi
\noexpand\nextmove
}

\def\modifymargins#1#2{
\newdimen\addtoh
\newdimen\addtow
\addtoh=#1
\addtow=#2

\advance\topmargin by -\addtoh
\multiply\addtoh by 2
\advance\textheight by \addtoh

\advance\oddsidemargin by -\addtow
\advance\evensidemargin by -\addtow
\multiply\addtow by 2
\advance\textwidth by \addtow
}

\begingroup
\catcode`\~=11
\gdef\centertilde#1{\lower #1pt\hbox{~}}
\endgroup

\newcount\currenttime
\newcount\hour
\newcount\minute

\def\printtime{%
\currenttime=\time
\hour=\currenttime
\divide\hour by 60
\minute=-\hour
\multiply\minute by 60
\advance\minute by \currenttime
\the\hour:\ifnum\minute<10 0\fi\the\minute
}

\begingroup
\makeatletter
\global\let\@@date=\@date
\gdef\@date{\@@date\ --- \printtime}
\endgroup

\def\oggi{\number\day\space 
\ifcase\month\or
Gennaio\or Febbraio\or Marzo\or Aprile\or Maggio\or Giugno\or
Luglio\or Agosto\or Settembre\or Ottobre\or Novembre\or Dicembre\fi
\space \number\year}

\newcounter{rmexample}

\def\proof{\noindent {\sl Proof.\ \ }}

\def\qed{\hfill{\boxit{}}
  \ifdim\lastskip<\medskipamount \removelastskip\penalty55\medskip\fi}
\def\qedn#1{\hfill{\boxit{}$_#1$}
  \ifdim\lastskip<\medskipamount \removelastskip\penalty55\medskip\fi}
\long\def\boxit#1{\vbox{\hrule\hbox{\vrule\kern3pt
                  \vbox{\kern3pt#1\kern3pt}\kern3pt\vrule}\hrule}}

\def\var{V\!ar}

\def\true{{\sf true}}
\def\false{{\sf false}}

\def\np{{\rm NP}}

\def\bh#1{\if#1{}{\rm BH}\else\mbox{BH$_{#1}$}\fi}
\def\S#1{\mbox{$\Sigma^p_{#1}$}}

\let\cedilla=\c
\def\c{\mbox{$\leadsto$}}

\def\profont{\sf}

\def\x3c{{\profont x3c}}

\def\possnewtheorem#1#2{
\expandafter\ifx\csname #1\endcsname\relax
\newtheorem{#1}{#2}
\fi
}

\def\possnewtheoremthree#1[#2]#3{
\expandafter\ifx\csname #1\endcsname\relax
\newtheorem{#1}[#2]{#3}
\fi
}

\possnewtheorem{theorem}{Theorem}
\possnewtheorem{corollary}{Corollary}
\possnewtheorem{lemma}{Lemma}
\possnewtheoremthree{proposition}[theorem]{Proposition}
\possnewtheorem{definition}{Definition}
\possnewtheorem{question}{Question}
\possnewtheorem{example}{Example}
\possnewtheorem{nontheorem}{Counterexample}
\possnewtheorem{property}{Property}
\possnewtheorem{assumption}{Assumption}
\possnewtheorem{conjecture}{Conjecture}
\possnewtheorem{notation}{Notation}
\newenvironment{theorem*}[1]{{\noindent \bf Theorem~#1}\begin{it}}{\end{it}\

}

\def\after#1#2{#1~{\sf after}~#2}

\let\citeyear=\cite
\expandafter\ifx\csname ttytty\endcsname\relax
\def\ttytex#1{#1\nop}			
\else

\modifymargins{0pt}{-10pt}
\def\frac#1#2{#1/#2}
\def\_{\char 95}
\expandafter\ifx\csname makeatletter\endcsname\relax\makeatletter\fi
\def\@ttyfig#1#2{\def\specialtext{\special{txt:#2}}\specialtext\egroup}
\def\ttyfig{\relax\bgroup\catcode`\^^M\active\let^^M
\let\-\relax\catcode`\ \active\@ttyfig}
\let\ttytex\ttyfig
\expandafter\ifx\csname makeatother\endcsname\relax\makeatletter\fi
\fi

\expandafter\ifx\csname ttytty\endcsname\relax\modifymargins{60pt}{40pt}%
\else\raggedright\parindent=1cm

\fi

\def\resolve{\mbox{resolve}}
\def\rescn{\mbox{ResCn}}

\possnewtheorem{algorithm}{Algorithm}
\possnewtheorem{counterexample}{Counterexample}

\title{Superredundancy:
A tool for Boolean formula minimization complexity analysis}
\author{Paolo Liberatore\thanks{%
DIAG, Sapienza University of Rome.
{\tt liberato@diag.uniroma1.it}
}}
\date{}

\begin{document}

\maketitle



\begin{abstract}

A superredundant clause is a clause that is redundant in the resolution closure
of a formula. The converse concept of superirredundancy ensures membership of
the clause in all minimal CNF formulae that are equivalent to the given one.
This allows for building formulae where some clauses are fixed when minimizing
size. An example are proofs of complexity hardness of the problems of minimal
formula size. Others are proofs of size when forgetting variables or revising a
formula. Most clauses can be made superirredundant by splitting them over a new
variable.

\end{abstract}

\section{Introduction}
\label{section-introduction}

Given a Boolean formula, the minimization problem is to find a formula of
minimal size that is equivalent to
it~\cite{mccl-56,rude-sang-87,coud-94,theo-etal-96,coud-sasa-02,uman-etal-06}.
The decision problem variant that is analyzed in computational complexity is to
check whether a Boolean formula is equivalent to one that is bounded in size by
a given number. This problem led to the creation of the polynomial
hierarchy~\cite{stoc-76}. Yet, it eluded a precise complexity characterization
for over twenty
years~\cite{hamm-koga-93,cepe-kuce-08,buch-uman-11,hema-schn-11}. Framing it
within a complexity class is easy, as it can be solved by a simple
guess-and-check algorithm: guess a formula of the given size, check its
equivalence with the given formula. The difficult part is proving
hardness~\cite{hamm-koga-93,cepe-kuce-08,buch-uman-11,hema-schn-11}.

An example is the proof of \np-hardness of checking whether a Horn formula can
be reduced size within a certain bound~\cite{hamm-koga-93,cepe-kuce-08}. An
hardness proof may start from an arbitrary CNF formula and produce a natural
number and a Horn formula that is equivalent to a formula of size bounded by
that number if and only if the CNF formula is satisfiable.

The Horn formula has to be related to the CNF formula by this condition, but
can otherwise be chosen. A choice is to include in the Horn formula some
essential prime implicates and some other clauses~\cite{hamm-koga-93}. The
first part is guaranteed to be in every minimal formula equivalent to the Horn
one. The size of such minimal formulae are then given by how much the second
part can be shrunk while maintaining equivalence. The equivalence between the
non-essential clauses is helped by the essential ones: equivalence is on the
whole formula, including the essential prime implicates.

In spite of its apparent simplicity, finding such a hardness proof turned out
difficult~\cite{hamm-koga-93,cepe-kuce-08,buch-uman-11,hema-schn-11}. Ensuring
essentiality is not easy when clauses are mixed. Making an example of an
essential prime implicate is trivial: a non-tautological clause is always
essential in a formula comprising it only. It may not be with other clauses.
Other clauses are necessary, if not for making the formula reducible in size or
not depending on the satisfiability of the CNF formula. Essentiality is
complicated to establish because is not a local property. It does not depend on
the clause only, but on the whole formula. A clause may be essential in a
formula, but the addition of a single other clause makes it no longer so.

This article shows an alternative mechanism for existence proofs, which involve
the construction of a formula, like hardness proofs. A clause of a formula is
superirredundant if it is irredundant in the resolution closure of a formula.
Such a clause is in all equivalent formulae of minimal size. Conceptually,
superirredundancy can be established by iteratively resolving all clauses of
the formula and then removing the given clause. If it follows from the others,
it is superredundant. Otherwise, it is superirredundant.

The superirredundant clauses of a formula are in all equivalent formulae of
minimal size, but not the other way around. Superirredundancy is sufficient,
but is not necessary. When minimizing a formula, it is not as useful as
essentiality as it may miss several clauses that are guaranteed in all minimal
formulae. Yet, it compensates this drawback by simple conditions to prove it
and simple ways to ensure it. A clause can often be proved superirredundant by
replacing some variables with truth values in the formula. A clause can often
be made superirredundant by splitting it on a new variable.

A hardness proof for a minimization problem can be built by assuming some
clauses superirredundant. Assuming, not ensuring. They are assumed to be
present in all minimal formulae, not proved so. The remaining clauses can be
removed or otherwise replaced by others while maintaining equivalence. The
clauses assumed superirredundant help in ensuring equivalence. They are
targeted to this aim. To this aim only, they are not built to be
superirredundant at the same time.

The Horn formula is built in such a way the other clauses can be reduced size
if and only if the CNF formula is satisfiable. Only now the clauses that were
assumed superirredundant are verified to be so. If they are not, they are split
to make them superirredundant. Granted, this is not always possible. Yet, it
often can. If it does, superirredundancy provides a simplification in finding
such a proof. Instead of ensuring the essentiality of some clauses and the
reducibility of the others at the same time, it allows concentrating on each
aim at time. First, the other clauses are proved to be reducible in size when
appropriate; second, the clauses that are supposed to be superirredundant are
made so.

A following example illustrates the details of how to build a reduction for the
problem of Horn minimization in a simple way---a way that did not take twenty
years to be developed. It is just an example. This problem is already known to
be \np-hard. Many other minimization problems have been framed exactly in the
polynomial hierarchy. Yet, some related problems are still open. For example,
given a formula, can forgetting a set of variables~\cite{lin-reit-94}, or
literals~\cite{lang-etal-03} or subformulae~\cite{fang-etal-18} be represented
within a certain bound? With fixed symbols~\cite{moin-07}? Can
revising~\cite{pepp-08} or updating~\cite{kats-mend-91} a formula be
represented within a certain bound? These are still minimization problems. They
are still open. Another article employs superirredundancy in the complexity
analysis of forgetting~\cite{libe-20}.

Superirredundancy differs from irredundancy, essentiality and membership in all
minimal formulae. This is shown by the clause $a$ in the formula
{} $F = \{a, \neg a \vee b, \neg b \vee a\}$:
it is not superirredundant, but is redundant, is an essential prime implicate
and is in all minimal formulae that are equivalent to $F$.

\begin{itemize}

\item the resolution closure of $F$ is
{} $\{a, b, \neg a \vee b, \neg b \vee a\}$,
where $a$ is redundant; therefore, $a$ is not superirredundant in $F$;

\item the clause $a$ is irredundant in $F$ since $F \backslash \{a\}$ is
{} $\{\neg a \vee b, \neg b \vee a\}$,
which does not entail $a$;

\item the prime implicates of $F$ are $a$ and $b$; the only CNF formula
equivalent to $F$ made only of prime implicates is $\{a, b\}$, which contains
$a$; therefore, $a$ is an essential prime implicate of $F$;

\item the only minimal-size formula equivalent to $F$ is $\{a,b\}$;
as a result, $a$ belongs to all minimal-size formulae equivalent to $F$.

\end{itemize}

Besides some preliminaries in Section~\ref{section-preliminaries}, the
technical content of the article is split in three parts:
Section~\ref{section-superredundancy} defines superredundancy and gives a
number of its necessary and sufficient conditions; Section~\ref{section-split}
shows how to make a clause superirredundant in most cases;
Section~\ref{section-example} shows the details of an example usage of
superirredundancy in a hardness proof. Concluding remarks are in
Section~\ref{section-conclusions}.

\section{Preliminaries}
\label{section-preliminaries}

\subsection{Formulae}

The formulae in this article are all propositional in conjunctive normal form
(CNF): they are sets of clauses, a clause being the disjunction of some literals
and a literal a propositional variable or its negation. This is not truly a
restriction, as every formula can be turned into CNF without changing its
semantics. A clause is sometimes identified with the set of literals it
contains. For example, a subclause is a subset of a clause.

If $l$ is a negative literal $\neg x$, its negation $\neg l$ is defined as $x$.

The variables a formula $A$ contains are denoted $\var(A)$.

\begin{definition}
\label{size}

The size $||A||$ of a formula $A$ is the number of variable occurrences it
contains.

\end{definition}

This is not the same as the cardinality of $\var(A)$ because a variable may
occur multiple times in a formula. For example, $A = \{a, \neg a \vee b, a \vee
\neg b\}$ has size five because it contains five literal occurrences even if
its variables are only two. The size is obtained by removing from the formula
all propositional connectives, commas and parentheses and counting the number
of symbols left.

Other definitions are possible but are not considered in this article. An
alternative measure of size is the total number of symbols a formula contains
(including conjunctions, disjunctions, negations and parenthesis). Another is
the number of clauses (regardless of their length).

The definition of size implies the definition of minimality: a formula is
minimal if it is equivalent to no formula smaller than it. Given a formula, a
minimal equivalent formula is a possibly different but equivalent formula that
is minimal. As an example, $A = \{a, \neg a \vee b, a \vee \neg b\}$ has size
five since it contains five literal occurrences; yet, it is equivalent to $B =
\{a,b\}$, which only contains two literal occurrences. No formula equivalent
to $A$ or $B$ is smaller than that: $B$ is minimal. Minimizing a formula means
obtaining a minimal equivalent formula. This problem has long been
studied~\cite{hema-schn-11,cepe-kuce-08}.

\begin{definition}
\label{clauses-literal}

The clauses of a formula $A$ that contain a literal $l$ are denoted by $A \cap
l = \{c \in A \mid l \in c\}$.

\end{definition}

This notation cannot cause confusion: when is between two sets, the symbol
$\cap$ denotes their intersection; when is between a set and a literal, it
denotes the clauses of the set that contain the literal. This is like seeing $A
\cap l$ as the shortening of $A \cap \mathrm{clauses}(l)$, where
$\mathrm{clauses}(l)$ is the set of all possible clauses that contain the
literal $l$.

When a formula entails a clause but none of its strict subclauses, the clause
is a prime implicate of the formula. Formally, $F \models c$ holds but $F
\models c'$ does not for any clause $c' \subset c$. Prime implicates are a
common tool in formula minimization~\cite{hema-schn-11,cepe-kuce-08}.

\subsection{Resolution}
\label{subsection-resolution}

Resolution is a syntactic derivation mechanism that produces a new clause that
is a consequence of two clauses: $c_1 \vee l, c_2 \vee \neg l \vdash c_1 \vee
c_2$. The result is implicitly removed repetitions.


Unless noted otherwise, tautologic clauses are excluded. Writing
{} $c_1 \vee a, c_2 \vee \neg a \vdash c_1 \vee c_2$
implicitly assumes that none of the three clauses is a tautology unless
explicitly stated. Two clauses that would resolve in a tautology are considered
not to resolve, which is not a limitation~\cite{love-14}.
Tautologic clauses are forbidden in formulae, which is not a limitation either
since tautologies are always satisfied. This assumption has normally little
importance, but is crucial to superredundancy, defined in the next section.

%

\begin{lemma}
\label{different}

A clause that is the result of resolving two clauses does not contain the
resolving variable and is different from both of them if none of them is a
tautology.

\end{lemma}

\proof A clause $c$ is the result of resolving two clauses only if they have
the form $c_1 \vee a$ and $c_2 \vee \neg a$ for some variable $a$, and $c$ is
$c_1 \vee c_2$. If $c_1 \vee c_2$ is equal to $c_1 \vee a$, it contains $a$.
Since clauses are sets of literals, they do not contain repeated elements. As a
result, $a \not\in c_1$ as otherwise $c_1 \vee a$ would contain $a$ twice.
Together with $a \in c_1 \vee c_2$ this implies $a \in c_2$, which makes $c_2
\vee \neg a$ a tautology. The case $c = c_2 \vee \neg a$ is similar.~\qed

A resolution proof $F \vdash G$ is a binary forest where the roots are the
clauses of $G$, the leaves the clauses of $F$ and every parent is the result of
resolving its two children.


\begin{definition}

The resolution closure of a formula $F$ is the set $\rescn(F) = \{c \mid F
\vdash c\}$ of all clauses that result from applying resolution zero or more
times from $F$.

\end{definition}

The clauses of $F$ are derivable by zero-step resolutions from $F$. Therefore,
$F \vdash c$ and $c \in \rescn(F)$ hold for every $c \in F$.

The resolution closure is similar to the deductive closure but not identical.
For example, $a \vee b \vee c$ is in the deductive closure of $F = \{a \vee
b\}$ but not in the resolution closure. It is a consequence of $F$ but is not
obtained by resolving clauses of $F$.

All clauses in the resolution closure $\rescn(F)$ are in the deductive closure
but not the other way around. The closures differ because resolution does not
expand clauses: $a \vee b \vee c$ is not a resolution consequence of $a \vee
b$. Adding expansion kills the difference~\cite{lee-67,slag-69}.

\[
F \models c \mbox{ if and only if }
c' \in \rescn(F) \mbox{ for some } c' \subseteq c
\]

That resolution does not include expansion may suggest that it cannot generate
any non-minimal clause. That would be too good to be true, since a clause would
be minimal just because it is obtained by resolution. In fact, it is not the
case. Expansion is only one of the reasons clauses may not be minimal, as seen
in the formula
{} $\{a \vee b \vee c, a \vee b \vee e, \neg e \vee c \vee d\}$:
the second and third clauses resolve to
{} $a \vee c \vee b \vee d$,
which is however not minimal: it contains the first clause of the formula,
{} $a \vee b \vee c$.

What is the case is that resolution generates all prime
implicates~\cite{lee-67,slag-69}, the minimal entailed clauses. The relation
between $\rescn(F)$ and the deductive closure of $F$ tells that if a clause is
entailed, a subset of it is generated by resolution; since the only entailed
subclause of a prime implicate is itself, it is the only one resolution may
generate. Removing all clauses that contain others from $\rescn(F)$ results in
the set of the prime implicates of $F$.

Minimal equivalent formulae are all made of minimal entailed clauses, as
otherwise literals could be removed from them. Since resolution allows deriving
all prime implicates of a formula~\cite{lee-67,slag-69}, it derives all clauses
of all minimal equivalent formulae.

\begin{property}
\label{minimal-resolution}

If $B$ is a minimal CNF formula equivalent to $A$,
then $B \subseteq \rescn(A)$.

\end{property}

While $\rescn(F)$ contains all clauses generated by an arbitrary number of
resolutions, some properties used in the following require the clauses obtained
by a single resolution step.

\begin{definition}
\label{resolve-function}

The resolution of two formulae is the set of clauses obtained by resolving each
clause of the first formula with each clause of the second:

\[
\resolve(A,B) = \{c \mid c',c'' \vdash c 
\mbox{ where } c' \in A \mbox{ and } c'' \in B\}
\]

If either of the two formulae comprises a single clause, the abbreviations
{} $\resolve(A,c) = \resolve(A, \{c\})$,
{} $\resolve(c,B) = \resolve(\{c\}, B)$ and
{} $\resolve(c,c') = \resolve(\{c\}, \{c'\})$
are used.

\end{definition}

This set contains only the clauses that results from resolving a single clause
of $A$ with a single clause of $B$. Exactly one resolution of one clause with
one clause. Not zero, not multiple ones. A clause of $A$ is not by itself in
$\resolve(A,B)$ unless it is also the resolvent of another clause of $A$ with a
clause of $B$.

\section{Superredundancy}
\label{section-superredundancy}

The running example presented in the introduction is a prototypical
minimization problem: given a Horn formula $A$ and a number $k$, decide whether
$A$ is equivalent to a Horn formula of size $k$ or less. Its \np{} membership
is easy to prove: guess a Horn formula $B$ of size at most $k$ and verify its
equivalence with $A$. Equivalence between Horn formulae can be checked in
polynomial time. Therefore, the whole problem is in \np.

The difficult part of classing it in the polynomial hierarchy is to establish
its hardness. This can be done by reducing the \np-hard problem of Boolean
satisfiability into it. Given a CNF formula $F$, the task is to produce a Horn
formula $A$ and a number $k$ such that $A$ is equivalent to a Horn formula $B$
of size bounded by $k$ if and only if $F$ is satisfiable. A similar reduction
would prove the hardness of other minimization problems. For example, releasing
the Horn constraint increases the complexity from \np{} to \S{2}. The
translation is then required to produce a formula $A$ that can be expressed in
size $k$ if and only the QBF $\forall X \exists Y . F$ is valid. For the sake
of this exposition, $A$ is assumed Horn and the condition is the satisfiability
of $F$.

A reduction could translate a satisfiable $F$ into
{} $\{a \vee \neg b, a \vee c\}$
and an unsatifiable $F$ into
{} $\{a \vee \neg b, a \vee c, l\}$.
The first formula has size $4$, the second $4+1$. The presence of $l$ provides
the required increase in size.

While the addition of a literal like $l$ works, it is not enough. A
counterexample is a reduction that translates a satisfiable $F$ into
{} $\{a \vee \neg b, a \vee c\}$
and an unsatifiable $F$ into
{} $\{\neg a \vee c, l\}$.
While $l$ is necessary in all formulae equivalent to the second, it does not
produce an increase in size: the first formula has size $4$, the second $2+1$.
The $+1$ is the second formula is overcome by the decrease of size of the other
clauses from $4$ to $2$.

The mechanism of having or not having $l$ in the formula only works if the rest
of the formula is fixed. Otherwise, the presence of $l$ may invalidate the
construction like the last addition to a house of cards crushes its lower
layers: the unsatisfiability of $F$ may force $l$ into the formula, but if it
also allows removing a clause of two literals the total size change is $-1$,
not the required $+1$.

The lower layers, the other clauses, are fixed in place by superirredundancy.
Its formal definition is below, but what counts is that a superirredundant
clause of a formula belongs to every minimal equivalent formula. The number of
its literals is never subtracted from the overall size of the formula. The
literals of the superirredundant clauses are fixed; the other clauses may
provide the required $+1$ addition in size.

This is a plan of this section: Section~\ref{superredundancy-definition}
presents the formal definition of superredundancy and its converse,
superirredundancy; Section~\ref{superirredundancy-minimality} shows how
superirredundancy relates to minimal formulae;
Section~\ref{equivalent-superredundancy} presents some equivalent conditions to
superredundancy, Section~\ref{sufficient-superredundancy} some necessary ones.

\subsection{Definition of superredundancy}
\label{superredundancy-definition}

Superirredundancy is based on resolution. Summarizing the notation introduced
in Section~\ref{section-preliminaries}: $\resolve(F,G)${\plural} are the
clauses obtained by resolving each clause of the first formula with each of the
second; $F \vdash G$ means that the clauses of the second formula are obtained
by repeatedly resolving the clauses of the first. This condition is formalized
by $G \subset \rescn(F)$ since $\rescn(F)$ is the set of all clauses obtained
by repeatedly resolving the clauses of a formula $F$.

Resolution does not just tell whether a clause is implied from a set. It also
tells why: the implied clause is consequence of other two, in turns consequence
of others and so on. This structure is necessary in many proofs below, and is
the very reason why they employ resolution $\vdash$ instead of entailment
$\models$.

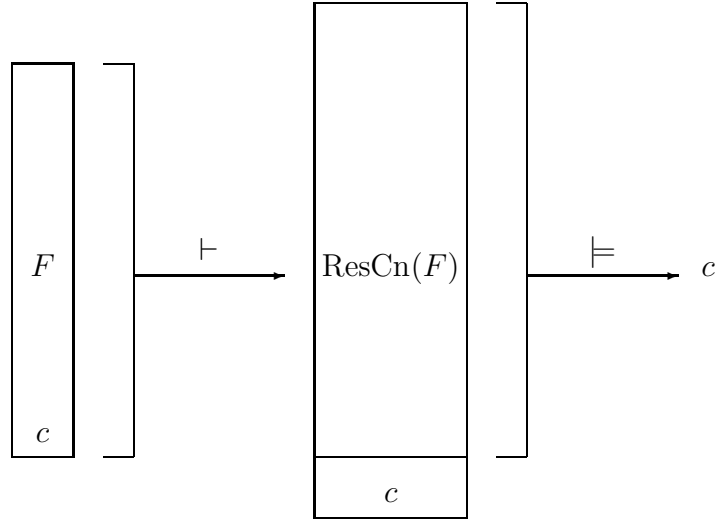
\begin{figure}
\begin{center}
\setlength{\unitlength}{5000sp}%
\begingroup\makeatletter\ifx\SetFigFont\undefined%
\gdef\SetFigFont#1#2#3#4#5{%
  \reset@font\fontsize{#1}{#2pt}%
  \fontfamily{#3}\fontseries{#4}\fontshape{#5}%
  \selectfont}%
\fi\endgroup%
\begin{picture}(3477,2574)(6439,-7273)
\thinlines
{\color[rgb]{0,0,0}\put(7951,-7261){\framebox(750,2550){}}
}%
{\color[rgb]{0,0,0}\put(7951,-6961){\line( 1, 0){750}}
}%
{\color[rgb]{0,0,0}\put(8851,-4711){\line( 1, 0){150}}
\put(9001,-4711){\line( 0,-1){2250}}
\put(9001,-6961){\line(-1, 0){150}}
}%
{\color[rgb]{0,0,0}\put(9001,-6061){\vector( 1, 0){750}}
}%
{\color[rgb]{0,0,0}\put(6451,-6961){\framebox(300,1950){}}
}%
{\color[rgb]{0,0,0}\put(6901,-5011){\line( 1, 0){150}}
\put(7051,-5011){\line( 0,-1){1950}}
\put(7051,-6961){\line(-1, 0){150}}
}%
{\color[rgb]{0,0,0}\put(7051,-6061){\vector( 1, 0){750}}
}%
\put(8326,-7186){\makebox(0,0)[b]{\smash{{\SetFigFont{12}{24.0}
{\rmdefault}{\mddefault}{\updefault}{\color[rgb]{0,0,0}$c$}%
}}}}
\put(8326,-6061){\makebox(0,0)[b]{\smash{{\SetFigFont{12}{24.0}
{\rmdefault}{\mddefault}{\updefault}{\color[rgb]{0,0,0}$\rescn(F)$}%
}}}}
\put(9376,-5986){\makebox(0,0)[b]{\smash{{\SetFigFont{12}{24.0}
{\rmdefault}{\mddefault}{\updefault}{\color[rgb]{0,0,0}$\models$}%
}}}}
\put(6601,-6061){\makebox(0,0)[b]{\smash{{\SetFigFont{12}{24.0}
{\rmdefault}{\mddefault}{\updefault}{\color[rgb]{0,0,0}$F$}%
}}}}
\put(6601,-6886){\makebox(0,0)[b]{\smash{{\SetFigFont{12}{24.0}
{\rmdefault}{\mddefault}{\updefault}{\color[rgb]{0,0,0}$c$}%
}}}}
\put(7426,-5986){\makebox(0,0)[b]{\smash{{\SetFigFont{12}{24.0}
{\rmdefault}{\mddefault}{\updefault}{\color[rgb]{0,0,0}$\vdash$}%
}}}}
\put(9901,-6061){\makebox(0,0)[b]{\smash{{\SetFigFont{12}{24.0}
{\rmdefault}{\mddefault}{\updefault}{\color[rgb]{0,0,0}$c$}%
}}}}
\end{picture}%
\nop{
F
  |- ResCn(F) |= c
     c
c
}
\caption{A superredundant clause}
\label{figure-superredundant}
\end{center}
\end{figure}

A clause $c$ of $F$ may or may not be redundant in $\rescn(F)$. It is redundant
if it is a consequence of $\rescn(F) \backslash \{c\}$, like in
Figure~\ref{figure-superredundant}.

An example is $a$ in
{} $\{a, \neg a \vee b, \neg b \vee a\}$.
The resolution closure of this formula is
{} $\rescn(\{a, \neg a \vee b, \neg b \vee a\}) =
{}  \{a, b, \neg a \vee b, \neg b \vee a\}$,
where $a$ is redundant since it is entailed by $b$ and $\neg b \vee a$.

Such a redundancy is not always the case. For example, the resolution closure
of $\{a,b\}$ is itself, since its two clauses do not resolve. As a result,
$a$ is not redundant in $\rescn(\{a,b\}) = \{a,b\}$.

If a clause is redundant~\cite{gott-ferm-93,libe-05-c}, it is also
superredundant, but not the other way around. For example, $a$ is
superredundant in $\{a, \neg a \vee b, \neg b \vee a\}$ because it is entailed
by the other clauses of the resolution closure of this formula but is not
redundant because it is not entailed by the other clauses of the formula
itself, which are $\neg a \vee b$ and $\neg b \vee a$.

When $c$ is redundant in the resolution closure, $\rescn(F) \backslash \{c\}$
is equivalent to $\rescn(F)$. Even if $c$ is not redundant in $F$, it is not
truly necessary as a formula $\rescn(F) \backslash \{c\}$ not containing it is
equivalent to $F$. In the first example, $a$ is not redundant in
{} $\{a, \neg a \vee b, \neg b \vee a\}$,
but this formula is equivalent to
{} $\rescn(\{a, \neg a \vee b, \neg b \vee a\}) \backslash \{a\} =
{}  \{b, \neg a \vee b, \neg b \vee a\}$,
which does not contain the clause $a$.


This is a weak version of redundancy: while $c$ may not be removed from $F$, it
can be replaced by other consequences of $F$. The converse is therefore a
strong version of minimality: $c$ cannot be removed even adding all other
resolution consequences of $F$. This is why it is called superirredundant. A
superirredundant clause cannot be removed even adding resolution consequences
in its place. It is irredundant even expanding $F$ this way, even switching to
such supersets of $F$.

The opposite to superirredundancy is unsurprisingly called superredundancy: a
clause is superredundant if it is redundant in the superset $\rescn(F)$ of $F$.
Such a clause can be replaced by other resolution consequences of $F$.

\begin{definition}

A clause $c$ of a formula $F$ is superredundant if it is redundant in the
resolution closure of the formula: $\rescn(F) \backslash \{c\} \models c$.
It is superirredundant if it is not superredundant.

\end{definition}

A superirredundant clause of a formula will be proved to be in all minimal
formulae equivalent to that formula. It is necessary in them. This is how
fixing a part of the formula is achieved: by ensuring that its clauses are
superirredundant. The opposite concept of superredundancy is introduced because
it simplifies a number of technical results.

Contrary to what it may look, creating superirredundant clauses is not
difficult.

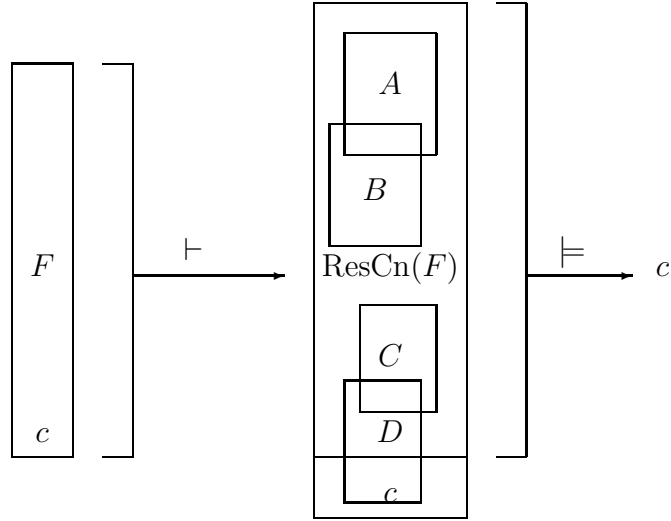
\begin{figure}
\begin{center}
\setlength{\unitlength}{5000sp}%
\begingroup\makeatletter\ifx\SetFigFont\undefined%
\gdef\SetFigFont#1#2#3#4#5{%
  \reset@font\fontsize{#1}{#2pt}%
  \fontfamily{#3}\fontseries{#4}\fontshape{#5}%
  \selectfont}%
\fi\endgroup%
\begin{picture}(3252,2574)(6439,-7273)
\thinlines
{\color[rgb]{0,0,0}\put(6451,-6961){\framebox(300,1950){}}
}%
{\color[rgb]{0,0,0}\put(7951,-7261){\framebox(750,2550){}}
}%
{\color[rgb]{0,0,0}\put(7951,-6961){\line( 1, 0){750}}
}%
{\color[rgb]{0,0,0}\put(8101,-5461){\framebox(450,600){}}
}%
{\color[rgb]{0,0,0}\put(8026,-5911){\framebox(450,600){}}
}%
{\color[rgb]{0,0,0}\put(8101,-7186){\framebox(375,600){}}
}%
{\color[rgb]{0,0,0}\put(8176,-6736){\framebox(375,525){}}
}%
{\color[rgb]{0,0,0}\put(6901,-5011){\line( 1, 0){150}}
\put(7051,-5011){\line( 0,-1){1950}}
\put(7051,-6961){\line(-1, 0){150}}
}%
{\color[rgb]{0,0,0}\put(8851,-4711){\line( 1, 0){150}}
\put(9001,-4711){\line( 0,-1){2250}}
\put(9001,-6961){\line(-1, 0){150}}
}%
{\color[rgb]{0,0,0}\put(7051,-6061){\vector( 1, 0){750}}
}%
{\color[rgb]{0,0,0}\put(9001,-6061){\vector( 1, 0){525}}
}%
\put(6601,-6061){\makebox(0,0)[b]{\smash{{\SetFigFont{12}{24.0}
{\rmdefault}{\mddefault}{\updefault}{\color[rgb]{0,0,0}$F$}%
}}}}
\put(6601,-6886){\makebox(0,0)[b]{\smash{{\SetFigFont{12}{24.0}
{\rmdefault}{\mddefault}{\updefault}{\color[rgb]{0,0,0}$c$}%
}}}}
\put(9676,-6061){\makebox(0,0)[b]{\smash{{\SetFigFont{12}{24.0}
{\rmdefault}{\mddefault}{\updefault}{\color[rgb]{0,0,0}$c$}%
}}}}
\put(8326,-7186){\makebox(0,0)[b]{\smash{{\SetFigFont{12}{24.0}
{\rmdefault}{\mddefault}{\updefault}{\color[rgb]{0,0,0}$c$}%
}}}}
\put(8326,-6061){\makebox(0,0)[b]{\smash{{\SetFigFont{12}{24.0}
{\rmdefault}{\mddefault}{\updefault}{\color[rgb]{0,0,0}$\rescn(F)$}%
}}}}
\put(8326,-5161){\makebox(0,0)[b]{\smash{{\SetFigFont{12}{24.0}
{\rmdefault}{\mddefault}{\updefault}{\color[rgb]{0,0,0}$A$}%
}}}}
\put(8251,-5686){\makebox(0,0)[b]{\smash{{\SetFigFont{12}{24.0}
{\rmdefault}{\mddefault}{\updefault}{\color[rgb]{0,0,0}$B$}%
}}}}
\put(8326,-6511){\makebox(0,0)[b]{\smash{{\SetFigFont{12}{24.0}
{\rmdefault}{\mddefault}{\updefault}{\color[rgb]{0,0,0}$C$}%
}}}}
\put(8326,-6886){\makebox(0,0)[b]{\smash{{\SetFigFont{12}{24.0}
{\rmdefault}{\mddefault}{\updefault}{\color[rgb]{0,0,0}$D$}%
}}}}
\put(9226,-5986){\makebox(0,0)[b]{\smash{{\SetFigFont{12}{24.0}
{\rmdefault}{\mddefault}{\updefault}{\color[rgb]{0,0,0}$\models$}%
}}}}
\put(7351,-5986){\makebox(0,0)[b]{\smash{{\SetFigFont{12}{24.0}
{\rmdefault}{\mddefault}{\updefault}{\color[rgb]{0,0,0}$\vdash$}%
}}}}
\end{picture}%
\nop{
F
      A      
  |-  B      |- c
      C
      c in D |/- c
c
}
\caption{A superredundant clause and some minimal equivalent formulae}
\label{figure-minimal}
\end{center}
\end{figure}

A typical situation with a superredundant clause $c$ is that $F$ has some
minimal equivalent formulae like $D$ which contains $c$, but also has other
minimal equivalent formulae $A$, $B$ and $C$ which do not contain $c$, like in
Figure~\ref{figure-minimal}. This is possible because of $\rescn(F) \backslash
\{c\} \models c$, which allows some subsets of $\rescn(F) \backslash \{c\}$
like $D$ to entail $c$. Some other subsets like $A$, $B$ and $C$ may still be
minimal even if they contain $c$.

Superredundancy is the same as redundancy in the resolution closure. It is not
the same as redundancy in the deductive closure: such a redundancy is always
the case unless the clause contains all variables in the alphabet. Otherwise,
if $a$ is a variable not in $c$, then $c \models c \vee a$ and $c \models c
\vee \neg a$; as a result, if $c \in F$ then $Cn(F) \backslash \{c\}$ contains
$c \vee a$ and $c \vee \neg a$, which imply $c$. The same argument does not
apply to resolution because neither $c \vee a$ nor $c \vee \neg a$ follow from
$c$ by resolution. This is why superredundancy is defined in terms of
resolution and not entailment.

Redundancy implies superredundancy: if $c$ follows from $F \backslash \{c\}$ it
also follows from $\rescn(F) \backslash \{c\}$ by monotonicity of entailment.
Not the other way around. For example, $a$ is irredundant in $F = \{a, \neg a
\vee b, \neg b \vee a\}$ but is superredundant: $a$ and $\neg a \vee b$ resolve
to $b$, which resolves with $\neg b \vee a$ back to $a$.

The formula in this example is not minimal, as it is equivalent to $\{a,b\}$.
It shows that a non-minimal formula may contain a superredundant clause. This
will be proved to be always the case. In reverse: a formula entirely made of
superirredundant clauses is minimal.

\subsection{Superirredundancy and minimality}
\label{superirredundancy-minimality}

The aim of superirredundancy is to prove that a clause belongs to all formulae
that are equivalent to the given one. The following lemma proves this.

\begin{lemma}
\label{all}

If $c$ is a superirredundant clause of $F$,
it is contained in every minimal CNF formula equivalent to $F$.

\end{lemma}

\proof Let $B$ be a minimal formula that is equivalent to $F$. By
Property~\ref{minimal-resolution}, $B \subseteq \rescn(F)$. If $B$ does not
contain $c$, this containment strengthens to $B \subseteq \rescn(F) \backslash
\{c\}$. A consequence of the equivalence between $B$ and $F$ is that $B$
implies every clause of $F$, including $c$. Since $B \models c$ and $B
\subseteq \rescn(F) \backslash \{c\}$, by monotonicity $\rescn(F) \backslash
\{c\} \models c$ follows. This is the opposite of the assumed superirredundancy
of $c$.~\qed

This lemma provides a sufficient condition for a clause being in all minimal
formulae equivalent to the given one. Not a necessary one, though. A clause
that is not superirredundant may still be in all minimal formulae. A
counterexample only requires three clauses.

\begin{counterexample}
\label{not-all}

The first clause of
{} $F = \{a, \neg a \vee b, a \vee \neg b\}$
is superredundant but is in all minimal formulae equivalent to $F$.

\end{counterexample}

\proof Its first clause is $a$. It resolves with $\neg a \vee b$ into $b$. The
resolution closure of $F$ contains $a$, $b$ and $a \vee \neg b$. The first is
redundant as it is entailed by the second and the third. This proves it
superredundant.

Yet, the only minimal formula equivalent to $F$ is $F' = \{a, b\}$, which
contains $a$ in spite of its superredundancy. That $\{a,b\}$ is minimal is
proved by the superirredundancy of its clauses: its resolution closure is
$\{a,b\}$ itself since its clauses do not resolve; none of the two clauses is
redundant in it.~\qed


The proof of this counterexample relies on the syntactic dependency of
superredundancy: $a$ is superredundant in
{} $\{a, \neg a \vee b, a \vee \neg b\}$
but superirredundant in its minimal equivalent formula $\{a,b\}$.

\begin{lemma}
\label{syntax-depend}

There exist two equivalent formulae such that:
a.\ both formulae contain the same clause, and
b.\ that clause is superredundant in one formula but not in the other.

\end{lemma}

\proof The formulae are
{} $F = \{a, \neg a \vee b, a \vee \neg b\}$
and
{} $F' = \{a, b\}$.

The resolution closure of the first is
{} $\rescn(F) = \{a, b, \neg a \vee b, a \vee \neg b\}$
since the only possible resolutions in $F$ are
{} $a, \neg a \vee b \vdash b$ and
{} $b, a \vee \neg b \vdash a$.
While
{} $\neg a \vee b$ and $a \vee \neg b$
have opposite literals, they would only resolve into a tautology. Since the
clauses of $F'$ do not resolve, its resolution closure is $F'$ itself:
$\rescn(F') = \{a,b\}$.

The clause $a$ is redundant in
{} $\rescn(F) = \{a, b, \neg a \vee b, a \vee \neg b\}$
since it is entailed by $b$ and $a \vee \neg b$. It is not redundant in
$\rescn(F') = \{a,b\}$ since it is not entailed by $b$.~\qed

Being dependent on the syntax of the formula, superredundancy and
superirredundancy differ from every condition that is independent of the
syntax, such as:

\begin{itemize}

\item redundancy in the set of prime implicates, which is also employed in
formula minimization~\cite{hamm-koga-93};

\item essentiality of prime implicates, defined as containment of a prime
implicate in all prime CNFs equivalent to the formula~\cite{hamm-koga-93}

\item presence in all minimal CNF formulae equivalent to $F$.

\end{itemize}

These conditions are independent on the syntax since the prime implicates, the
prime equivalent CNFs and the minimal equivalent formulae are the same for two
equivalent formulae. Being independent on the syntax, they are not the same as
superredundancy or superirredundancy, which are proved dependent on the syntax
by Lemma~\ref{syntax-depend}.



Lemma~\ref{all} does not contradict the inequality of superirredundancy and
presence in all minimal equivalent formulae. It only proves that a
superirredundant clause of a formula is in all minimal formulae equivalent to
it. Not the other way around. A clause may be in all minimal formulae while not
being superirredundant.

\begin{lemma}
\label{superirredundant-not-minimal}

There exists a formula that contains a superredundant clause
that is in all its minimal CNF equivalent formulae.

\end{lemma}

\proof Let $c$ be a clause that is superredundant in $F$ and superirredundant
in $F'$ with $F \equiv F'$. Such a condition is possible according to
Lemma~\ref{syntax-depend}.

Since $c$ is superirredundant in $F'$, it is contained in all minimal formulae
equivalent to $F'$. Since this formula is equivalent to $F$, their minimal
equivalent formulae are the same.~\qed

Superirredundancy in a formula is a strictly stronger condition than membership
in all its minimal equivalent formulae. It implies that, but is not implied.

This is to be kept in mind when superirredundancy is used as a precondition.
Some lemmas below prove that if a clause is superirredundant in a formula then
it is superirredundant in a similar formula. The precondition of such a result
is the superirredundancy of the clause; its membership in all minimal
equivalent formulae may not be enough. Superirredundancy is required.
Membership to all minimal equivalent formulae is a consequence of both the
premise and the conclusion, not a substitute of the premise.

An application of superirredundancy is to guarantee a formula to be minimal.


\begin{lemma}
\label{minimal}

If a formula contains only superirredundant clauses, it is minimal.

\end{lemma}

\proof Let $B$ be a minimal formula equivalent to $A$. By Lemma~\ref{all}, the
superirredundant clauses of $A$ are in all minimal formulae that are equivalent
to $A$. Therefore, $B$ contains all of them. If $A$ only comprises
superirredundant clauses, $B$ contains all of them: $A \subseteq B$. The only
case where $B$ could not be the same as $A$ is when this containment is strict,
but that would imply that $B$ is not minimal since $A$ is equivalent but
smaller.~\qed

If a formula is built so that all its clauses are superirredundant, it is
guaranteed to be minimal. Not the other way around. Rather the opposite: a
minimal formula may be made only of superredundant clauses. An example is
{} $\{\neg a \vee b, \neg b \vee c, \neg c \vee a\}$. Resolving the first two
clauses $\neg a \vee b$ and $\neg b \vee c$ generates $\neg a \vee c$.
Resolving the other pairs produces the opposite cycle of clauses
{} $\{\neg a \vee c, \neg c \vee b, \neg b\vee a\}$,
which is equivalent to the original. The clauses of the original are therefore
entailed by some of their resolution consequences. Yet, the original is
minimal.

\subsection{Equivalent conditions to superredundancy}
\label{equivalent-superredundancy}

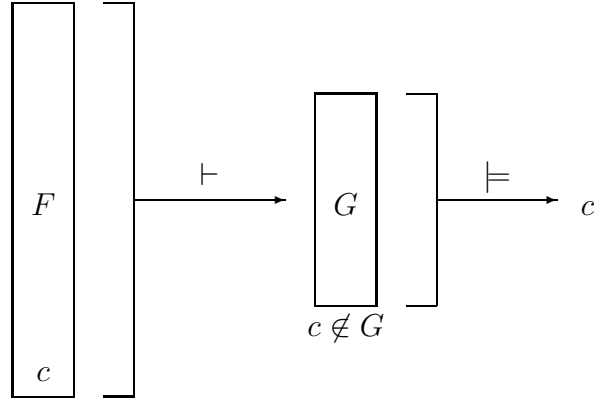
\begin{figure}
\begin{center}
\setlength{\unitlength}{5000sp}%
\begingroup\makeatletter\ifx\SetFigFont\undefined%
\gdef\SetFigFont#1#2#3#4#5{%
  \reset@font\fontsize{#1}{#2pt}%
  \fontfamily{#3}\fontseries{#4}\fontshape{#5}%
  \selectfont}%
\fi\endgroup%
\begin{picture}(2877,1974)(6439,-6973)
\thinlines
{\color[rgb]{0,0,0}\put(6451,-6961){\framebox(300,1950){}}
}%
{\color[rgb]{0,0,0}\put(7951,-6511){\framebox(300,1050){}}
}%
{\color[rgb]{0,0,0}\put(6901,-5011){\line( 1, 0){150}}
\put(7051,-5011){\line( 0,-1){1950}}
\put(7051,-6961){\line(-1, 0){150}}
}%
{\color[rgb]{0,0,0}\put(8401,-5461){\line( 1, 0){150}}
\put(8551,-5461){\line( 0,-1){1050}}
\put(8551,-6511){\line(-1, 0){150}}
}%
{\color[rgb]{0,0,0}\put(7051,-5986){\vector( 1, 0){750}}
}%
{\color[rgb]{0,0,0}\put(8551,-5986){\vector( 1, 0){600}}
}%
\put(6601,-6061){\makebox(0,0)[b]{\smash{{\SetFigFont{12}{24.0}
{\rmdefault}{\mddefault}{\updefault}{\color[rgb]{0,0,0}$F$}%
}}}}
\put(8101,-6061){\makebox(0,0)[b]{\smash{{\SetFigFont{12}{24.0}
{\rmdefault}{\mddefault}{\updefault}{\color[rgb]{0,0,0}$G$}%
}}}}
\put(9301,-6061){\makebox(0,0)[b]{\smash{{\SetFigFont{12}{24.0}
{\rmdefault}{\mddefault}{\updefault}{\color[rgb]{0,0,0}$c$}%
}}}}
\put(6601,-6886){\makebox(0,0)[b]{\smash{{\SetFigFont{12}{24.0}
{\rmdefault}{\mddefault}{\updefault}{\color[rgb]{0,0,0}$c$}%
}}}}
\put(8101,-6661){\makebox(0,0)[b]{\smash{{\SetFigFont{12}{24.0}
{\rmdefault}{\mddefault}{\updefault}{\color[rgb]{0,0,0}$c \not\in G$}%
}}}}
\put(8851,-5911){\makebox(0,0)[b]{\smash{{\SetFigFont{12}{24.0}
{\rmdefault}{\mddefault}{\updefault}{\color[rgb]{0,0,0}$\models$}%
}}}}
\put(7426,-5911){\makebox(0,0)[b]{\smash{{\SetFigFont{12}{24.0}
{\rmdefault}{\mddefault}{\updefault}{\color[rgb]{0,0,0}$\vdash$}%
}}}}
\end{picture}%
\nop{
F |- G |= c
}
\caption{An example of a superredundant clause}
\label{exemplifies}
\end{center}
\end{figure}

Superirredundancy differs from membership in all minimal equivalent formula,
but is easier than that to ensure. Simplicity is its motivation. Therefore, it
makes sense to simplify it further. The following lemma gives a number of
equivalent conditions, all based on $F$ deriving another formula $G$ that in
turn derives $c$. The first equivalent condition is exemplified by
Figure~\ref{exemplifies}. Instead of proving that $c$ is a consequence of
$\rescn(F) \backslash \{c\}$, these equivalent conditions allow proving the
existence of a possibly smaller formula $G$ that entails $c$.

\begin{lemma}
\label{equivalent}

A clause $c$ of a formula $F$ is superredundant if and only if a formula $G$
satisfying either one of the following conditions exists:

\begin{enumerate}

\item $F \vdash G$ and $G \models c$ where $c \not\in G$

\item $F \vdash G$ and $G \vdash c'$ where $c \not\in G$ and $c' \subseteq c$

\item $F \vdash G$ and $G \vdash c$ with $c \not\in G$ or
$F \vdash c'$ with $c' \subset c$

\item $F \vdash G$ and $G \vdash c$ with $c \not\in G$ or
$F \models c'$ with $c' \subset c$

\end{enumerate}

\end{lemma}

\proof Equivalence with the first condition is proved in the two directions. A
clause $c$ of $F$ is superredundant if and only if $\rescn(F) \backslash \{c\}
\models c$. If this condition is true, the claim holds with $G = \rescn(F)
\backslash \{c\}$, since $c$ is not in $\rescn(F) \backslash \{c\}$ by
construction and the definition of resolution closure implies $F \vdash c'$ for
every $c' \in \rescn(F)$. In the other direction, if $F \vdash G$ then $G
\subseteq \rescn(F)$. Since $c \not\in G$, this containment strengthens to $G
\subseteq \rescn(F) \backslash \{c\}$. Since $G \models c$, it follows
$\rescn(F) \backslash \{c\} \models c$ by monotonicity.

The first condition is equivalent to the second because $G \models c$ is the
same as $G \vdash c'$ with $c' \subseteq c$.

The second condition is proved equivalent to the third considering the two
directions separately. The second condition includes $c' \subseteq c$, which
comprises two cases: $c' = c$ and $c' \subset c$. If $c' = c$, the second
condition becomes $F \vdash G \vdash c$ with $c \not\in G$, the same as the
first alternative of the third condition. If $c' \subset c$, the second
condition is $F \vdash G \vdash c'$, which implies $F \vdash c'$ with $c'
\subset c$; this is the second alternative of the third condition. In the other
direction, the first alternative of the third condition is $F \vdash G \vdash
c$ with $c \not\in G$, which is the same as the second condition with $c' = c$.
The second alternative is $F \vdash c'$ with $c' \subset c$; the second
condition holds with $G = \{c'\}$.

Equivalence with the fourth condition holds because $F \vdash c'$ implies $F
\models c'$, and in the other direction $F \models c'$ implies $F \vdash c''$
with $c'' \subset c$, and the third equivalent condition holds with $c''$ in
place of $c'$.~\qed

A clause $c$ is superredundant if it follows from $F$ by a resolution proof
that contains a set of clauses $G$ sufficient to prove $c$. As such, $G$ is a
sort of ``cut'' in a resolution tree $F \vdash c$, separating $c$ from $F$.
This cut can be next to the root, next to the leaves, or somewhere in between.
In practice, it is useful at the first resolution steps (next to the leaves) or
at the last (next to the root).

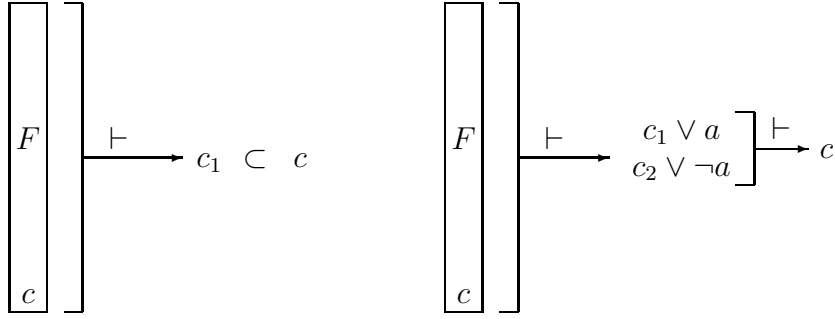
\begin{figure}
\begin{center}
\setlength{\unitlength}{5000sp}%
\begingroup\makeatletter\ifx\SetFigFont\undefined%
\gdef\SetFigFont#1#2#3#4#5{%
  \reset@font\fontsize{#1}{#2pt}%
  \fontfamily{#3}\fontseries{#4}\fontshape{#5}%
  \selectfont}%
\fi\endgroup%
\begin{picture}(4077,1554)(5119,-4663)
\thinlines
{\color[rgb]{0,0,0}\put(5131,-4651){\framebox(180,1530){}}
}%
{\color[rgb]{0,0,0}\put(7291,-4651){\framebox(180,1530){}}
}%
{\color[rgb]{0,0,0}\put(5401,-3121){\line( 1, 0){ 90}}
\put(5491,-3121){\line( 0,-1){1530}}
\put(5491,-4651){\line(-1, 0){ 90}}
}%
{\color[rgb]{0,0,0}\put(7561,-3121){\line( 1, 0){ 90}}
\put(7651,-3121){\line( 0,-1){1530}}
\put(7651,-4651){\line(-1, 0){ 90}}
}%
{\color[rgb]{0,0,0}\put(8731,-3661){\line( 1, 0){ 90}}
\put(8821,-3661){\line( 0,-1){360}}
\put(8821,-4021){\line(-1, 0){ 90}}
}%
{\color[rgb]{0,0,0}\put(5491,-3886){\vector( 1, 0){495}}
}%
{\color[rgb]{0,0,0}\put(7651,-3886){\vector( 1, 0){450}}
}%
{\color[rgb]{0,0,0}\put(8821,-3841){\vector( 1, 0){270}}
}%
\put(5221,-3841){\makebox(0,0)[b]{\smash{{\SetFigFont{12}{24.0}
{\rmdefault}{\mddefault}{\updefault}{\color[rgb]{0,0,0}$F$}%
}}}}
\put(5221,-4606){\makebox(0,0)[b]{\smash{{\SetFigFont{12}{24.0}
{\rmdefault}{\mddefault}{\updefault}{\color[rgb]{0,0,0}$c$}%
}}}}
\put(7381,-3841){\makebox(0,0)[b]{\smash{{\SetFigFont{12}{24.0}
{\rmdefault}{\mddefault}{\updefault}{\color[rgb]{0,0,0}$F$}%
}}}}
\put(7381,-4606){\makebox(0,0)[b]{\smash{{\SetFigFont{12}{24.0}
{\rmdefault}{\mddefault}{\updefault}{\color[rgb]{0,0,0}$c$}%
}}}}
\put(6571,-3931){\makebox(0,0)[b]{\smash{{\SetFigFont{12}{24.0}
{\rmdefault}{\mddefault}{\updefault}{\color[rgb]{0,0,0}$c$}%
}}}}
\put(8461,-3796){\makebox(0,0)[b]{\smash{{\SetFigFont{12}{24.0}
{\rmdefault}{\mddefault}{\updefault}{\color[rgb]{0,0,0}$c_1  \vee  a$}%
}}}}
\put(8461,-3976){\makebox(0,0)[b]{\smash{{\SetFigFont{12}{24.0}
{\rmdefault}{\mddefault}{\updefault}{\color[rgb]{0,0,0}$c_2  \vee  \neg a$}%
}}}}
\put(9181,-3886){\makebox(0,0)[b]{\smash{{\SetFigFont{12}{24.0}
{\rmdefault}{\mddefault}{\updefault}{\color[rgb]{0,0,0}$c$}%
}}}}
\put(5671,-3841){\makebox(0,0)[b]{\smash{{\SetFigFont{12}{24.0}
{\rmdefault}{\mddefault}{\updefault}{\color[rgb]{0,0,0}$\vdash$}%
}}}}
\put(7831,-3841){\makebox(0,0)[b]{\smash{{\SetFigFont{12}{24.0}
{\rmdefault}{\mddefault}{\updefault}{\color[rgb]{0,0,0}$\vdash$}%
}}}}
\put(8956,-3796){\makebox(0,0)[b]{\smash{{\SetFigFont{12}{24.0}
{\rmdefault}{\mddefault}{\updefault}{\color[rgb]{0,0,0}$\vdash$}%
}}}}
\put(6121,-3931){\makebox(0,0)[b]{\smash{{\SetFigFont{12}{24.0}
{\rmdefault}{\mddefault}{\updefault}{\color[rgb]{0,0,0}$c_1$}%
}}}}
\put(6346,-3931){\makebox(0,0)[b]{\smash{{\SetFigFont{12}{24.0}
{\rmdefault}{\mddefault}{\updefault}{\color[rgb]{0,0,0}$\subset$}%
}}}}
\end{picture}%
\nop{
                                F
F                                    c1 v a
  |- c1 c c                       |-         |- c
c                                    c2 v -a 
                                c
}
\caption{Superredundancy proved by the last step of resolution}
\label{figure-depicted}
\end{center}
\end{figure}

The next equivalent condition to superredundancy cuts the resolution tree at
its very last point, one step short of regenerating $c$. It comprises two
alternatives, depicted in Figure~\ref{figure-depicted}. They are due to the two
possibilities contemplated by the third condition of Lemma~\ref{equivalent}:
either $F$ implies a proper subset of $c$ or $c$ itself with a resolution proof
cut by a set $G$.

\begin{lemma}
\label{one-two}

A clause $c$ of a formula $F$ is superredundant if and only if either
{} $F \vdash c_1$ where $c_1 \subset c$
or
{} $F \vdash c_1 \vee a, c_2 \vee \neg a$
for some variable $a$ not occurring in $c$ and clauses $c_1$ and $c_2$ such
that $c = c_1 \vee c_2$.

\end{lemma}

\proof By Lemma~\ref{equivalent}, superredundancy is equivalent to $F \vdash G
\vdash c$ with $c \not\in G$ or $F \vdash c'$ with $c' \subset c$.

The second part of this condition is the same as $F \vdash c_1$ and $c_1
\subset c$ with $c_1 = c'$, the first alternative in the statement of the
lemma.

The first part $F \vdash G \vdash c$ with $c \not\in G$ is now proved to be the
same as the second alternative in the statement of the lemma:
{} $F \vdash c_1 \vee a, c_2 \vee \neg a$
where $a \not\in c$ and $c = c_1 \vee c_2$.

If $F \vdash G \vdash c$ with $c \not\in G$, since $c$ is not in $G$, the
derivation $G \vdash c$ contains at least a resolution step. Let $c'$ and $c''$
be the two clauses that resolve to $c$ in this derivation. Since they resolve
to $c$, they have the form $c' = c_1 \vee a$ and $c'' = c_2 \vee \neg a$ for
some variable $a$. Their resolution $c = c_1 \vee c_2$ does not contain $a$ by
Lemma~\ref{different}. These two clauses are obtained by resolution from $G$.
Since $F \vdash G$, they also derive by resolution from $F$.

In the other direction, $F \vdash c_1 \vee a, c_2 \vee \neg a$ implies
superredundancy. This is proved with $G = \{c_1 \vee a, c_2 \vee \neg a\}$. The
first condition $G \vdash c$ holds because $c_1 \vee a$ and $c_2 \vee \neg a$
resolve into $c$. The second condition $c \not\in G$ holds because $c$ does not
contain $a$ by assumption while both clauses of $G$ both do.~\qed

This lemma says that looking at all possible sets of clauses $G$ when checking
superredundancy is a waste of time. The sets comprising pairs of clauses
containing an opposite literal suffice. Their form provides an even further
simplification: they are obtained by splitting the clause in two and adding an
opposite literal to each. Superredundancy is the same as resolution deriving
either such a pair or a proper subset of the clause.

A clause is proved superredundant by such a splitting. Yet, proving
superredundancy is not the final goal. Proving the presence in all minimal
equivalent formula is. It follows from superirredundancy, not superredundancy.
The lemma helps in this. Instead of checking all possible sets of clauses $G$,
it allows concentrating only on the pairs obtained by splitting the clause.

When the clauses of $F$ do not resolve, the second alternative offered by
Lemma~\ref{one-two} never materializes: a clause is superredundant if and only
if it is a proper superset of a clause of $F$. This quite trivial
specialization looks pointless, but turns essential when paired with the
subsequent Lemma~\ref{set-value}.

\begin{lemma}
\label{no-resolution}

If no two clauses of $F$ resolve,
then
a clause of $F$ is superredundant
if and only if
$F$ contains a clause that is a strict subset of it.

\end{lemma}

\proof By Lemma~\ref{one-two}, $c \in F$ is superredundant if and only if $F
\vdash c_1$ with $c_1 \subset c$ or $F \vdash c_1 \vee a, c_2 \vee \neg a$ with
$c = c_1 \vee c_2$. The second condition implies $c_1 \vee a, c_2 \vee \neg a
\in F$ since the clauses of $F$ do not resolve; this contradicts the assumption
since these two clauses resolve. As a result, the only actual possibility is
the first: $F \vdash c_1$ with $c_1 \subset c$. Since the clauses of $F$ do not
resolve, $c_1$ cannot be the result of resolving clauses. Therefore, it is in
$F$.~\qed

\

\begin{figure}
\begin{center}
\setlength{\unitlength}{5000sp}%
\begingroup\makeatletter\ifx\SetFigFont\undefined%
\gdef\SetFigFont#1#2#3#4#5{%
  \reset@font\fontsize{#1}{#2pt}%
  \fontfamily{#3}\fontseries{#4}\fontshape{#5}%
  \selectfont}%
\fi\endgroup%
\begin{picture}(2277,1977)(6439,-6976)
\thinlines
{\color[rgb]{0,0,0}\put(6451,-6661){\line( 1, 0){300}}
}%
{\color[rgb]{0,0,0}\put(6451,-6961){\framebox(300,1950){}}
}%
{\color[rgb]{0,0,0}\put(6901,-6061){\line( 1, 0){150}}
\put(7051,-6061){\line( 0,-1){900}}
\put(7051,-6961){\line(-1, 0){150}}
}%
{\color[rgb]{0,0,0}\put(7051,-6811){\vector( 1, 0){225}}
}%
{\color[rgb]{0,0,0}\put(7126,-5011){\line( 1, 0){150}}
\put(7276,-5011){\line( 0,-1){1650}}
\put(7276,-6661){\line(-1, 0){150}}
}%
{\color[rgb]{0,0,0}\put(7351,-6661){\line( 0, 1){150}}
\put(7351,-6511){\line( 1, 0){900}}
\put(8251,-6511){\line( 0,-1){150}}
}%
{\color[rgb]{0,0,0}\put(7276,-5836){\vector( 1, 0){1275}}
}%
{\color[rgb]{0,0,0}\put(7801,-6511){\line( 0, 1){675}}
}%
\put(6601,-6061){\makebox(0,0)[b]{\smash{{\SetFigFont{12}{24.0}
{\rmdefault}{\mddefault}{\updefault}{\color[rgb]{0,0,0}$F$}%
}}}}
\put(6601,-6886){\makebox(0,0)[b]{\smash{{\SetFigFont{12}{24.0}
{\rmdefault}{\mddefault}{\updefault}{\color[rgb]{0,0,0}$c$}%
}}}}
\put(7801,-6886){\makebox(0,0)[b]{\smash{{\SetFigFont{12}{24.0}
{\rmdefault}{\mddefault}{\updefault}{\color[rgb]{0,0,0}$\resolve(F,c)$}%
}}}}
\put(7201,-6961){\makebox(0,0)[b]{\smash{{\SetFigFont{12}{24.0}
{\rmdefault}{\mddefault}{\updefault}{\color[rgb]{0,0,0}$\vdash$}%
}}}}
\put(8101,-5761){\makebox(0,0)[b]{\smash{{\SetFigFont{12}{24.0}
{\rmdefault}{\mddefault}{\updefault}{\color[rgb]{0,0,0}$\models$}%
}}}}
\put(8701,-5836){\makebox(0,0)[b]{\smash{{\SetFigFont{12}{24.0}
{\rmdefault}{\mddefault}{\updefault}{\color[rgb]{0,0,0}$c$}%
}}}}
\end{picture}%
\nop{
     ------------------------\                          .
F\ c                           \                        .
     ---\                        ||- c
          |- resolve(F,c) -----/
   c ---/
}
\caption{Superredundancy proved by immediate resolution consequences}
\label{figure-early}
\end{center}
\end{figure}
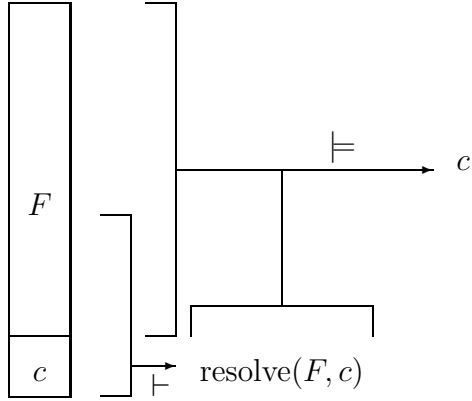

Formula $G$ of Lemma~\ref{equivalent} can be seen as a cut in a resolution tree
from $F$ to $c$. It separates all occurrences of $c \in F$ in the leaves from
$c$ in the root. Lemma~\ref{one-two} places the cut next to the root. The
following places it next to the leaves. It proves that resolving $c$ with
clauses of $F$ only is enough. The clauses obtained by these resolutions are
$\resolve(c,F)$, according to Definition~\ref{resolve-function}. This situation
is shown by Figure~\ref{figure-early}.

\begin{lemma}
\label{first-step}

A clause $c$ of $F$ is superredundant if and only if
{} $F \backslash \{c\} \cup \resolve(c,F) \models c$.

\end{lemma}

\proof The first equivalent condition to superredundancy offered by
Lemma~\ref{equivalent} is the existence of a set $G$ such that $F \vdash G$, $G
\models c$ and $c \not\in G$. The proof is composed of two parts: the first is
that $F \backslash \{c\} \cup \resolve(c,F)$ is such a set $G$ if it entails
$c$; the second is that if such a set $G$ exists, the derivation of $F \vdash
G$ can be rearranged so that $c$ resolves only with other clauses of $F$. The
first is almost trivial, the second is not because $c$ may resolve with clauses
obtained by resolution in $F \vdash G$. The rearranged derivation begins with a
batch of resolutions of $c$ with other clauses of $F$, and $c$ is then no
longer used. The resolvents of these first resolutions and the rest of $F$
makes the required set $G$.

\

If $F \backslash \{c\} \cup \resolve(c,F) \models c$, then $G = F \backslash
\{c\} \cup \resolve(c,F)$ proves $c$ superredundant: $F \vdash G$, $G \models
c$ and $c \not\in G$. The first condition $F \vdash G$ holds because the only
clauses of $G$ that are not in $F$ are the result of resolving $c \in F$ with a
clause of $F$; the second condition $G \models c$ holds by assumption; the
third condition is that $G$ does not contain $c$, and it holds because $G$ is
the union of $F \backslash \{c\}$ and $\resolve(c,F)$, where $F \backslash
\{c\}$ does not contain $c$ by construction and $\resolve(c,F)$ because
resolving a clause does not generate the clause itself by
Lemma~\ref{different}.

\

The rest of the proof is devoted to proving the converse:
{} $F \vdash G$, $G \models c$ and $c \not\in G$ imply
{} $F \backslash \{c\} \cup \resolve(c,F) \models c$.

The claim is proved by repeatedly modifying $G$ until it becomes a subset of $F
\backslash \{c\} \cup \resolve(c,F)$ while still maintaining its properties $F
\vdash G$, $G \models c$ and $c \not\in G$.

This process ends because a measure defined on the derivation $F \vdash G$
decreases until reaching zero. This measure is the almost-size of the
derivation $F \vdash G$ plus the rise of $c$ in it. Both are based on the size
of the subtrees of $F \vdash G$: each clause in the derivation is generated
independently of the others, and is therefore the root of its own tree.

The size of the derivation $F \vdash G$ is the number of clauses it contains.
Its almost-size is the number of clauses except the roots.

\begin{figure}
\begin{center}
\setlength{\unitlength}{5000sp}%
\begingroup\makeatletter\ifx\SetFigFont\undefined%
\gdef\SetFigFont#1#2#3#4#5{%
  \reset@font\fontsize{#1}{#2pt}%
  \fontfamily{#3}\fontseries{#4}\fontshape{#5}%
  \selectfont}%
\fi\endgroup%
\begin{picture}(1695,1644)(5116,-4843)
\thinlines
{\color[rgb]{0,0,0}\put(5311,-3301){\line( 3,-1){540}}
\put(5851,-3481){\line( 0,-1){180}}
\put(5851,-3661){\line(-3,-1){540}}
\put(5311,-3841){\line( 0, 1){540}}
}%
{\color[rgb]{0,0,0}\put(5311,-4291){\line( 3,-1){540}}
\put(5851,-4471){\line( 0,-1){180}}
\put(5851,-4651){\line(-3,-1){540}}
\put(5311,-4831){\line( 0, 1){540}}
}%
{\color[rgb]{0,0,0}\put(5941,-3211){\line( 0,-1){540}}
}%
{\color[rgb]{0,0,0}\put(5941,-3481){\line( 1, 0){ 90}}
}%
{\color[rgb]{0,0,0}\put(5941,-4201){\line( 0,-1){540}}
}%
{\color[rgb]{0,0,0}\put(5941,-4471){\line( 1, 0){ 90}}
}%
{\color[rgb]{0,0,0}\put(6121,-3391){\line( 0,-1){1170}}
\put(6121,-4561){\line( 3, 1){540}}
\put(6661,-4381){\line( 0, 1){810}}
\put(6661,-3571){\line(-3, 1){540}}
}%
\put(5806,-4336){\makebox(0,0)[b]{\smash{{\SetFigFont{12}{24.0}
{\rmdefault}{\mddefault}{\updefault}{\color[rgb]{0,0,0}$c$}%
}}}}
\put(5806,-3346){\makebox(0,0)[b]{\smash{{\SetFigFont{12}{24.0}
{\rmdefault}{\mddefault}{\updefault}{\color[rgb]{0,0,0}$c$}%
}}}}
\put(6796,-4021){\makebox(0,0)[b]{\smash{{\SetFigFont{12}{24.0}
{\rmdefault}{\mddefault}{\updefault}{\color[rgb]{0,0,0}$G$}%
}}}}
\put(5131,-4066){\makebox(0,0)[b]{\smash{{\SetFigFont{12}{24.0}
{\rmdefault}{\mddefault}{\updefault}{\color[rgb]{0,0,0}$F$}%
}}}}
\put(5536,-4606){\makebox(0,0)[b]{\smash{{\SetFigFont{12}{24.0}
{\rmdefault}{\mddefault}{\updefault}{\color[rgb]{0,0,0}$9$}%
}}}}
\put(5536,-3616){\makebox(0,0)[b]{\smash{{\SetFigFont{12}{24.0}
{\rmdefault}{\mddefault}{\updefault}{\color[rgb]{0,0,0}$5$}%
}}}}
\end{picture}%
\nop{
                    c |
                      |- ...
     (5 nodes) |- ... |
F                             |- ... G
                    c |
                      |- ...
     (9 nodes) |- ... |                          rise of c is 5+9 = 14
}
\caption{An example of the rise of a clause}
\label{figure-rise}
\end{center}
\end{figure}
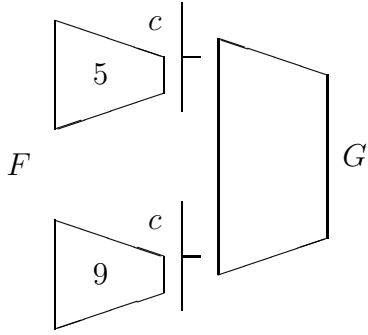

The derivation $F \vdash G$ may contain some resolutions of $c$ with other
clauses. The rise of an individual resolution of $c$ with a clause $c''$ in $F
\vdash G$ is the number of nodes in the tree rooted at $c''$ minus one. The
total rise of $c$ in $F \vdash G$ is the sum of all resolutions of $c$ in it.
It measures the overall distance of $c$ from the other leaves of the tree, its
elevation from the ground. Figure~\ref{figure-rise} shows an example.

Both the almost-size and the rise of $c$ are not negative. They are sums, each
addend being the size of a nonempty tree minus one; since each tree is not
empty, its size is at least one; the addend is at least zero. Their sums are at
least zero.

\

If $G$ is a subset of $F \cup \resolve(c,F)$, then $c \not\in G$ implies it is
also a subset of $F \backslash \{c\} \cup \resolve(c,F)$. Since $G$ implies
$c$, also does its superset $F \backslash \{c\} \cup \resolve(c,F)$. This is
the claim.

Otherwise, $G$ is not a subset of $F \cup \resolve(c,F)$. This means that $G$
contains a clause $c' \in G$ that is not in $F$ and is not the result of
resolving $c$ with a clause of $F$. Since $c'$ is not in $F$, it is the result
of resolving two clauses $c''$ and $c'''$. One of them may be $c$ or not; if it
is, the other one is not in $F$ and is therefore the result of resolving two
other clauses.

If neither $c''$ nor $c'''$ is equal to $c$, the modified set
{} $G' = G \backslash \{c'\} \cup \{c'', c'''\}$
has the same properties of $G$ that prove $c$ superredundant: $F \vdash G'$,
$G' \models c$ and $c \not\in G'$. Let $a$ and $b$ be the size of the trees
rooted in $c''$ and $c'''$ in $F \vdash G$. Since $F \vdash G$ has $c'$ as a
root, its almost-size includes $a+b+1-1 = a+b$. Instead, $F \vdash G'$ has
$c''$ and $c''$ as roots in place of $c'$; therefore, its almost-size includes
$(a-1) + (b-1) = a+b-2$. The almost-size of $F \vdash G'$ is smaller than $F
\vdash G$. The rise of $c$ is the same, since the resolutions of $c$ are the
same in the two derivations. Summarizing, almost-size decreases while rise
maintains its value.

If either $c''$ or $c'''$ is equal to $c$, the same set
{} $G' = G \backslash \{c'\} \cup \{c'', c'''\}$
does not work because it does not maintain $c \not\in G'$. Since the two cases
$c''=c$ and $c'''=c$ are symmetric, only the second is analyzed: $c' \in G$ is
generated by $c, c'' \vdash c'$ in $F \vdash G$. If $c''$ is in $F$, then $c'$
is in $\resolve(c,F)$ because it is the result of resolving $c$ with a clause
of $F$. Otherwise, $c''$ is a clause obtained by resolving two other clauses.

These two clauses resolve in $c''$, which resolves with $c$. Two resolutions,
two pairs of opposite literals. Let $l$ be the literal of $c$ that is negated
in $c''$ and $l'$ the literal that is resolved upon in the resolution that
generates $c''$. At least one of the two clauses that generate $c''$ contains
$\neg l$ since $c''$ does. At least means either one or both.

\

The first case is that both clauses that resolve in $c''$ contain $\neg l$.
Since they also contain $l'$ and $\neg l'$, they can be written $\neg l' \vee
\neg l \vee c_2$ and $l' \vee \neg l \vee c_3$. They resolve in $c'' = \neg l
\vee c_2 \vee c_3$. Since $c$ contains $l$, it can be written $l \vee c_1$. It
resolves with $c'' = \neg l \vee c_2 \vee c_3$ to $c' = c_1 \vee c_2 \vee c_3$.

\begin{center}
\setlength{\unitlength}{5000sp}%
\begingroup\makeatletter\ifx\SetFigFont\undefined%
\gdef\SetFigFont#1#2#3#4#5{%
  \reset@font\fontsize{#1}{#2pt}%
  \fontfamily{#3}\fontseries{#4}\fontshape{#5}%
  \selectfont}%
\fi\endgroup%
\begin{picture}(2682,1005)(4489,-4306)
\put(5491,-3481){\makebox(0,0)[rb]{\smash{{\SetFigFont{12}{24.0}
{\rmdefault}{\mddefault}{\updefault}{\color[rgb]{0,0,0}$l \vee c_1$}%
}}}}
\put(5491,-3751){\makebox(0,0)[rb]{\smash{{\SetFigFont{12}{24.0}
{\rmdefault}{\mddefault}{\updefault}{\color[rgb]{0,0,0}$\neg l' \vee \neg l \vee c_2$}%
}}}}
\put(5491,-4021){\makebox(0,0)[rb]{\smash{{\SetFigFont{12}{24.0}
{\rmdefault}{\mddefault}{\updefault}{\color[rgb]{0,0,0}$l' \vee \neg l \vee c_3$}%
}}}}
\thinlines
{\color[rgb]{0,0,0}\put(5581,-3706){\line( 4,-1){317.647}}
}%
{\color[rgb]{0,0,0}\put(5581,-3976){\line( 4, 1){317.647}}
}%
{\color[rgb]{0,0,0}\put(5581,-3436){\line( 6,-1){1481.351}}
}%
{\color[rgb]{0,0,0}\put(6706,-3841){\line( 4, 1){360}}
}%
\put(7156,-3751){\makebox(0,0)[lb]{\smash{{\SetFigFont{12}{24.0}
{\rmdefault}{\mddefault}{\updefault}{\color[rgb]{0,0,0}$c_1 \vee c_2 \vee c_3$}%
}}}}
\put(6301,-3886){\makebox(0,0)[b]{\smash{{\SetFigFont{12}{24.0}
{\rmdefault}{\mddefault}{\updefault}{\color[rgb]{0,0,0}$\neg l \vee c_2 \vee c_3$}%
}}}}
\end{picture}%
\nop{
   l v c1 -----------------------------\                       .
                                        |
  -l' v -l v c2 ---\                    +-- c1 v c2 v c3
                    +--- -l v c2 vc3 --/
   l' v -l v c3 ---/
}
\end{center}

A different derivation from the same clauses resolves $l \vee c_1$ with $\neg
l' \vee \neg l \vee c_2$, producing $c_1 \vee \neg l' \vee c_2$, and with $l'
\vee \neg l \vee c_3$, producing $c_1 \vee l' \vee c_3$. The produced clauses
$c_1 \vee \neg l' \vee c_2$ and $c_1 \vee l' \vee c_3$ resolve to $c_1 \vee c_2
\vee c_3$, the same conclusion of the original derivation. This is a valid
derivation, with an exception discussed below.

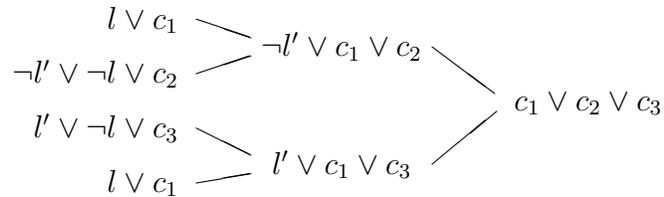
\begin{center}
\setlength{\unitlength}{5000sp}%
\begingroup\makeatletter\ifx\SetFigFont\undefined%
\gdef\SetFigFont#1#2#3#4#5{%
  \reset@font\fontsize{#1}{#2pt}%
  \fontfamily{#3}\fontseries{#4}\fontshape{#5}%
  \selectfont}%
\fi\endgroup%
\begin{picture}(2682,1005)(4489,-4306)
\put(5491,-3481){\makebox(0,0)[rb]{\smash{{\SetFigFont{12}{24.0}
{\rmdefault}{\mddefault}{\updefault}{\color[rgb]{0,0,0}$l \vee c_1$}%
}}}}
\put(5491,-3751){\makebox(0,0)[rb]{\smash{{\SetFigFont{12}{24.0}
{\rmdefault}{\mddefault}{\updefault}{\color[rgb]{0,0,0}$\neg l' \vee \neg l \vee c_2$}%
}}}}
\put(5491,-4021){\makebox(0,0)[rb]{\smash{{\SetFigFont{12}{24.0}
{\rmdefault}{\mddefault}{\updefault}{\color[rgb]{0,0,0}$l' \vee \neg l \vee c_3$}%
}}}}
\thinlines
{\color[rgb]{0,0,0}\put(6751,-3571){\line( 4,-3){309.600}}
}%
{\color[rgb]{0,0,0}\put(6751,-4156){\line( 6, 5){318.688}}
}%
{\color[rgb]{0,0,0}\put(5581,-3436){\line( 3,-1){270}}
}%
{\color[rgb]{0,0,0}\put(5581,-3706){\line( 3, 1){270}}
}%
{\color[rgb]{0,0,0}\put(5581,-3976){\line( 2,-1){270}}
}%
{\color[rgb]{0,0,0}\put(5581,-4246){\line( 6, 1){270}}
}%
\put(7156,-3886){\makebox(0,0)[lb]{\smash{{\SetFigFont{12}{24.0}
{\rmdefault}{\mddefault}{\updefault}{\color[rgb]{0,0,0}$c_1 \vee c_2 \vee c_3$}%
}}}}
\put(5491,-4291){\makebox(0,0)[rb]{\smash{{\SetFigFont{12}{24.0}
{\rmdefault}{\mddefault}{\updefault}{\color[rgb]{0,0,0}$l \vee c_1$}%
}}}}
\put(6301,-3616){\makebox(0,0)[b]{\smash{{\SetFigFont{12}{24.0}
{\rmdefault}{\mddefault}{\updefault}{\color[rgb]{0,0,0}$\neg l' \vee c_1 \vee c_2$}%
}}}}
\put(6301,-4201){\makebox(0,0)[b]{\smash{{\SetFigFont{12}{24.0}
{\rmdefault}{\mddefault}{\updefault}{\color[rgb]{0,0,0}$l' \vee c_1 \vee c_3$}%
}}}}
\end{picture}%
\nop{
   l v c1 ---------\                                             .
                    +--- c1 v -l' v c2 ---\                      .
  -l' v -l v c2 ---/                       |
                                           +--- c1 v c2 v c3
   l v c1 ---------\                       |
                    +--- c1 v  l' v c3 ---/
   l' v -l v c3 ---/
}
\end{center}

This derivation proves $F \vdash G'$ where
{} $G' = G \backslash
{}  \{c_1 \vee c_2 \vee c_3\} \cup
{}  \{c_1 \vee \neg l' \vee c_2, c_1 \vee l' \vee c_3\}$.
The only clause of $G$ that $G'$ does not contain is $c_1 \vee c_2 \vee c_3$,
which is implied by resolution from its clauses $c_1 \vee \neg l' \vee c_2$ and
$c_1 \vee l' \vee c_3$. Therefore, $G' \models G$. This implies $G' \models c$
since $G \models c$. Since $c_1 \vee \neg l' \vee c_2$ is obtained by resolving
two clauses over $l$, it does not contain $l$ by Lemma~\ref{different}. The
same applies to $c_1 \vee l' \vee c_3$. Since $c$ contains $l$, it is not equal
to either of these two clauses. It is not equal to any other clause of $G'$
either, since these are also clauses of $G$ and $c$ is not in $G$. This proves
that $G'$ has the same properties that prove $c$ superredundant: $F \vdash G'$,
$G' \models c$ and $c \not\in G'$.

Since $\neg l' \vee \neg l \vee c_2$ and $l' \vee \neg l \vee c_3$ are
generated by resolution from $F$, each is the root of a resolution tree. Let
$a$ and $b$ be their size. The almost-size of the original derivation $F \vdash
G$ includes $a+b+2$, that of $F \vdash G'$ has $a+b+2$ in its place.
Almost-size does not change.

The rise of resolving $c = l \vee c_1$ with $c'' = \neg l \vee c_2 \vee c_3$ in
the original derivation is one less the size of the tree rooted in $c''$. This
tree comprises $c''$ and the trees rooted in $\neg l' \vee \neg l \vee c_2$ and
$l' \vee \neg l \vee c_3$. Its size is therefore $a+b+1$. The rise of $c$ is
therefore $a+b$. The two resolutions of $c = l \vee c_1$ in the modified
derivation are with $\neg l' \vee \neg l \vee c_2$ and $l' \vee \neg l \vee
c_3$. The rise of $c$ in the first is the size of tree rooted in $\neg l' \vee
\neg l \vee c_2$ minus one: $a - 1$; the rise of $c$ in the second is $b - 1$.
Their sum is $a + b - 2$, which is strictly less than $a+b$.

In summary, switching from $F \vdash G$ to $F \vdash G'$ maintains the
almost-size and decreases the rise of $c$.

The exception mentioned above is that the new resolutions may generate
tautologies, which are not allowed. Since $\neg l' \vee \neg l \vee c_2$, $l'
\vee \neg l \vee c_3$ and $c_1 \vee c_2 \vee c_3$ are in the original
derivation, they are not tautologies. As a result, $c_1$, $c_2$ and $c_3$ do
not contain opposite literals, $c_2$ does not contain $l'$ and $c_3$ does not
contain $\neg l'$. The new clause $c_1 \vee \neg l' \vee c_2$ is tautological
only if $l'$ is in $c_1$, and $c_1 \vee l' \vee c_3$ only if $\neg l'$ is in
$c_1$. By symmetry, only the second case is considered: $c_1 = \neg l' \vee
c_1'$.

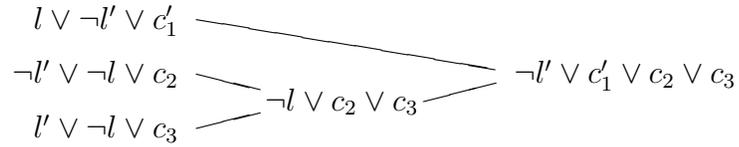
\begin{center}
\setlength{\unitlength}{5000sp}%
\begingroup\makeatletter\ifx\SetFigFont\undefined%
\gdef\SetFigFont#1#2#3#4#5{%
  \reset@font\fontsize{#1}{#2pt}%
  \fontfamily{#3}\fontseries{#4}\fontshape{#5}%
  \selectfont}%
\fi\endgroup%
\begin{picture}(2682,735)(4489,-4036)
\put(5491,-4021){\makebox(0,0)[rb]{\smash{{\SetFigFont{12}{24.0}
{\rmdefault}{\mddefault}{\updefault}{\color[rgb]{0,0,0}$l' \vee \neg l \vee c_3$}%
}}}}
\put(5491,-3481){\makebox(0,0)[rb]{\smash{{\SetFigFont{12}{24.0}
{\rmdefault}{\mddefault}{\updefault}{\color[rgb]{0,0,0}$l \vee \neg l' \vee c_1'$}%
}}}}
\put(5491,-3751){\makebox(0,0)[rb]{\smash{{\SetFigFont{12}{24.0}
{\rmdefault}{\mddefault}{\updefault}{\color[rgb]{0,0,0}$\neg l' \vee \neg l \vee c_2$}%
}}}}
\thinlines
{\color[rgb]{0,0,0}\put(5581,-3706){\line( 4,-1){317.647}}
}%
{\color[rgb]{0,0,0}\put(5581,-3976){\line( 4, 1){317.647}}
}%
{\color[rgb]{0,0,0}\put(5581,-3436){\line( 6,-1){1481.351}}
}%
{\color[rgb]{0,0,0}\put(6706,-3841){\line( 4, 1){360}}
}%
\put(6301,-3886){\makebox(0,0)[b]{\smash{{\SetFigFont{12}{24.0}
{\rmdefault}{\mddefault}{\updefault}{\color[rgb]{0,0,0}$\neg l \vee c_2 \vee c_3$}%
}}}}
\put(7156,-3751){\makebox(0,0)[lb]{\smash{{\SetFigFont{12}{24.0}
{\rmdefault}{\mddefault}{\updefault}{\color[rgb]{0,0,0}$\neg l' \vee c_1' \vee c_2 \vee c_3$}%
}}}}
\end{picture}%
\nop{
   l v -l' v c1'-----------------------\                       .
                                        |
  -l' v -l v c2 ---\                    +-- -l' v c1' v c2 v c3
                    +--- -l v c2 vc3 --/
   l' v -l v c3 ---/
}
\end{center}

An alternative derivation resolves $l \vee \neg l' \vee c_1$ with $\neg l' \vee
\neg l \vee c_2$, resulting in $\neg l' \vee c_1 \vee c_2$, a subset of the
original result $\neg l' \vee c_1 \vee c_2 \vee c_2$.

\begin{center}
\setlength{\unitlength}{5000sp}%
\begingroup\makeatletter\ifx\SetFigFont\undefined%
\gdef\SetFigFont#1#2#3#4#5{%
  \reset@font\fontsize{#1}{#2pt}%
  \fontfamily{#3}\fontseries{#4}\fontshape{#5}%
  \selectfont}%
\fi\endgroup%
\begin{picture}(2682,735)(4489,-4036)
\put(5491,-4021){\makebox(0,0)[rb]{\smash{{\SetFigFont{12}{24.0}
{\rmdefault}{\mddefault}{\updefault}{\color[rgb]{0,0,0}$l' \vee \neg l \vee c_3$}%
}}}}
\put(5491,-3481){\makebox(0,0)[rb]{\smash{{\SetFigFont{12}{24.0}
{\rmdefault}{\mddefault}{\updefault}{\color[rgb]{0,0,0}$l \vee \neg l' \vee c_1'$}%
}}}}
\put(5491,-3751){\makebox(0,0)[rb]{\smash{{\SetFigFont{12}{24.0}
{\rmdefault}{\mddefault}{\updefault}{\color[rgb]{0,0,0}$\neg l' \vee \neg l \vee c_2$}%
}}}}
\thinlines
{\color[rgb]{0,0,0}\put(5581,-3436){\line( 3,-1){270}}
}%
{\color[rgb]{0,0,0}\put(5581,-3706){\line( 3, 1){270}}
}%
\put(6301,-3616){\makebox(0,0)[b]{\smash{{\SetFigFont{12}{24.0}
{\rmdefault}{\mddefault}{\updefault}{\color[rgb]{0,0,0}$\neg l' \vee c_1' \vee c_2$}%
}}}}
\end{picture}%
\nop{
  l v -l' v c1' ---\                                          .
                    +--- -l' v c1' v c2
  -l' v -l v c2 ---/
                                                              .
   l' v -l v c3
}
\end{center}

Since $\neg l' \vee c_1' \vee c_2 \subseteq \neg l' \vee c_1' \vee c_2 \vee
c_3$, it holds $\neg l' \vee c_1' \vee c_2 \models \neg l' \vee c_1' \vee c_2
\vee c_3$, which implies $G' \models G$ where
{} $G' = G \backslash
{}	\{\neg l' \vee c_1' \vee c_2 \vee c_3\} \cup
{}	\{\neg l' \vee c_1' \vee c_2\}$,
which in turns implies $G' \models c$ since $G \models c$. This set $G'$ is
still obtained by $F$ by resolution. It does not contain $c$ because $c \not\in
G$ and the added clause $\neg l' \vee c_1' \vee c_2$ is not $c$. It is not $c$
because it does not contain $l$ while $c$ does, and it does not contain $l$ by
Lemma~\ref{different} because it is the result of resolving two clauses over
$l$. This proves that $G'$ inherit from $G$ all properties that prove $c$
superredundant: $F \vdash G'$, $G' \models c$ and $c \not\in G'$.

If the tree rooted in $\neg l' \vee \neg l \vee c_2$ has size $a$ and the tree
rooted in $l' \vee \neg l \vee c_3$ has size $b$, the derivation $F \vdash G$
includes $a+b+2$ in its almost-size. The derivation $F \vdash G'$ has $a+1$ in
its place, a decrease in almost-size. The rise of this resolution of $c$ in $F
\vdash G$ is $a+b$. In $F \vdash G'$, it is $a-1$. Both almost-size and rise of
$c$ decrease.

\

The second case is that only one of the clauses that resolve into $c''$
contains $\neg l$. Their resolution literal is still denoted $l'$; therefore,
they can be written $\neg l' \vee \neg l \vee c_2$ and $l' \vee c_3$. The
result of resolving them is $c'' = \neg l \vee c_2 \vee c_3$, which resolves
with $c = l \vee c_1$ to generate $c' = c_1 \vee c_2 \vee c_3$.

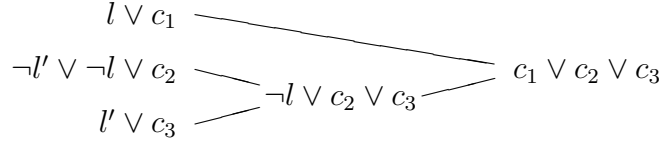
\begin{center}
\setlength{\unitlength}{5000sp}%
\begingroup\makeatletter\ifx\SetFigFont\undefined%
\gdef\SetFigFont#1#2#3#4#5{%
  \reset@font\fontsize{#1}{#2pt}%
  \fontfamily{#3}\fontseries{#4}\fontshape{#5}%
  \selectfont}%
\fi\endgroup%
\begin{picture}(2592,735)(4579,-4036)
\put(7156,-3751){\makebox(0,0)[lb]{\smash{{\SetFigFont{12}{24.0}
{\rmdefault}{\mddefault}{\updefault}{\color[rgb]{0,0,0}$c_1 \vee c_2 \vee c_3$}%
}}}}
\put(5491,-3481){\makebox(0,0)[rb]{\smash{{\SetFigFont{12}{24.0}
{\rmdefault}{\mddefault}{\updefault}{\color[rgb]{0,0,0}$l \vee c_1$}%
}}}}
\put(5491,-3751){\makebox(0,0)[rb]{\smash{{\SetFigFont{12}{24.0}
{\rmdefault}{\mddefault}{\updefault}{\color[rgb]{0,0,0}$\neg l' \vee \neg l \vee c_2$}%
}}}}
\put(5491,-4021){\makebox(0,0)[rb]{\smash{{\SetFigFont{12}{24.0}
{\rmdefault}{\mddefault}{\updefault}{\color[rgb]{0,0,0}$l' \vee c_3$}%
}}}}
\thinlines
{\color[rgb]{0,0,0}\put(5581,-3706){\line( 4,-1){317.647}}
}%
{\color[rgb]{0,0,0}\put(5581,-3976){\line( 4, 1){317.647}}
}%
{\color[rgb]{0,0,0}\put(5581,-3436){\line( 6,-1){1481.351}}
}%
{\color[rgb]{0,0,0}\put(6706,-3841){\line( 4, 1){360}}
}%
\put(6301,-3886){\makebox(0,0)[b]{\smash{{\SetFigFont{12}{24.0}
{\rmdefault}{\mddefault}{\updefault}{\color[rgb]{0,0,0}$\neg l \vee c_2 \vee c_3$}%
}}}}
\end{picture}%
\nop{
   l v c1 -----------------------------\                          .
                                        |
  -l' v -l v c2 ---\                    +-- c1 v c2 v c3
                    +--- -l v c2 vc3 --/
   l' v c3 --------/
}
\end{center}

Since $l \vee c_1$ and $\neg l' \vee \neg l \vee c_2$ oppose on $l$, they
resolve. The result is $c_1 \vee \neg l' \vee c_2$, which resolves with $l'
\vee c_3$ into $c_1 \vee c_2 \vee c_3$. The same three clauses produce the same
clause. This is a valid derivation with an exception discussed below.

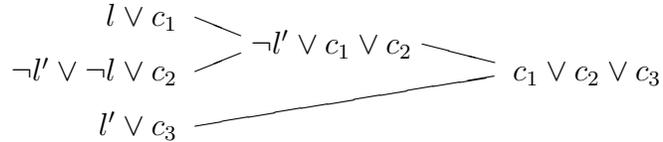
\begin{center}
\setlength{\unitlength}{5000sp}%
\begingroup\makeatletter\ifx\SetFigFont\undefined%
\gdef\SetFigFont#1#2#3#4#5{%
  \reset@font\fontsize{#1}{#2pt}%
  \fontfamily{#3}\fontseries{#4}\fontshape{#5}%
  \selectfont}%
\fi\endgroup%
\begin{picture}(2592,735)(4579,-4036)
\put(7156,-3751){\makebox(0,0)[lb]{\smash{{\SetFigFont{12}{24.0}
{\rmdefault}{\mddefault}{\updefault}{\color[rgb]{0,0,0}$c_1 \vee c_2 \vee c_3$}%
}}}}
\put(5491,-3481){\makebox(0,0)[rb]{\smash{{\SetFigFont{12}{24.0}
{\rmdefault}{\mddefault}{\updefault}{\color[rgb]{0,0,0}$l \vee c_1$}%
}}}}
\put(5491,-3751){\makebox(0,0)[rb]{\smash{{\SetFigFont{12}{24.0}
{\rmdefault}{\mddefault}{\updefault}{\color[rgb]{0,0,0}$\neg l' \vee \neg l \vee c_2$}%
}}}}
\put(5491,-4021){\makebox(0,0)[rb]{\smash{{\SetFigFont{12}{24.0}
{\rmdefault}{\mddefault}{\updefault}{\color[rgb]{0,0,0}$l' \vee c_3$}%
}}}}
\thinlines
{\color[rgb]{0,0,0}\put(5581,-3436){\line( 5,-2){225}}
}%
{\color[rgb]{0,0,0}\put(5581,-3706){\line( 5, 2){225}}
}%
{\color[rgb]{0,0,0}\put(6706,-3571){\line( 4,-1){360}}
}%
{\color[rgb]{0,0,0}\put(5581,-3976){\line( 6, 1){1481.351}}
}%
\put(6256,-3616){\makebox(0,0)[b]{\smash{{\SetFigFont{12}{24.0}
{\rmdefault}{\mddefault}{\updefault}{\color[rgb]{0,0,0}$\neg l' \vee c_1 \vee c_2$}%
}}}}
\end{picture}%
\nop{
   l v c1 ---------\                                              .
                    +--- c1 v -l' v c2 ---\                       .
  -l' v -l v c2 ---/                       +--- c1 v c2 v c3
                                           |
   l' v c3 -------------------------------/
}
\end{center}

This derivation proves $F \vdash G'$ where
{} $G' = G \backslash \{c_1 \vee c_2 \vee c_3\}
{}       \cup \{c_1 \vee \neg l' \vee c_2, l' \vee c_3\}$.
The only clause of $G$ that $G'$ does not contain is $c_1 \vee c_2 \vee c_3$,
but this is the result of resolving $c_1 \vee \neg l' \vee c_2$ and $l' \vee
c_3$, two clauses of $G'$. As a result, $G' \models G$. This proves $G' \models
c$ since $G \models c$. Finally, $c$ is not in $G'$. Since $c \not\in G$,
suffices to prove that $c$ is not any of the two clauses that $G'$ contains
while $G$ does not. This is the case because $c = l \vee c_1$ contains $l$
while the two clauses does not. The original derivation contains $c_1 \vee c_2
\vee c_3$ as the result of resolving two clauses over $l$;
Lemma~\ref{different} tells that $l$ is not in $c_1 \vee c_2 \vee c_3$. As a
result, $l$ is in
{} $c_1 \vee \neg l' \vee c_2$ or $l' \vee c_3$
only if either $l = \neg l'$ or $l = l'$. That implies that $\neg l \vee c_2
\vee c_3$ contains $l'$ while it is obtained in the original derivation by
resolving two clauses over $l'$, contradicting Lemma~\ref{different}. This
proves that $G'$ inherits all properties that prove $c$ superredundant: $F
\vdash G'$, $G' \models c$ and $c \not\in G'$.

Since $\neg l' \vee \neg l \vee c_2$ and $l' \vee c_3$ are obtained by
resolution from $F$, each is the root of a resolution tree. Let $a$ and $b$ be
their size.

The almost-size of $F \vdash G$ includes a part for the tree rooted in $c_1
\vee c_2 \vee c_3$; that part is $a+b+2$. The derivation $F \vdash G'$ is the
same except that it has the parts for $c_1 \vee \neg l' \vee c_2$ and $l' \vee
c_3$ instead: $a+1$ and $b-1$. The almost-size decreases from $a+b+2$ to $a+b$.

The rise of the resolution of $c$ with $\neg l \vee c_2 \vee c_3$ in $F \vdash
G$ is the size of the tree rooted in $\neg l \vee c_2 \vee c_3$; this tree
contains $a+b+1$ nodes; the rise in $F \vdash G$ is therefore $a+b$. The rise
of the resolution of $c$ with $\neg l' \vee \neg l \vee c_2$ in $F \vdash G'$
is instead the size of tree rooted in $\neg l' \vee \neg l \vee c_2$ minus one:
$a-1$. This is smaller than $a+b$ since $b$ is nonnegative.

Summarizing, the change in the derivation strictly decreases both its overall
size and its rise of $c$.

The exception that makes the new derivation invalid is when the new clause $c_1
\vee \neg l' \vee c_2$ is a tautology. Valid resolution derivations do not
contain tautologies. This also applies to the original one. Since $c_1 \vee c_2
\vee c_3$ is not a tautology, $c_1 \vee c_2$ is neither. Since $\neg l' \vee
\neg l \vee c_2$ is not a tautology, $c_2$ does not contain $l'$. The new
clause $c_1 \vee \neg l' \vee c_2$ is a tautology only if $l'$ is in $c_1$.
Equivalently, $c_1 = l' \vee c_1'$ for some $c_1'$.

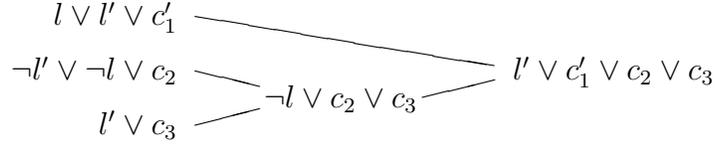
\begin{center}
\setlength{\unitlength}{5000sp}%
\begingroup\makeatletter\ifx\SetFigFont\undefined%
\gdef\SetFigFont#1#2#3#4#5{%
  \reset@font\fontsize{#1}{#2pt}%
  \fontfamily{#3}\fontseries{#4}\fontshape{#5}%
  \selectfont}%
\fi\endgroup%
\begin{picture}(2592,735)(4579,-4036)
\put(5491,-4021){\makebox(0,0)[rb]{\smash{{\SetFigFont{12}{24.0}
{\rmdefault}{\mddefault}{\updefault}{\color[rgb]{0,0,0}$l' \vee c_3$}%
}}}}
\put(5491,-3481){\makebox(0,0)[rb]{\smash{{\SetFigFont{12}{24.0}
{\rmdefault}{\mddefault}{\updefault}{\color[rgb]{0,0,0}$l \vee l' \vee c_1'$}%
}}}}
\put(5491,-3751){\makebox(0,0)[rb]{\smash{{\SetFigFont{12}{24.0}
{\rmdefault}{\mddefault}{\updefault}{\color[rgb]{0,0,0}$\neg l' \vee \neg l \vee c_2$}%
}}}}
\put(7156,-3751){\makebox(0,0)[lb]{\smash{{\SetFigFont{12}{24.0}
{\rmdefault}{\mddefault}{\updefault}{\color[rgb]{0,0,0}$l' \vee c_1' \vee c_2 \vee c_3$}%
}}}}
\thinlines
{\color[rgb]{0,0,0}\put(5581,-3706){\line( 4,-1){317.647}}
}%
{\color[rgb]{0,0,0}\put(5581,-3976){\line( 4, 1){317.647}}
}%
{\color[rgb]{0,0,0}\put(5581,-3436){\line( 6,-1){1481.351}}
}%
{\color[rgb]{0,0,0}\put(6706,-3841){\line( 4, 1){360}}
}%
\put(6301,-3886){\makebox(0,0)[b]{\smash{{\SetFigFont{12}{24.0}
{\rmdefault}{\mddefault}{\updefault}{\color[rgb]{0,0,0}$\neg l \vee c_2 \vee c_3$}%
}}}}
\end{picture}%
\nop{
   l v l' v c1' -----------------------\                          .
                                        |
  -l' v -l v c2 ---\                    +-- l' v c1' v c2 v c3
                    +--- -l v c2 vc3 --/
   l' v c3 --------/
}
\end{center}

The root of the derivation is $l' \vee c_1 \vee c_2 \vee c_3$ in this case.
This is a superset of its grandchild $l' \vee c_3$. Since subclauses imply
superclauses, $G' = G \backslash \{c_1 \vee c_2 \vee c_3\} \cup \{l' \vee
c_3\}$ implies $G$. By transitivity, it implies $c$.

The clause $l' \vee c_1' \vee c_2 \vee c_3$ does not contain $l$ by
Lemma~\ref{different} because it is the result of resolving two clauses upon
$l$ in the original derivation. As a result, its subset $l' \vee c_3$ does not
contain $l$ either. It therefore differs from $c$, which contains $l$. This
implies $c \not\in G'$ since $c \not\in G$ and $c \not= l' \vee c_3$.

Since $F \vdash G$ includes the derivation of $l' \vee c_3$, also $F \vdash G'$
holds.

All three properties of $G$ are inherited by $G'$: $F \vdash G'$, $G' \models
c$ and $c \not\in G'$.

Let $a$ and $b$ be the size of the trees rooted at $\neg l' \vee \neg l \vee
c_2$ and $l' \vee c_3$. The almost-size of $F \vdash G$ has a component for its
root $l' \vee c_1 \vee c_2 \vee c_3$ of value $a+b+2$. The derivation $F \vdash
G'$ has the root $l' \vee c_3$ in its place, contributing only $a-1$ to the
almost-size. The almost-size decreases. The rise of $c$ also decreases. In the
original derivation $c$ resolves with $\neg l \vee c_2 \vee c_3$, contributing
$a+b$ to the overall rise. In $F \vdash G'$ this resolution is absent,
contributing $0$ to the overall rise. Since $a$ is the size of a non-empty
tree, it is greater than zero. The same holds for $b$. The rise of $c$
decreases by $a+b$, which is at least $2$.

\

All of this proves that if $G$ is not a subset of $F \cup \resolve(c,F)$ it can
be changed to decrease its overall measure while still proving $c$
superredundant. This change preserves $F \vdash G$, $G \models c$ and $c
\not\in G$ and strictly decreases the measure of $F \vdash G$, defined as the
sum of its almost-size and rise of $c$.

This change can be iterated as long as $G$ is not a subset of $F \cup
\resolve(c,F)$. It terminates because the measure strictly decreases at each
step but is never negative as proved above. When it terminates, $G$ is a subset
of $F \cup \resolve(c,F)$ because otherwise it could be iterated. Since $c
\not\in G$ is one of the preserved properties, $G$ is also a subset of $F
\backslash \{c\} \cup \resolve(c,F)$. Since $G$ implies $c$, its superset $F
\backslash \{c\} \cup \resolve(c,F)$ implies $c$ too. This is the claim.~\qed

Lemma~\ref{first-step} allows for a simple algorithm for checking
superredundancy: resolve $c$ with all other clauses of $F$ and then remove it.
If $c$ is still entailed, it is superredundant. This proves that checking
superredundancy is polynomial-time for example in the Horn and Krom cases (the
latter is that all clauses contain two literals at most). More generally, it is
polynomial in all restrictions that are closed under resolution and where
inference can be checked in polynomial time.

\begin{theorem}
\label{horn-krom}

Checking superredundancy is polynomial-time in the Horn and Krom cases.

\end{theorem}

\proof The clauses in $\resolve(F,c)$ can be generated in polynomial time by
resolving each clause of $F$ with $c$. The result is still Horn or Krom because
resolving two Horn clauses or two Krom clauses respectively generates a Horn
and Krom clause. Checking $F \backslash \{c\} \cup \resolve(F,c) \models c$
therefore only takes polynomial time.~\qed

If $c$ is a single literal $l$, it only resolves with clauses containing $\neg
l$. The condition is very simple in this case. 

\begin{theorem}
\label{first-step-literal}

A single-literal clause $l$ of $F$ is superredundant if and only if

\[
\{c \in F \mid \neg l \not\in c ,~ c \not= l\} \cup
\{c \mid c \vee \neg l \in F\} \models l
\]

\end{theorem}

\proof When a clause $c$ comprises a single literal $l$, Lemma~\ref{first-step}
equates its superredundancy to $F \backslash \{l\} \cup \resolve(F,l) \models
l$. The first part $F \backslash \{l\}$ of the formula comprises clauses that
resolve with $l$ and clauses that do not:

\[
F \backslash \{l\}
=
\{c \in F \mid \neg l \not\in c ,~ c \not= l \}
\cup
\{c \in F \mid \neg l \in c\}
\]

If $c$ is in the second set, then $\resolve(F,l)$ contains $\resolve(c,l) = c
\backslash \{\neg l\}$, which implies $c$. Therefore,
{} $F \backslash \{l\} \cup \resolve(F,l)$
is equivalent to
{} $\{c \in F \mid \neg l \not\in c ,~ c \not= l \} \cup \resolve(F,l)$.
The clauses that resolve with $l$ are all of the form $c \vee \neg l$, and the
result of the resolution is $c$. Therefore, $\resolve(F,l)$ can be rewritten as
$\{c \mid c \vee \neg l \in F\}$. This proves the claim.~\qed

This theorem shows how to check the superredundancy of a single-literal clause
of a formula. All it takes is a simple transformation of the formula: the unit
clause $l$ is removed, and the literal $\neg l$ is deleted from all clauses
containing it. If what remains imply $l$, then $l$ is superredundant.

This condition can be further simplified when not only the clause to be checked
is a literal, but its converse does not even occur in the formula. Such
literals are usually called pure in the automated reasoning
field~\cite{joha-05}.

\begin{theorem}
\label{pure}

If $\neg l$ does not occur in $F$, the single-literal clause $l$ of $F \cup
\{l\}$ is superredundant if and only if $F \models l$.

\end{theorem}

\proof A clause $l$ of $F \cup \{l\}$ is superredundant if and only if $F'
\models l$ where
{} $F' = \{c \in F \cup \{l\} \mid \neg l \not\in c ,~ c \not= l\} \cup
{}       \{c \mid c \vee \neg l \in F \cup \{l\}\} \models l$
thanks to Theorem~\ref{first-step-literal} applied to $F \cup \{l\}$. If
$\neg l$ does not occur in $F$, then $F'$ simplifies as follows.

\begin{eqnarray*}
F' &=&
\{c \in F \cup \{l\} \mid \neg l \not\in c ,~ c \not= l\} \cup
\{c \mid c \vee \neg l \in F \cup \{l\}\} \\
&=&
\{c \in F \cup \{l\} \mid c \not= l\} \\
&=&
F
\end{eqnarray*}

This proves that $l$ is superredundant in $F \cup \{l\}$
if and only if $F \models l$ in this case.~\qed

The formula where $l$ is superredundant is $F \cup \{l\}$. Alternatively, $l$
is superredundant in $F$ if and only if $F \backslash \{l\} \models l$,
provided that $\neg l$ does not occur in $F$.

Talking about pure literals, another equivalent condition exists. A variable
may always occur with the same sign in a formula; if so, the clauses containing
it are irrelevant to the superredundancy of the others.

\begin{lemma}
\label{pure-other}

If $\neg l$ does not occur in $F$, $l \not\in c$ and $l \in c'$ for some clause
$c'$, the clause $c$ of $F$ is superredundant if and only if it is
superredundant in $F \backslash \{c'\}$.

\end{lemma}

\proof No derivation $F \vdash c$ involves $c'$. This is proved by
contradiction. Clause $c'$ may resolve with other clauses of $F$, but the
resolving variable cannot be $l$ since no clause of $F$ contains $\neg l$.
Therefore, the resulting clauses all contain $l$. The same applies to them:
they may resolve, but the result contains $l$. Inductively, this proves that
$l$ is also in the root of the derivation tree. The root is $c$, which does not
contain $l$ by assumption. This contradiction proves that $c'$ is not involved
in $F \vdash c$.

A derivation $F \vdash G \vdash c$ is a resolution tree with leaves $F$ and
root $c$. It is a particular case of $F \vdash c$. Therefore, it does not
contain $c'$. As a result, it can be rewritten $F \backslash \{c'\} \vdash G
\vdash c$.

A derivation $F \vdash c''$ with $c'' \subset c$ does not contain $c'$ because
$c''$ does not contain $l$. Therefore, $F \vdash c''$ is the same as $F
\backslash \{c\} \vdash c''$ for every subset $c''$ of $c$.

The third equivalent condition to superredundancy in Lemma~\ref{equivalent} is:
$F \vdash G \vdash c$ with $c \not\in G$ or $F \vdash c''$ with $c'' \subset
c$. These conditions are respectively equivalent to $F \backslash \{c'\} \vdash
G \vdash c$ and $F \backslash \{c'\} \vdash c''$, the third equivalent
condition to the superredundancy of $c$ in $F \backslash \{c'\}$ of
Lemma~\ref{equivalent}.~\qed

A particular case meeting the assumption of the lemma is when a variable only
occurs in one clause. The lemma specializes as follows.

\begin{corollary}
\label{only-one}

If a variable occurs in $F$ only in the clause $c$, then $c' \not= c$ is
superredundant in $F$ if and only if it is superredundant in $F \backslash
\{c\}$.

\end{corollary}

\subsection{Sufficient conditions to superredundancy}
\label{sufficient-superredundancy}

A number of equivalent conditions have been provided. Time to turn to
sufficient conditions. The next corollary shows a sufficient condition to 
superredundancy. The following results are about superirredundancy.

\begin{corollary}
\label{contains}

If $F \models c'$ and $c' \subset c \in F$,
then $c$ is superredundant in $F$.

\end{corollary}

\proof Immediate consequence of the fourth equivalent condition of
Lemma~\ref{equivalent}.~\qed

While proving superredundancy is sometimes useful, its main aim is to prove a
clause in all minimal equivalent formulae, which is the case if it is
superirredundant. This is why much effort is devoted to proving
superirredundancy.

A way to prove superirredundancy is by first simplifying the formula and then
proving superirredundancy on the result. Of course, not all simplifications
work. Proving the superirredundancy of $a \vee b \vee c$ in
{} $F = \{a \vee b, a \vee b \vee c\}$
is a counterexample. Removing the first clause from $F$ makes $a \vee b \vee c$
superirredundant in what remains, $F' = \{a \vee b \vee c\}$. Yet, $a \vee b
\vee c$ is not superirredundant in $F$.

A simplification works only if the superirredundancy of the clause in the
simplified formula implies its superirredundancy in the original formula. In
the other way around, superredundancy in the original implies superredundancy
in the simplification.

What is required is that ``$c$ superredundant in $F$'' implies ``$c$
superredundant in the simplified $F$''. One direction is enough. ``Implies'',
not ``if and only if''.

The following lemmas are formulated in the direction where superredundancy
implies superredundancy. This simplifies their formulation and their proofs,
but they are mostly used in reverse: superirredundancy implies
superirredundancy.


\begin{lemma}
\label{superset}

If a clause $c$ of $F$ is superredundant,
it is also superredundant in $F \cup \{c'\}$.

\end{lemma}

\proof The assumed superredundancy of a clause $c$ of $F$ is by definition
$\rescn(F) \backslash \{c\} \models c$. The derivations by resolution from $F$
are also valid from $F \cup \{c'\}$, where $c'$ is just not used. Therefore,
$\rescn(F) \subseteq \rescn(F \cup \{c'\})$. This implies
{} $\rescn(F) \backslash \{c\} \subseteq
{}  \rescn(F \cup \{c'\}) \backslash \{c\}$,
which implies
{} $\rescn(F \cup \{c'\}) \backslash \{c\} \models
{}  \rescn(F) \backslash \{c\}$.
By transitivity of entailment, the claim follows: $\rescn(F \cup \{c'\})
\backslash \{c\} \models c$.~\qed

How this lemma is used: a formula $F$ may simplify when a clause is added to
it, making superirredundancy easy to prove. For example, adding the
single-literal clause $x$ allows removing from $F$ all clauses that contain
$x$. If all clauses but $c$ contain $x$, the superirredundancy of $c$ in $F$
follows from the superirredundancy of $c$ in $\{c, x\}$.

This example of adding a single-literal clause extends to a full-fledged
sufficient condition to superredundancy. Adding $x$ to $F$ has the same effect
of replacing $x$ with $\true$ and simplifying the formula. This transformation
is defined as follows.

\begin{eqnarray*}
F[\true/x] &=& \{c[\true/x] \mid c \in F ,~ c[\true/x] \not= \top \} \\
c[\true/x] &=&
\left\{
\begin{array}{ll}
c \backslash \{\neg x\}	&	\mbox{if } x \not\in c	\\
\top			&	\mbox{otherwise}
\end{array}
\right.
\end{eqnarray*}

The symbol $\top$ used in the definition does not occur in the final formula
since clauses that are turned into $\top$ are removed from $F$. In other words,
$F[\true/x]$ is a formula built over variables and propositional connectives;
it does not contain any special symbol for $\true$ or $\false$.

Swapping $x$ and $\neg x$ turns the definition of $F[\true/x]$ into
$F[\false/x]$.

The next lemma shows that such substitutions often preserve superredundancy.
This would be obvious if superirredundancy were the same as clause primality or
a similar semantical notion, but it has been proved not to be by
Lemma~\ref{syntax-depend}.

\begin{lemma}
\label{set-value}

A clause $c$ of $F[\true/x]$ is superredundant
if it is superredundant in $F$, 
{} it contains neither $x$ nor $\neg x$
{} and $F$ does not contain $c \vee \neg x$.
The same holds for $F[\false/x]$ if $F$ does not contain $c \vee x$.

\end{lemma}

\proof The claim is proved for $x=\true$. It holds for $x = \false$ by
symmetry.

The assumption that $c$ is superredundant in $F$ is equivalent to
{} $F \backslash \{c\} \cup \resolve(F,c) \models c$
thanks to Lemma~\ref{first-step}.
The claim is the superredundancy of $c$ in $F[\true/x]$, which is equivalent to
{} $F[\true/x] \backslash \{c\} \cup \resolve(F[\true/x],c)) \models c$
still thanks to Lemma~\ref{first-step}.

The claim is the last of a chain of properties that follow from the assumption
{} $F \backslash \{c\} \cup \resolve(F,c) \models c$.

\begin{enumerate}

\item $(F \backslash \{c\} \cup \resolve(F,c))[\true/x] \models c$


Let $H = F \backslash \{c\} \cup \resolve(F,c)$. The assumption is $H \models
c$, the claim $H[\true/x] \models c$. By Boole's expansion
theorem~\cite{bool-54}, $H$ is equivalent to
{} $x \wedge H[\true/x] \vee \neg x \wedge H[\false/x]$.
The assumption $H \models c$ is therefore the same as
{} $x \wedge H[\true/x] \vee \neg x \wedge H[\false/x] \models c$.
Since a disjunction is implied by any of its disjuncts,
{} $x \wedge H[\true/x] \models c$
follows. Since neither $c$ nor $H[\true/x]$ contain $x$, this is the same as
$H[\true/x] \models c$.

\item $(F \backslash \{c\} \cup \resolve(F,c))[\true/x] =
(F \backslash \{c\})[\true/x] \cup \resolve(F,c)[\true/x]$


Formula $F \backslash \{c\} \cup \resolve(F,c)$ is a union. The substitution
therefore applies to each of its sets.

The two parts of the formula are considered separately.

\item $(F \backslash \{c\})[\true/x] = F[\true/x] \backslash \{c\}$
if $c \vee \neg x \not\in F$



Both formulae are made of some clauses of $F$ with the substitution $[\true/x]$
applied to them. They differ on whether $c$ is subtracted before or after the
substitution. The claim is proved by showing that for every $c' \in F$, the
clause $c'[\true/x]$ is in $(F \backslash \{c\})[\true/x]$ if and only if it is
in $F[\true/x] \backslash \{c\}$. The two cases $x \in c'$ and $x \not\in c'$
are considered separately.

If $c'$ contains $x$, then $c'[\true/x] = \top$. As a result, $F[\true/x]$ does
not contain $c'[\true/x]$; its subset $F[\true/x] \backslash \{c\}$ does not
either. Neither does $(F \backslash \{c\})[\true/x]$; indeed, $c' \in F
\backslash \{c\}$ since $c'$ contains $x$ while $c$ does not, but still
$c'[\true/x]$ is equal to $\top$ because it contains $x$, and is not therefore
in $(F \backslash \{c\})[\true/x]$.

If $c'$ does not contain $x$, then $c'[\true/x] = c' \backslash \{\neg x\}$.
This clause is equal to $c$ if and only if $c'$ is either $c$ or $c \vee \neg
x$; the second cannot be the case since $F$ by assumption contains $c'$ but not
$c \vee \neg x$. As a result, $c'[\true/x] = c$ if and only if $c' = c$. If $c'
= c$, then $c'[\true/x] = c$ is removed from $F[\true/x]$ when subtracting $c$
and $c'$ is removed from $F$ when subtracting $c$. If $c' \not= c$, then
$c'[\true/x] = c$ is not removed from $F[\true/x]$ and is therefore in
$F[\true/x] \backslash \{c\}$; also $c'$ is not removed from $F$ and is
therefore in $F \backslash \{c\}$, which means that $c'[\true/x]$ is in $(F
\backslash \{c\})[\true/x]$. In both cases, either $c'[\true/x]$ is in both
sets or in none.

\item $\resolve(F,c)[\true/x] = \resolve(F[\true/x],c)$


Expanding the definitions of $\resolve(F,c)$ and $\resolve(F[\true/x],c)$ shows
that the claim is:

\ttytex{
\[
(\bigcup_{c' \in F} \resolve(c',c))[\true/x] =
\bigcup_{c'' \in F[\true/x]} \resolve(c'',c)
\]
}{
(U{c' in F} resolve(c',c))[true/x] = U{c'' in F[true/x]} resolve(c'',c)
}

The first substitution is applied to a union, and can therefore equivalently
be applied to each of its members:

\ttytex{
\[
\bigcup_{c' \in F} (\resolve(c',c)[\true/x]) =
\bigcup_{c'' \in F[\true/x]} \resolve(c'',c)
\]
}{
U{c' in F} (resolve(c',c)[true/x]) = U{c'' in F[true/x]} resolve(c'',c)
}

If $x$ is in $c'$, it is also in the result of resolving $c'$ with $c$ since
$c$ does not contain $\neg x$ by assumption. As a result,
$\resolve(c',c)[\true/x] = \top$: the clauses $c'$ that contain $x$ do not
contribute to the first union. The claim therefore becomes:

\ttytex{
\[
\bigcup_{c' \in F ,~ x \not\in x} (\resolve(c',c)[\true/x]) =
\bigcup_{c'' \in F[\true/x]} \resolve(c'',c)
\]
}{
U{c' in F, x not in c'} (resolve(c',c)[true/x]) =
U{c'' in F[true/x]} resolve(c'',c)
}

The formula $F[\true/x]$ comprises by definition the clauses $c'' =
c'[\true/x]$ such that $c' \in F$ and $c'[\true/x] \not= \top$. The second
condition $c'[\true/x] \not= \top$ is false if $x \in c'$. The clauses
containing $x$ do not contribute to the second union either.

\ttytex{
\[
\bigcup_{c' \in F ,~ x \not\in x} (\resolve(c',c)[\true/x]) =
\bigcup_{c' \in F, ~ x \not\in x} \resolve(c'[\true],c)
\]
}{
U{c' in F, x not in c'} resolve(c',c))[true/x] =
U{c' in F, x not in c'} resolve(c'[true/x],c) =
}

This equality is proved as a consequence of the pairwise equality of the
elements of the unions.

\ttytex{
\[
\resolve(c',c)[\true/x] = \resolve(c'[\true/x],c)
\mbox{ for every } c' \in F \mbox{ such that } x \not\in c'
\]
}{
resolve(c',c)[true/x] = resolve(c'[true/x],c)
for every c' in F that does not contain x
}

Neither $c$ nor $c'$ contain $x$: the first by the assumption of the lemma, the
second because of the restriction in the above equality. Since $\resolve(c',c)$
only contains literals of $c$ and $c'$, it does not contain $x$ either.
Replacing $x$ with $\true$ in a clause that does not contain $x$ is the same as
removing $\neg x$.

\ttytex{
\[
\resolve(c',c) \backslash \{\neg x\} = \resolve(c' \backslash \{\neg x\},c)
\mbox{ for every } c' \in F \mbox{ such that } x \not\in c'
\]
}{
resolve(c',c) \ {-x} = resolve(c' \ {-x},c)
for every c' in F that does not contain x
}

If $l$ is a literal of $c'$ such that $\neg l \in c$, then $l \not= \neg x$
because $c$ does not contain $x$. As a result, $l \in c'$ implies $l \in c'
\backslash \{\neg x\}$. The converse also holds because of set containment.
This proves that $c$ resolves with $c'$ on a literal if and only it resolves
with $c' \backslash \{\neg x\}$ on the same literal.

If these clauses do not resolve, both sides of the equality are empty and
therefore equal. Otherwise, both $c'$ and $c' \backslash \{\neg x\}$ resolve
with $c$ on the same literal $l$. The result of resolving $c'$ with $c$ is
{} $\resolve(c',c) = c \cup c' \backslash \{l,\neg l\}$;
as a result, the left-hand side of the equality is
{} $\resolve(c',c) \backslash \{\neg x\} =
{}  c \cup c' \backslash \{l,\neg l\} \backslash \{\neg x\}$.
This is the same as the right hand side
{} $\resolve(c' \backslash \{x\},c) =
{}  c \cup (c' \backslash \{\neg x\}) \backslash \{l,\neg l\}$
since $c$ does not contain $\neg x$ by assumption.

\end{enumerate}


Summing up, the superredundancy of $c$ in $F$ expressed as
{} $F \backslash \{c\} \cup \resolve(F,c) \models c$
thanks to Lemma~\ref{first-step} implies
{} $(F \backslash \{c\} \cup \resolve(F,c))[\true/x] \models c$,
and the formula in this entailment is the same as
{} $F[\true/x] \backslash \{c\} \cup \resolve(F[\true/x],c)$.
The conclusion
{} $F[\true/x] \backslash \{c\} \cup \resolve(F[\true/x],c) \models c$
is equivalent to the superredundancy of $c$ in $F[\true/x]$ thanks to
Lemma~\ref{first-step}.~\qed

Is the assumption $c \vee \neg x \not\in F$ necessary? A counterexample
disproves the claim of lemma without this assumption: the clause $a$ is
superredundant in $F = \{a \vee \neg x, a, x\}$ because it is redundant, but is
superirredundant in $F[\true/x] = \{a\}$.

A way to prove superirredundancy is by applying Lemma~\ref{set-value} coupled
with Lemma~\ref{no-resolution}. A suitable evaluation of the variables not in
$c$ removes or simplifies the other clauses of $F$ to the point they do not
resolve, where Lemma~\ref{no-resolution} shows that $c$ is superirredundant.
Lemma~\ref{set-value} proves that $c$ is also superirredundant in $F$.

An example is
{} $F = \{a \vee b, b \vee c, \neg b \vee \neg d, \neg c \vee d \vee e\}$.
Replacing $c$ with $\true$ and $d$ with $\false$ deletes the second and third
clause and simplifies the fourth, leaving
{} $F[\true/c][\false/d] = \{a \vee b, e\}$,
where $a \vee b$ is superirredundant because no clause resolve in this formula.
This proves that $a \vee b$ is also superirredundant in $F$.

A substitution may not prevent all resolutions, but still breaks the formula in
small unlinked parts. Such parts can be worked on separately.

\begin{lemma}
\label{separate}

If $F'$ does not share variables with $F$ and $F'$ is satisfiable, a clause $c$
of $F$ is superredundant if and only if it is superredundant in $F \cup F'$.

\end{lemma}

\proof Since $c$ is in $F$, it is also in $F \cup F'$. By
Lemma~\ref{first-step}, the superredundancy of $c$ in $F \cup F'$ is equivalent
to
{} $(F \cup F' \cup \resolve(F \cup F', c)) \backslash \{c\} \models c$.
Since $c$ is in $F$ and $F$ does not share variables with $F'$, the clause $c$
does not share variables with $F'$ and therefore does not resolve with any
clause in $F'$. This proves $\resolve(F \cup F',c) = \resolve(F,c)$. The
entailment becomes
{} $(F \cup F' \cup \resolve(F, c)) \backslash \{c\} \models c$,
and also
{} $(F \cup \resolve(F, c)) \backslash \{c\} \cup F' \models c$
since $c \not\in F'$. This is the same as
{} $(F \cup \resolve(F, c)) \backslash \{c\} \models c$
because $F'$ is satisfiable and because of the separation of the variables.
This is equivalent to the superredundancy of $c$ in $F$ by
Lemma~\ref{first-step}.~\qed

Short guide on using Lemma~\ref{set-value} and Lemma~\ref{separate}: to prove
$c$ superirredundant in $F$, all clauses $c'$ that resolve with it are found
and all their variables not in $c$ collected; these variables are set to values
that satisfy as many clauses $c'$ as possible. All these clauses link $c$ with
the rest of $F$, and removing them makes $c$ isolated and therefore
superirredundant.

For example, the clause
{} $c = a \vee b$
is proved superirredundant in
{} $F = \{a \vee b, \neg a \vee c \vee d, \neg b \vee \neg c \vee \neg f,
{}	\neg d \vee f \vee g, d \vee h\}$
by a substitution that removes the clauses of $F$ that share variables with
$c$.

The clauses of $F$ that share variables with $c = a \vee b$ are
{} $\neg a \vee c \vee d$ and $\neg b \vee \neg c \vee \neg f$.
They are to be removed by substituting variables other than $a$ and $b$. For
example, $c=\true$ removes the first and simplifies the second into $\neg b
\vee \neg f$, which is removed by $f=\false$. This substitution turns $F$ into
{} $F[c/\true][f/\false] = \{a \vee b, \neg d \vee g, d \vee h\}$.
Since its clause $c = a \vee b $ does not share variables with the other two
clauses, it is superirredundant. As a result, it is also superredundant in $F$.

\

A final sufficient condition to superirredundancy is given by the following
lemma. It is specular to Theorem~\ref{pure}: that result applies when $\neg
l$ is not in $F$, this one when $l$ is not in $F$.


\begin{lemma}
\label{no-positive}

If $l$ does not occur in $F$, then $l$ is superirredundant in $F \cup \{l\}$ if
this formula is satisfiable.

\end{lemma}

\proof By Theorem~\ref{first-step-literal}, the superredundancy of $l$ in $F
\cup \{l\}$ is the same as $F' \models l$, where $F'$ is:

\begin{eqnarray*}
F' &=&
\{c \in F \cup \{l\} \mid \neg l \not\in c ,~ c \not= l\} \cup
\{c \mid c \vee \neg l \in F \cup \{l\}\}				\\
&=&
\{c \in F \mid \neg l \not\in c\} \cup \{c \mid c \vee \neg l \in F\}	\\
\end{eqnarray*}

The first part of the union is a subset of $F$; the second comprises only
subclauses of $F$. Since $F$ does not contain $l$, this union $F'$ does not
contain $l$ either. Therefore, $F'$ entails $l$ only if it is unsatisfiable.

The unsatisfiability of $F'$ is proved to contradict the assumption of the
lemma. The first part of $F'$ is a subset of $F$; each clause $c$ of its second
part is a consequence of $c \vee \neg l \in F$ and $l$, and is therefore
entailed by $F \cup \{l\}$. Therefore, this union is entailed by $F \cup
\{l\}$. It unsatisfiability implies the unsatisfiability of $F \cup \{l\}$,
which is contrary to an assumption of the lemma.~\qed

\section{Ensuring superirredundancy}
\label{section-split}

The intended application of superirredundancy is in existence proofs: produce a
formula satisfying certain conditions, some involving its minimal equivalent
formulae. Superirredundancy fixes a part of these minimal equivalent formulae.
The other conditions are ensured separately. The key is ``separately''. The
formula can be built to meet these other conditions and then some of its
clauses turned superirredundant.

The example in the next section shows how to create a reduction from Boolean
satisfiability to the problem of Horn minimality. It does not attempt to build
a Horn formula that can be shrunk over a certain limit if and only if a given
CNF formula is satisfiable. Rather, it builds the Horn formula so that some of
its clauses can be removed under the same condition. The other clauses are then
fixed by turning them superirredundant. This simplifies the process of creating
the reduction, as the target formula needs not to satisfy all required
conditions right from the beginning.

If $a \vee b$ is not superirredundant but is required in all minimal equivalent
formulae, it is split into $a \vee x$ and $b \vee \neg x$, where $x$ is a new
variable. The two resulting clauses are superirredundant in most cases.

The replacing clauses $a \vee x$ and $b \vee \neg x$ imply the original clause
$a \vee b$ by resolution. They are however not exactly equivalent to it because
of the new variable. They are only when restricting to all variables but $x$.
This restriction defines the concept of
forgetting~\cite{lin-reit-94,lang-etal-03}.

\begin{definition}
\label{express-forget}

A formula $B$ expresses forgetting all variables from $A$ except $Y$ if and
only if $\var(B) \subseteq Y$ and $B \models C$ is the same as $A \models C$
for all formulae $C$ such that $\var(C) \subseteq Y$.

\end{definition}

The formula
{} $B = A \backslash \{a \vee b\} \cup \{a \vee x, b \vee \neg x\}$
expresses forgetting $x$ from $A$. It is equivalent to it when disregarding the
new variable $x$. An alternative definition is that the models of $A$ and $B$
are the same when neglecting the evaluation of $x$. The following theorem
proves it, where $A \cap l${\plural} are the clauses of $A$ that contain the
literal $l$.

\begin{theorem}[{\cite[Theorem~6]{wang-15},\cite[Theorem~6]{delg-17}}]
\label{resolve-out}

The formula
{} $A \backslash (A \cap x) \backslash (A \cap \neg x) \cup
{}  \resolve(A \cap x, A \cap \neg x)$
expresses forgetting $x$ from $A$.

\end{theorem}

In the example, $A$ is the formula after splitting $a \vee b$ into $a \vee x$
and $b \vee \neg x$. Therefore, $A \cap x$ is $\{a \vee x\}$ and $A \cap \neg
x$ is $\{b \vee \neg x\}$. As a result,
{} $A \backslash (A \cap x) \backslash (A \cap \neg x) \cup
{}  \resolve(A \cap x, A \cap \neg x)$
is exactly the formula before splitting.

This is the first requisite on turning a clause superirredundant: the change
does not alter the semantics of the formula. Introducing a new variable makes
exact equivalence impossible, but forgetting is the close enough.

Another requirement is that turning $a \vee b$ superirredundant does not make
other clauses superredundant. That would make the change work only for a single
clause, not for all clauses that are required to be superirredundant.

In summary, splitting a clause works if:

\begin{itemize}

\item the resulting formula is similar enough to the original;

\item the split clause is superirredundant;

\item the other clauses remain superirredundant.

\end{itemize}

The first point is formalized as: the original formula expresses forgetting the
new variable from the generated formula. The second and the third points have
additional requirements, they are not always the case. They are proved in
reverse, by showing the consequences of superredundancy.

The following section shows an example of the mechanism, the subsequent ones
illustrate each of the points above.

\subsection{An example of making a clause superirredundant}

An example illustrates the method. The first clause is superredundant in the
following formula, as can be proved by computing the resolution closure.


\[
\{
	a \vee b \vee c,
	\neg a \vee d, \neg c \vee d, \neg d \vee a \vee c
\}
\]

The last three clauses are the same as $a \vee c \equiv d$. They make $a \vee
c$ equivalent to $d$. Consequently, the first clause $a \vee b \vee c$ is
replaceable by $d \vee b$ and therefore superredundant.

If it is required to be superirredundant, it can be made so by splitting it
into $a \vee x$ and $\neg x \vee b \vee c$.

\[
\{
	a \vee x, \neg x \vee b \vee c,
	\neg a \vee d, \neg c \vee d, \neg d \vee a \vee c
\}
\]

The last three clauses still make $a \vee c$ equivalent to $d$, but this is no
longer a problem because $a$ and $c$ are now separated: $a$ is in $a \vee x$
and $c$ is in $\neg x \vee b \vee c$. The only way to join them back to apply
their equivalence to $d$ is to resolve the two parts into $a \vee b \vee c$.
This removes $x$ and $\neg x$, which are necessary to derive the two clauses $a
\vee x$ and $\neg x \vee b \vee c$ back. All clauses in this formula are
superirredundant, which can be checked by computing the resolution closure.



\subsection{Preserving the semantics of the formula}

The first point to prove is that splitting a clause does not change the meaning
of the formula. The semantics changes slightly since the original formula does
not mention $x$ at all while the modified one does. In the example, $a=\false$,
$b=\true$ and $x=\false$ satisfy the original clause $a \vee b \vee c$ but not
its part $a \vee x$. The modified formula cannot be equivalent to the original
since it contains the new variable. Yet, it is equivalent apart from it. This
is what forgetting does: it removes a variable while semantically preserving
everything else.

\begin{lemma}
\label{split-forget}

Every CNF formula $F$ that contains a clause $c_1 \vee c_2$, where $c_1$ and
$c_2$ are two clauses, and does not mention $x$ expresses forgetting $x$ from
{} $F \backslash \{c_1 \vee c_2\} \cup \{c_1 \vee x, c_2 \vee \neg x\}$.

\end{lemma}

\proof The formula
{} $F \backslash \{c_1 \vee c_2\} \cup \{c_1 \vee x, c_2 \vee \neg x\}$
in the statement of the lemma is denoted $F''$. Theorem~\ref{resolve-out}
proves that a formula expresses forgetting $x$ from it is
{} $F'' \backslash (F'' \cap x) \backslash (F'' \cap \neg x) \cup
{}  \resolve(F'' \cap x, F'' \cap \neg x)$.
The claim is proved by showing that this formula is $F$.


The only clause of $F''$ containing $x$ is $c_1 \vee x$ and the only clause
containing $\neg x$ is $c_2 \vee \neg x$. Therefore,
{} $F'' \cap x$ is $\{c_1 \vee x\}$ and
{} $F'' \cap \neg x$ is $\{c_2 \vee \neg x\}$.
The formula that expresses forgetting is therefore
{} $F'' \backslash \{c_1 \vee x\} \backslash \{c_2 \vee \neg x\} \cup
{}  \resolve(c_1 \vee x, c_2 \vee \neg x)$,
which is equal to
{} $F'' \backslash \{c_1 \vee x\} \backslash \{c_2 \vee \neg x\} \cup
{}  \{c_1 \vee c_2\}$
since $\resolve(c_1 \vee x, c_2 \vee \neg x) = \{c_1 \vee c_2\}$. Replacing
$F''$ with its definition turns this formula into
{} $F \backslash \{c_1 \vee c_2\} \cup \{c_1 \vee x, c_2 \vee \neg x\} 
{} \backslash \{c_1 \vee x\} \backslash \{c_2 \vee \neg x\} \cup
{} \{c_1 \vee c_2\}$.
Computing unions and set subtractions shows that this formula is $F$.~\qed

This lemma tells that
{} $F \backslash \{c_1 \vee c_2\} \cup \{c_1 \vee x, c_2 \vee \neg x\}$
is like $F$ apart from $x$. This is the basic requirement for the split: it
preserves the semantics as much as possible. The modified formula has the same
consequences of the original that do not involve $x$.

\subsection{Making a clause superirredundant}

The aim of the split is not just to preserve the semantics but also to make the
two pieces of the split clause superirredundant.

The addition of $x$ and $\neg x$, more than the split itself, is what creates
superirredundancy. The two parts contain $x$ and $\neg x$, and are the only
clauses containing them. They are necessary to derive every other clause
containing them, including themselves.

An exception is when the part containing $x$ derives another containing $x$
which derives it back. The presence of $x$ in the whole derivation sequence
ensures that removing $x$ everywhere does not invalidate the derivation. The
result is a derivation from a part of the original clause (without $x$ added)
to other clauses and back. It proves the superredundancy of that part. This
explains the exception: superirredundancy is only obtained if none of the two
parts of the clause is superredundant by itself.

\begin{lemma}
\label{make-irredundant}

If $c_1 \vee x$ is superredundant in
{} $F \backslash \{c_1 \vee c_2\} \cup \{c_1 \vee x, c_2 \vee \neg x\}$,
then $c_1$ is superredundant in $F \cup \{c_1\}$,
provided that:

\begin{itemize}
\item $c_1 \vee c_2$ is in $F$;
\item $c_1$ is not in $F$; and
\item $x$ does not occur in $F$.
\end{itemize}

\end{lemma}

\proof Lemma~\ref{first-step} reformulates the superredundancy in the
assumption and in the claim as entailments.

\begin{eqnarray*}
&&
F \backslash \{c_1 \vee c_2\} \cup \{c_1 \vee x, c_2 \vee \neg x\}
	\backslash \{c_1 \vee x\} \cup
	\resolve(c_1 \vee x, F \backslash \{c_1 \vee c_2\} \cup
			\{c_1 \vee x, c_2 \vee \neg x\}) \\
&& ~~~~~ \models c_1 \vee x \\
&& F \cup \{c_1\} \backslash \{c_1\} \cup \resolve(c_1, F \cup \{c_1\}) \\
&& ~~~~~ \models c_1
\end{eqnarray*}

The claim is proved if the first entailment implies the second. Since $F$
contains $c_1 \vee c_2$, it is the same as
{} $F \backslash \{c_1 \vee c_2\} \cup \{c_1 \vee c_2\}$.
This allows reformulating the second entailment in terms of
{} $F' = F \backslash \{c_1 \vee c_2\}$;
the first can be as well.

\begin{eqnarray*}
F' \cup \{c_1 \vee x, c_2 \vee \neg x\} \backslash \{c_1 \vee x\} \cup
	\resolve(c_1 \vee x, F' \cup \{c_1 \vee x, c_2 \vee \neg x\})
		& \models & c_1 \vee x \\
F' \cup \{c_1 \vee c_2\} \cup \{c_1\} \backslash \{c_1\} \cup
	\resolve(c_1, F' \cup \{c_1 \vee c_2\} \cup \{c_1\})
		\models c_1
\end{eqnarray*}

The set subtractions can be computed immediately.

In the first formula,
{} $F' \cup \{c_1 \vee x, c_2 \vee \neg x\} \backslash \{c_1 \vee x\}$
is equal to
{} $F' \cup \{c_2 \vee \neg x\}$
since neither $c_2 \vee \neg x$ nor any clause in $F'$ is equal to $c_1 \vee
x$. The former is not because it contains $\neg x$, the latter are not because
$F'$ is a subset of $F$, which does not mention $x$.

In the second formula,
{} $F' \cup \{c_1 \vee c_2\} \cup \{c_1\} \backslash \{c_1\}$
is equal to $F' \cup \{c_1 \vee c_2\}$ since neither $c_1 \vee c_2$ nor any
clause in $F'$ is equal to $c_1$. The first is not because it is in $F$ while
$c_1$ is not, the second are not because $F'$ is a subset of $F$, which does
not contain $c_1$.

\begin{eqnarray*}
F' \cup \{c_2 \vee \neg x\} \cup
	\resolve(c_1 \vee x, F' \cup \{c_1 \vee x, c_2 \vee \neg x\})
		& \models & c_1 \vee x \\
F' \cup \{c_1 \vee c_2\} \cup
	\resolve(c_1, F' \cup \{c_1 \vee c_2\} \cup \{c_1\})
		& \models & c_1
\end{eqnarray*}

Both entailments contain the resolution of a clause with a union. This is the
same as the resolution of the clause with each component of the union.

\begin{eqnarray*}
F' \cup \{c_2 \vee \neg x\} \cup
	\resolve(c_1 \vee x, F') \cup
	\resolve(c_1 \vee x, \{c_1 \vee x\}) \cup
	\resolve(c_1 \vee x, \{c_2 \vee \neg x\})
		& \models & c_1 \vee x \\
F' \cup \{c_1 \vee c_2\} \cup
	\resolve(c_1, F') \cup
	\resolve(c_1, \{c_1 \vee c_2\}) \cup
	\resolve(c_1, \{c_1\})
		& \models & c_1
\end{eqnarray*}

Some parts of these formulae are empty because clauses do not resolve with
themselves or with their superclauses.

\begin{eqnarray*}
F' \cup \{c_2 \vee \neg x\} \cup
	\resolve(c_1 \vee x, F') \cup
	\resolve(c_1 \vee x, \{c_2 \vee \neg x\})
		& \models & c_1 \vee x \\
F' \cup \{c_1 \vee c_2\} \cup
	\resolve(c_1, F')
		& \models & c_1
\end{eqnarray*}

Since $c_1 \vee c_2$ is in $F$ and formulae are assumed not to contain
tautologies, the two subclauses $c_1$ and $c_2$ do not contain opposite
literals. Therefore, resolving $c_1 \vee x$ and $c_2 \vee \neg x$ only
generates $c_1 \vee c_2$, which is not a tautology. This simplifies
{} $\resolve(c_1 \vee x, \{c_2 \vee \neg x\})$ into $\{c_1 \vee c_2\}$.


\begin{eqnarray*}
F' \cup \{c_2 \vee \neg x\} \cup
	\resolve(c_1 \vee x, F') \cup
	\{c_1 \vee c_2\}
		& \models & c_1 \vee x \\
F' \cup \{c_1 \vee c_2\} \cup
	\resolve(c_1, F')
		& \models & c_1
\end{eqnarray*}

The set $\resolve(c_1 \vee x, F')$ contains the result of resolving $c_1 \vee
x$ with the clauses of $F'$. Since $F'$ is a subset of $F$, it does not contain
$x$. Therefore, the resolving literal of $c_1 \vee x$ with a clause of $c'' \in
F'$ is not $x$ if any. If $c_1 \vee x$ resolves with $c'' \in F$, then $c_1$
does as well. Adding $x$ to the resolvent generates the resolvent of $c_1 \vee
x$ and $c''$. Formally,
{} $\resolve(c_1 \vee x, F') = \{c' \vee x \mid c' \in \resolve(c_1, F')\}$.


\begin{eqnarray*}
F' \cup \{c_2 \vee \neg x\} \cup
	\{c' \vee x \mid c' \in \resolve(c_1, F')\} \cup
	\{c_1 \vee c_2\}
		& \models & c_1 \vee x \\
F' \cup \{c_1 \vee c_2\} \cup
	\resolve(c_1, F')
		& \models & c_1
\end{eqnarray*}


Replacing $x$ with $\false$ in the first entailment results in
{} $F' \cup \{c_2 \vee \neg \false\} \cup
{}	\{c' \vee \false \mid c' \in \resolve(c_1, F')\} \cup
{}	\{c_1 \vee c_2\}
{}		\models c_1 \vee \false$.
Simplifying according to the rules of propositional logic $\true \vee G =
\true$ and $\false \vee G = G$ turns this entailment into
{} $F' \cup \{\true\} \cup
{}	\{c' \mid c' \in \resolve(c_1, F')\} \cup
{}	\{c_1 \vee c_2\}
{}		\models c_1$,
which is the same as
{} $F' \cup
{}	\resolve(c_1, F') \cup
{}	\{c_1 \vee c_2\}
{}		\models c_1$,
the second entailment.

This proves that the first entailment implies the second: the assumption
implies the claim.~\qed

The intended usage of the lemma is to split a clause $c_1 \vee c_2$ of $F$ into
$c_1 \vee x$ and $c_2 \vee \neg x$, where $x$ is a new variable. Being new, $x$
does not occur in the rest of the formula. If the lemma is used this way, its
first and last assumptions are met. The second may not, and the claim may not
hold if $F$ contains $c_1$. Actually, the claim may not hold if any of its
three assumptions does not hold.

If $c_1 \vee c_2$ is not in $F$, the claim may not hold. A counterexample is
{} $c_1 = a \vee b$,
{} $c_2 = a$ and
{} $F = \emptyset$.
{}
The other preconditions of the lemma are satisfied: $c_1$ is not in $F$, where
$x$ does not occur; $c_1 \vee x$ is superredundant in
{} $F \backslash \{c_1 \vee c_2\} \cup \{c_1 \vee x, c_2 \vee \neg x\}$
since this formula is
{} $\{a \vee b \vee x, a \vee \neg x\}$,
whose two clauses resolve in $a \vee b$, which entails
{} $c_1 \vee x = a \vee b \vee x$.
The conclusion that $c_1 = a \vee b$ is superredundant in
{} $F \cup \{c_1\} = \emptyset \cup \{a \vee b\} = \{a \vee b\}$
is false since this formula allows no resolution.

If $c_1$ is in $F$, the claim may not hold. A counterexample is
{} $c_1 = a \vee b$,
{} $c_2 = a$ and
{} $F = \{a \vee b\}$.
{}
The other preconditions of the lemma are satisfied: $c_1 \vee c_2$ is in $F$,
where $x$ does not occur; $c_1 \vee x$ is superredundant in
{} $F \backslash \{c_1 \vee c_2\} \cup \{c_1 \vee x, c_2 \vee \neg x\}$
since this formula is
{} $\{a \vee b \vee x, a \vee \neg x\}$,
whose two clauses resolve in $a \vee b$, which entails
{} $c_1 \vee x = a \vee b \vee x$.
The conclusion that $c_1 = a \vee b$ is superredundant in
{} $F \cup \{c_1\} = \{a \vee b\} \cup \{a \vee b\} = \{a \vee b\}$
is false since this formula allows no resolution.

If $F$ mentions $x$, the claim may not hold. A counterexample is
{} $c_1 = a$,
{} $c_2 = b$ and
{} $F = \{a \vee b, x\}$.
The other preconditions of the lemma are satisfied: $c_1 \vee c_2 = a \vee b$
is in $F$, while $c_1$ is not; $c_1 \vee x$ is superredundant in
{} $F \backslash \{c_1 \vee c_2\} \cup \{c_1 \vee x, c_2 \vee \neg x\}$,
which is
{} $\{a \vee b, x\} \backslash \{a \vee b\} \cup \{a \vee x, b \vee \neg x\}$,
which is the same as
{} $\{x\} \cup \{a \vee x, b \vee \neg x\}$,
where $c_1 \vee x = a \vee x$ is superredundant because it is entailed by $x$.
The conclusion that $c_1 = a$ is superredundant in $F \cup \{c_1\}$ is false
since this formula is
{} $\{a \vee b, x\} \cup \{a\} = \{a \vee b, x, a\}$,
where no clauses resolve.

The three assumptions do not hinder the intended usage of the lemma: make a
clause $c_1 \vee c_2$ of $F$ superirredundant by splitting it into $c_1 \vee x$
and $c_2 \vee \neg x$ on a new variable $x$. The first assumption is met
because $c_1 \vee c_2$ is a clause of $F$ to be made superirredundant. The
third is met because $x$ is new. The second is met in the sense that $c_1 \vee
c_2$ can just be removed if $c_1$ is also in the formula.

When the three assumptions are met, the lemma tells that $c_1$ is
superredundant in $F \cup \{c_1\}$ if $c_1 \vee x$ is superredundant in $F
\backslash \{c_1 \vee c_2\} \cup \{c_1 \vee x, c_2 \vee \neg x\}$. This
implication is useful in reverse: $c_1 \vee x$ is superirredundant in
{} $F \backslash \{c_1 \vee c_2\} \cup \{c_1 \vee x, c_2 \vee \neg x\}$
unless $c_1$ is superredundant in $F \cup \{c_1\}$. The goal of making $c_1
\vee x$ superirredundant is hit, but only if $c_1$ is superirredundant in $F
\cup \{c_1\}$.


This condition is necessary. The following example shows it cannot be lifted.

\[
F = \{a \vee b, \neg a \vee c, a \vee \neg c\}
\]

The last two clauses are equivalent to $a \equiv c$. They make the first clause
superredundant because $a \vee b$ derives $c \vee b$ which derives $a \vee b$
back. Splitting does not make the clause superirredundant: $a \vee x$ still
derives $c \vee x$, which derives $a \vee x$ back. Removing $x$ from this
derivation results in $a$ that derives $c$ that derives $a$ back. Splitting $a
\vee b$ does not work because $a$ alone is already superredundant. Adding a new
variable $x$ does not change the situation.

\subsection{Maintaining the superirredundancy of the other clauses}

Preserving the semantics of the formula and making a clause superirredundant is
not enough. The other clauses must remain superirredundant. Otherwise, the
process may go on forever. Even attempting to have two clauses superirredundant
would fail if making one so makes the other not. The final requirement of
clause splitting is that the other clauses remain superirredundant. This is
mostly the case, with an exception that is discussed after the proof of the
lemma.

The lemma is formulated in reverse. Instead of ``superirredundancy is
maintained except in this condition'', it states ``superredundancy is generated
only in this condition''.

\begin{lemma}
\label{leave-irredundant}

If $c$ and $c_1 \vee c_2$ are two different clauses of $F$ and $c$ is
superredundant in
{} $F \backslash \{c_1 \vee c_2\} \cup \{c_1 \vee x, c_2 \vee \neg x\}$
and $x$ does not occur in $F$ then either:

\begin{itemize}

\item $c$ resolves with both $c_1$ and $c_2$; or

\item $c$ is superredundant in $F$.

\end{itemize}

\end{lemma}

\proof Lemma~\ref{first-step} reformulates the superredundancy in the
assumption and in the claim.

\begin{eqnarray*}
F \backslash \{c_1 \vee c_2\} \cup \{c_1 \vee x, c_2 \vee \neg x\}
\backslash \{c\} \cup
\resolve(c, F \backslash \{c_1 \vee c_2\} \cup \{c_1 \vee x, c_2 \vee \neg x\})
& \models & c \\
F \backslash \{c\} \cup \resolve(c, F)
& \models & c
\end{eqnarray*}

Since $F$ contains $c_1 \vee c_2$, it is the same as
{} $F \backslash \{c_1 \vee c_2\} \cup \{c_1 \vee c_2\}$.
Both this expression and
{} $F \backslash \{c_1 \vee c_2\} \cup \{c_1 \vee x, c_2 \vee \neg x\}$
contain
{} $F' = F \backslash \{c_1 \vee c_2\}$.

\begin{eqnarray*}
F' \cup \{c_1 \vee x, c_2 \vee \neg x\} \backslash \{c\} \cup
	\resolve(c, F' \cup \{c_1 \vee x, c_2 \vee \neg x\}) &
		\models & c	\\
F' \cup \{c_1 \vee c_2\} \backslash \{c\} \cup
	\resolve(c, F' \cup \{c_1 \vee c_2\}) &
		\models & c
\end{eqnarray*}

Resolving a clause with a set is the resolution of the clause with each
clause in the set.

\begin{eqnarray*}
F' \cup \{c_1 \vee x, c_2 \vee \neg x\} \backslash \{c\} \cup
	\resolve(c, F') \cup
	\resolve(c, \{c_1 \vee x\}) \cup
	\resolve(c, \{c_2 \vee \neg x\})
	& \models & c	\\
F' \cup \{c_1 \vee c_2\} \backslash \{c\} \cup
	\resolve(c, F') \cup
	\resolve(c, \{c_1 \vee c_2\})
	& \models & c
\end{eqnarray*}

The clauses in these entailments are the same except for:

\begin{itemize}

\item the first formula contains
{} $c_1 \vee x$,
{} $c_2 \vee \neg x$,
{} $\resolve(c,c_1 \vee x)$ and
{} $\resolve(c,c_2 \vee \neg x)$

\item the second formula contains
{} $c_1 \vee c_2$ and
{} $\resolve(c,c_1 \vee c_2)$

\end{itemize}

The difference depends on whether $c$ resolves with $c_1 \vee x$, $c_2 \vee
\neg x$ and $c_1 \vee c_2$. Since $c$ does not contain $x$, it resolves with
$c_1 \vee x$ if and only if it resolves with $c_1$, and the same for $c_2$. It
resolves with $c_1 \vee c_2$ if it resolves with either $c_1$ or $c_2$. All
depends on whether $c$ resolves with $c_1$ or with $c_2$.


Four cases are possible. Apart from the last case, the claim is proved by
removing all clauses containing $x$ from the first formula and adding their
resolution, which produces the second formula. Theorem~\ref{resolve-out} proves
that this procedure generates a formula that expresses forgetting $x$ and
therefore entails the same consequences that do not contain $x$, such as $c$.

\begin{enumerate}

\item $c$ resolves with neither $c_1$ nor $c_2$

The three sets $\resolve(c,c_1 \vee c_2)$, $\resolve(c,c_1 \vee x)$ and
$\resolve(c,c_2 \vee \neg x)$ are empty. The only other differing clauses are
{} $c_1 \vee x$ and $c_2 \vee \neg x$ in the first formula and
{} $c_1 \vee c_2$ in the second.
The first two are the only clauses containing $x$. Resolving them results in
the third. This proves the claim by Theorem~\ref{resolve-out}.


\item $c$ resolves with $c_1$ but not with $c_2$

The two sets $\resolve(c,c_1 \vee c_2$) and $\resolve(c,c_1 \vee x)$ contain a
clause, but $\resolve(c,c_2 \vee \neg x)$ does not since $c$ does not contain
$x$. The only differing clauses between the two formulae are
{} $c_1 \vee x$, $c_2 \vee \neg x$ and $\resolve(c,c_1 \vee x)$ in the first
and
{} $c_1 \vee c_2$ and $\resolve(c,c_1 \vee c_2)$ in the second.

Only two pairs of clauses contain $x$ with opposite sign: the first is
{} $c_1 \vee x$ and $c_2 \vee \neg x$,
the second is
{} $\resolve(c,c_1 \vee x)$ and $c_2 \vee \neg x$.

The first pair resolves into $c_1 \vee c_2$, the first differing clause in the
second formula.

The second pair is shown to resolve in the second differing clause,
$\resolve(c,c_1 \vee c_2)$. If $l$ is the resolving literal $l$ between $c$ and
$c_1 \vee x$, then
{} $\resolve(c,c_1 \vee x)$ is $c \vee c_1 \vee x \backslash \{l, \neg l\}$.
Since $c$ does not contain $x$, the resolving literal $l$ cannot be $x$. As a
result, this clause contains $x$. It therefore resolves with $c_2 \vee \neg x$
into
{} $c \vee c_1 \vee x \backslash \{l, \neg l\} \vee
{}  c_2 \vee \neg x \backslash \{x,\neg x\} =
{}  c \vee c_1 \backslash \{l, \neg l\} \vee c_2$.
Since $c_2$ does not resolve with $c$, it does not contain $\neg l$. It does
not contain $l$ either since otherwise $c_1 \vee c_2$ would be tautological.
The clause is therefore the same as
{} $c \vee c_1 \vee c_2 \backslash \{l, \neg l\}$.
This is $\resolve(c,c_1 \vee c_2)$, the second differing clause in the second
formula.


This proves that replacing all clauses containing $x$ in the first formula with
their resolution produces the second. This proves the claim by
Theorem~\ref{resolve-out}.

\item $c$ resolves with $c_2$ but not with $c_1$

Same as the previous case by symmetry.

\item $c$ resolves with both $c_1$ and $c_2$

The claim is proved because its first alternative is exactly that $c$ resolves
with both $c_1$ and $c_2$.

\end{enumerate}

All of this proves the claim in all four cases. In the first three, replacing
all clauses containing $x$ with their resolution in the first formula produces
the second; this implies that the two formulae have the same consequences that
do not contain $x$, such as $c$. This is the first alternative of the claim.
The fourth case coincides with the second alternative of the claim.~\qed

Ideally, all clauses would maintain their superirredundancy. This is the case
for most but not all. The exception is the clauses that resolve with both parts
of the clause that is split. Such clauses invalidate the proof. That raises the
question: could the proof be improved to include them? Or do they falsify the
statement of the lemma instead? The following example proves the latter.

\begin{eqnarray*}
F &=&	\{d, c_1 \vee c_2, a \vee e, \neg e \vee \neg a \vee \neg d\}	\\
F'' &=&	\{d, c_1 \vee x, \neg x \vee c_2,
          a \vee e, \neg e \vee \neg a \vee \neg d\}			\\
c &=&	a \vee b \vee d \vee e						\\
c_1 &=&	\neg a \vee b							\\
c_2 &=&	\neg d \vee e
\end{eqnarray*}

The formula obtained by splitting $c_1 \vee c_2$ is denoted $F''$. The clause
$c_1 \vee c_2 = \neg a \vee b \vee \neg d \vee e$ is superredundant in $F$
because it resolves with $a \vee e$ into $e \vee b \vee \neg d \vee e$, which
resolves with $\neg e \vee \neg a \vee \neg d$ back into $\neg a \vee b \vee
\neg d \vee e$. To make this clause superirredundant, it is split. However,
that makes the first clause $c = a \vee b \vee d \vee e$ superredundant.

That $c$ is superirredundant in $F$ is proved replacing $e$ with $\true$ and
simplifying the formula. That removes $a \vee e$ and turns $\neg e \vee \neg a
\vee \neg d$ into $\neg a \vee \neg d$. What remains is
{} $F[\true/e] = \{a \vee b \vee d \vee e, \neg a \vee b \vee \neg d \vee e,
{}               \neg a \vee \neg d\}$.
The first clause resolves both with the second and the third, but the result is
a tautology in both cases: $F[\true/e] \backslash \{c\} \cup \resolve(c,F)$ is
equivalent to $F[\true/e] \backslash \{c\}$, which does not entail $c$.
Lemma~\ref{first-step} proves that $c$ is not superredundant in $F[\true/e]$.
Since $c$ contains neither $x$ nor $\neg x$ and $F$ does not contain $c \vee
\neg e$, Lemma~\ref{set-value} ensures that $c$ would be superredundant in
$F[\true/e]$ if it were in $F$. But $c$ is not superredundant in $F[\true/e]$.
As a result, it is not superredundant in $F$.

Yet, $c$ is superredundant in $F''$, the formula after the split:
{} $c = a \vee b \vee d \vee e$ resolves with
{} $c_1 \vee x = \neg a \vee b \vee x$ into
{} $x \vee b \vee d \vee e$;
it also resolves with
{} $\neg x \vee c_2 = \neg x \vee \neg d \vee e$ into
{} $\neg x \vee a \vee b \vee e$;
the resulting two clauses resolve into $c$, and therefore imply it. They are
the set $G$ that proves $c$ superredundant according to Lemma~\ref{equivalent},
since they are obtained from $F''$ by resolution, none of them is $c$, and they
imply $c$.

If the target was to make both $c$ and $c_1 \vee c_2$ superirredundant,
Lemma~\ref{leave-irredundant} misses it. Yet, a second shot gets it: $c_1 \vee
x$ and $\neg x \vee c_2$ are now superirredundant, but $c$ no longer is;
splitting it makes it so:

\[
F''' = \{
	a \vee b \vee y, \neg y \vee d \vee e,
	\neg a \vee b \vee x, \neg x \vee \neg d \vee e,
	a \vee e, \neg e \vee \neg a \vee \neg d
\}
\]

The split separates $c$ into $a \vee b \vee y$ and $\neg y \vee d \vee e$. Both
parts resolve with $c_1 \vee c_2$, but this is not a problem because $c_1 \vee
c_2$ is no longer in the formula. It has already been split into $c_1 \vee x$
and $\neg x \vee c_2$. The first resolves with $a \vee b$ but not with $d \vee
e$, the second with $d \vee e$ but not with $a \vee b$. The original clause
$c_1 \vee c_2$ would be made superredundant by this splitting, but its two
parts $c_1 \vee x$ and $\neg x \vee c_2$ are not. By first splitting a clause
and then the other, both are made superirredundant.

\

Mission accomplished: if a clause is not superirredundant but should be,
splitting it on a new variable makes it so. Subclauses and clauses that resolve
with both parts are to be watched out, but the mechanism mostly works.

Making clauses superirredundant nails them to the formula. It forces them in
all minimal equivalent CNF formulae. Every CNF formula $F$ is $F' \cup F''$,
where $F'$ are its superirredundant clauses; every minimal formula equivalent
to $F$ is $F' \cup F'''$. The superirredundant clauses $F'$ are always there.
They provide the basement over which the other clauses build upon. They are the
skeleton, with its hard bones but also its flexible joins. The muscles, the
other clauses, may move it not by bending the bones but by rotating them at the
joins. The superirredundant clauses are fixed but may still leave space for
other clauses to change. Minimizing $F' \cup F''$ is altering $F''$ while
keeping $F'$. Is finding a minimal version of $F''$ that is equivalent to the
original formula when $F'$ is always present.

A way to ensure superirredundancy is to make clauses superirredundant. The
three lemmas in this section do this:

\begin{description}

\item Lemma~\ref{split-forget} proves that splitting a clause $c_1 \vee c_2$
into $c_1 \vee x$ and $c_2 \vee \neg x$ does not change the meaning of the
formula except for the new variable $x$;

\item Lemma~\ref{make-irredundant} proves that the two parts $c_1 \vee x$ and
$c_2 \vee \neg x$ are superirredundant unless $c_1$ or $c_2$ are superredundant
when added to the formula;

\item Lemma~\ref{leave-irredundant} proves that the other clauses remain
superirredundant after the split; the exception are the clauses that resolve
with both parts of the split clause; these are made superredundant, but can
themselves be split.

\end{description}

\section{Example}
\label{section-example}

Superirredundancy is applied to finding a proof of \np-hardness of deciding
whether a Horn formula can be compressed in a given size. This problem is known
to be \np-complete~\cite{hamm-koga-93,cepe-kuce-08,hamm-koga-95}. The new proof
shows that a reduction can be found progressively, by first building a
simplified version where some clauses are fixed and then making them
superirredundant.

Technically, an instance of the problem comprises a Horn formula $A$ and an
integer $k$; the question is whether a formula $B$ equivalent to $A$ exists
with $||B|| \leq k$.

It is proved \np-hard by a reduction from propositional satisfiability: given a
CNF formula $F$, the reduction builds an instance comprising $A$ and $k$ such
that $A$ is equivalent to another formula $B$ of size bounded by $k$ if and
only if $F$ is satisfiable.

The proof based on superirredundancy simplifies the task of finding such a
reduction by assuming that a part $A'$ of $A$ is fixed, that is, is also in
every equivalent $B$. This way, the question turns from the compressibility of
$A$ into the compressibility of $A'' = A \backslash A'$.

\begin{itemize}

\item $A''$ comprises a clause $x_i \vee \neg q$ and a clause $e_i \vee \neg q$
for every variable in $F$; this way, every propositional interpretation over
the alphabet of $F$ corresponds to a subset of $A''$, the one containing $x_i
\vee \neg q$ if $x_i$ is true and $e_i \vee \neg q$ if false;

\item $A'$ ensures that such a subset of $A''$ entails the rest of $A''$ if and
only if the propositional interpretation satisfies $F$.

\end{itemize}

From this roadmap, finding the reduction itself is almost trivial: a clause
$t_i \vee \neg q$ is entailed if and only if the subset of $A''$ includes
either $x_i \vee \neg q$ or $e_i \vee q$; another $c_j \vee \neg q$ is entailed
if and only if the $j$-th clause of $F$ is satisfied, which means that the
subset of $A''$ includes the clause that corresponds to a literal of the
clause; if all these clauses are entailed, all clauses $x_i \vee \neg q$ and
$e_i \vee \neg q$ are entailed.

The clauses allowing these entailments are assumed superirredundant. An example
clause of $F$ may be $x_1 \vee x_2$. It is satisfied by setting either $x_1$ or
$x_2$ to true. The subsets of $A''$ corresponding to these evaluations
respectively include $x_1 \vee \neg q$ and $x_2 \vee \neg q$. The
superirredundant clauses $\neg x_1 \vee c_1$ and $\neg x_2 \vee c_1$ allow the
derivation of $c_1 \vee \neg q$ by resolution from them.

The following figure shows how the missing clause $e_1 \vee \neg q$ is derived
when the formula also contains a second clause $\neg x_1 \vee \neg x_2$. The
clauses translate into
{} $\{
{}	\neg x_1 \vee c_1,
{}	\neg x_2 \vee c_1,
{}	\neg e_1 \vee c_2,
{}	\neg e_2 \vee c_2
{} \}$.
The clauses not included in the subset of $A''$ are crossed.

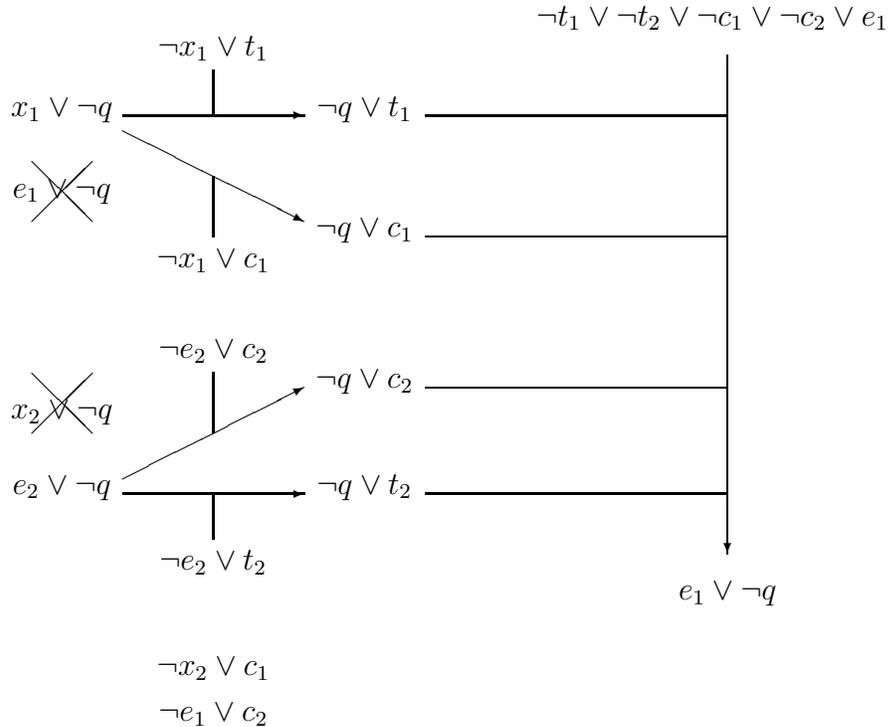
\begin{center}
\setlength{\unitlength}{5000sp}%
\begingroup\makeatletter\ifx\SetFigFont\undefined%
\gdef\SetFigFont#1#2#3#4#5{%
  \reset@font\fontsize{#1}{#2pt}%
  \fontfamily{#3}\fontseries{#4}\fontshape{#5}%
  \selectfont}%
\fi\endgroup%
\begin{picture}(3504,3675)(6589,-7726)
\thinlines
{\color[rgb]{0,0,0}\put(7501,-4486){\line( 0,-1){225}}
}%
{\color[rgb]{0,0,0}\put(7501,-5011){\line( 0,-1){300}}
}%
{\color[rgb]{0,0,0}\put(7501,-5986){\line( 0,-1){300}}
}%
{\color[rgb]{0,0,0}\put(6601,-4936){\line( 1,-1){300}}
}%
{\color[rgb]{0,0,0}\put(6601,-5236){\line( 1, 1){300}}
}%
{\color[rgb]{0,0,0}\put(7501,-6586){\line( 0,-1){225}}
}%
{\color[rgb]{0,0,0}\put(8551,-5311){\line( 1, 0){1500}}
}%
{\color[rgb]{0,0,0}\put(8551,-6061){\line( 1, 0){1500}}
}%
{\color[rgb]{0,0,0}\put(8551,-6586){\line( 1, 0){1500}}
}%
{\color[rgb]{0,0,0}\put(8551,-4711){\line( 1, 0){1500}}
}%
{\color[rgb]{0,0,0}\put(6601,-5986){\line( 1,-1){300}}
}%
{\color[rgb]{0,0,0}\put(6601,-6286){\line( 1, 1){300}}
}%
{\color[rgb]{0,0,0}\put(10051,-4411){\vector( 0,-1){2475}}
}%
{\color[rgb]{0,0,0}\put(7051,-4711){\vector( 1, 0){900}}
}%
{\color[rgb]{0,0,0}\put(7051,-4786){\vector( 2,-1){900}}
}%
{\color[rgb]{0,0,0}\put(7051,-6511){\vector( 2, 1){900}}
}%
{\color[rgb]{0,0,0}\put(7051,-6586){\vector( 1, 0){900}}
}%
\put(7501,-4411){\makebox(0,0)[b]{\smash{{\SetFigFont{12}{24.0}
{\rmdefault}{\mddefault}{\updefault}{\color[rgb]{0,0,0}$\neg x_1 \vee t_1$}%
}}}}
\put(7501,-6961){\makebox(0,0)[b]{\smash{{\SetFigFont{12}{24.0}
{\rmdefault}{\mddefault}{\updefault}{\color[rgb]{0,0,0}$\neg e_2 \vee t_2$}%
}}}}
\put(6751,-4711){\makebox(0,0)[b]{\smash{{\SetFigFont{12}{24.0}
{\rmdefault}{\mddefault}{\updefault}{\color[rgb]{0,0,0}$x_1 \vee \neg q$}%
}}}}
\put(6751,-5116){\makebox(0,0)[b]{\smash{{\SetFigFont{12}{24.0}
{\rmdefault}{\mddefault}{\updefault}{\color[rgb]{0,0,0}$e_1 \vee \neg q$}%
}}}}
\put(6751,-6211){\makebox(0,0)[b]{\smash{{\SetFigFont{12}{24.0}
{\rmdefault}{\mddefault}{\updefault}{\color[rgb]{0,0,0}$x_2 \vee \neg q$}%
}}}}
\put(6751,-6586){\makebox(0,0)[b]{\smash{{\SetFigFont{12}{24.0}
{\rmdefault}{\mddefault}{\updefault}{\color[rgb]{0,0,0}$e_2 \vee \neg q$}%
}}}}
\put(8251,-4711){\makebox(0,0)[b]{\smash{{\SetFigFont{12}{24.0}
{\rmdefault}{\mddefault}{\updefault}{\color[rgb]{0,0,0}$\neg q \vee t_1$}%
}}}}
\put(8251,-6586){\makebox(0,0)[b]{\smash{{\SetFigFont{12}{24.0}
{\rmdefault}{\mddefault}{\updefault}{\color[rgb]{0,0,0}$\neg q \vee t_2$}%
}}}}
\put(7501,-5461){\makebox(0,0)[b]{\smash{{\SetFigFont{12}{24.0}
{\rmdefault}{\mddefault}{\updefault}{\color[rgb]{0,0,0}$\neg x_1 \vee c_1$}%
}}}}
\put(7501,-7711){\makebox(0,0)[b]{\smash{{\SetFigFont{12}{24.0}
{\rmdefault}{\mddefault}{\updefault}{\color[rgb]{0,0,0}$\neg e_1 \vee c_2$}%
}}}}
\put(7501,-7486){\makebox(0,0)[b]{\smash{{\SetFigFont{12}{24.0}
{\rmdefault}{\mddefault}{\updefault}{\color[rgb]{0,0,0}$\neg x_2 \vee c_1$}%
}}}}
\put(8251,-5311){\makebox(0,0)[b]{\smash{{\SetFigFont{12}{24.0}
{\rmdefault}{\mddefault}{\updefault}{\color[rgb]{0,0,0}$\neg q \vee c_1$}%
}}}}
\put(7501,-5911){\makebox(0,0)[b]{\smash{{\SetFigFont{12}{24.0}
{\rmdefault}{\mddefault}{\updefault}{\color[rgb]{0,0,0}$\neg e_2 \vee c_2$}%
}}}}
\put(8251,-6061){\makebox(0,0)[b]{\smash{{\SetFigFont{12}{24.0}
{\rmdefault}{\mddefault}{\updefault}{\color[rgb]{0,0,0}$\neg q \vee c_2$}%
}}}}
\put(9976,-4261){\makebox(0,0)[b]{\smash{{\SetFigFont{12}{24.0}
{\rmdefault}{\mddefault}{\updefault}{\color[rgb]{0,0,0}$\neg t_1 \vee \neg t_2 \vee \neg c_1 \vee \neg c_2 \vee e_1$}%
}}}}
\put(10051,-7111){\makebox(0,0)[b]{\smash{{\SetFigFont{12}{24.0}
{\rmdefault}{\mddefault}{\updefault}{\color[rgb]{0,0,0}$e_1 \vee \neg q$}%
}}}}
\end{picture}%
\nop{
      -x1 v t1           -t1 v -t2 v -c1 v -c2 v e1
          |                    |
       ---+---> -q v t1 -------+
x1 v -q                        |
       ---+---> -q v c2 -------+
          |                    |
      -x1 v c2                 |
                               |
(e1 v -q)                      |
                               |
                               |
(x2 v -q)                      |
                               |
      -e2 v c1                 |
          |                    |
      ----+---> -q v c1 -------+
e2 v -q                        |
      ----+---> -q v t2 -------+
          |                    |
      -e2 v t2                 V
                            e1 v -q
      -x2 v c1
      -n1 v c2
}
\end{center}

The clauses $\neg x_1 \vee t_1$ and $\neg e_2 \vee t_1$ ensure that either $x_1
\vee \neg q$ or $e_1 \vee \neg q$ is included for each index $i$. Otherwise, a
subset of the same size could contain neither while including both $x_2 \vee
\neg q$ and $e_2 \vee \neg q$, which is invalid because it corresponds to
evaluating $x_2$ to both true and false.

The clause
{} $\neg t_1 \vee \neg t_2 \vee \neg c_1 \vee \neg c_2 \vee e_1$
completes the derivation. If the subset of $A''$ corresponds to a propositional
interpretation that satisfies both clauses of $F$, then all clauses $t_i \vee
\neg q$ and $c_j \vee \neg q$ are derived. All these clauses resolve into
{} $e_2 \vee \neg q$,
which was missing in the subset of $A''$.

The complete reduction from $F$ to $A$ and $k$ is as follows, where the formula
is $F = \{f_1,\ldots,f_m\}$ and $X=\{x_1,\ldots,x_n\}${\plural} are its
variables. The formula $A$ is built over an extended alphabet comprising $X$
and the additional variables
{} $E = \{e_1,\ldots,e_n\}$,
{} $T = \{t_1,\ldots,t_n\}$,
{} $C = \{c_1,\ldots,c_m\}$ and
{} $q$.

\begin{eqnarray*}
A &=& A_F \cup A_T \cup A_C \cup A_B					\\
A_F &=&
\{x_i \vee \neg q \mid x_i \in X\} \cup
\{e_i \vee \neg q \mid x_i \in X\}					\\
A_T &=&
\{\neg x_i \vee t_i, \neg e_i \vee t_i \mid x_i \in X\}			\\
A_C &=&
\{\neg x_i \vee c_j \mid      x_i \in f_j ,~ f_j \in F\} \cup
\{\neg e_i \vee c_j \mid \neg x_i \in f_j ,~ f_j \in F\}		\\
A_B &=&
\{\neg t_1 \vee \cdots \vee \neg t_n \vee
  \neg c_1 \vee \cdots \vee \neg c_m \vee
  x_i \vee \neg q \mid x_i \in X\} \cup					\\
&&
\{\neg t_1 \vee \cdots \vee \neg t_n \vee
  \neg c_1 \vee \cdots \vee \neg c_m \vee
  e_i \vee \neg q \mid x_i \in X\}					\\
k &=& 2 \times n + ||A_T|| + ||A_C|| + ||A_B||
\end{eqnarray*}

The fixed clauses are all of them but $A_F$. This way, they are in all formulae
equivalent to $A$. For equivalence, these need to entail all clauses of $A_F$
they do not contain. They can do in size $k$ only by including a clause of $x_i
\vee \neg q$ and a clause $e_i \vee \neg q$ for every index $i$, and they do
only if this choice corresponds to a model of $F$.

This argument assumes that the clauses of $A_F$ are fixed. Superirredundancy
ensures that. Lemma~\ref{set-value} ensures superirredundancy. For example,
replacing $q$ with false and simplifying the result removes all clauses but
$A_T \cup A_C$. These clauses contain only the literals $\neg x_i$, $\neg e_i$,
$t_i$ and $c_j$. They do not contain their negation. Therefore, they do not
resolve. Since they are not contained in each other, Lemma~\ref{no-resolution}
proves them superirredundant.

The clauses of $A_B$ are not superirredundant, but can be turned so using the
technique of Section~\ref{section-split}: each clause
{} $\neg t_1 \vee \cdots \vee \neg t_n \vee
{}  \neg c_1 \vee \cdots \vee \neg c_m \vee
{}  x_i \vee \neg q$
is split by a new variable $r_i$. The result is the pair of clauses
{} $\neg t_1 \vee \cdots \vee \neg t_n \vee
{}  \neg c_1 \vee \cdots \vee \neg c_m \vee
{}  x_i \vee \neg r_i$
and
{} $r_i \vee \neg q$.

This completes the proof. Its construction was incremental. Superirredundancy
is initially assumed so that some clauses are considered fixed. They allow
deriving the clauses that are not included in the minimal Horn formula if and
only if $F$ is satisfiable. Only when the reduction is completed,
superirredundancy is actually ensured by splitting the clauses that are not so.

\section{Conclusions}
\label{section-conclusions}

Superirredundancy helps to build formulae including clauses that resist
minimization in size: a superirredundant clause is in all minimal-size versions
of the formula. An application is hardness proofs of minimization problems,
like checking whether a formula can be compressed within a certain size.
Superirredundancy is not aimed at the minimization itself, but at building
formulae that have certain properties, like the targets of hardness reductions.
An example shown in this article is an alternative proof of the \np-hardness of
the problem of checking whether a Horn formula can be squeezed within a certain
bound. Another application is proving the hardness of checking minimal size
after forgetting some variables from a Horn or CNF formula~\cite{libe-20}.

Superredundancy is defined in terms of resolution, not in terms of
minimization. The presence of a superirredundant clause in all formulae that
are minimal among the equivalent ones is a consequence, not its definition.
Superirredundancy is sufficient to that, not necessary. Yet, it is easier to
achieve than that. Some conditions that are equivalent to superirredundancy and
others that are necessary and still others that are sufficient are presented. A
mechanism that often make a clause superirredundant while preserving the
superirredundancy of the others is also shown. It allows building a formula
incrementally: first its semantics is established, then the clauses that have
to be superirredundant and made so.

The example application is an alternative proof of hardness. The claim is
already known via prime implicate essentiality instead of
superirredundancy~\cite{hamm-koga-93}. Yet, while this proof surfaced some
twenty years after the problem was open, the one based on superirredundancy was
very simple to come up with. Its proof of correctness is not much shorter that
the previous one, but neither was this its aim. Building the reduction was, not
proving it correct.

Why bothering introducing a new notion just for proving again something that
was already known? Boolean minimization has been computationally framed in many
variants depending on the restriction on the formula and the definition of
minimality. Yet, it is not closed. An example open problem is the complexity of
checking whether forgetting some variables from a formula is expressed by a
formula of a certain size; four hardness proofs are obtained by applying
superirredundancy in a separate article~\cite{libe-20}. Another example where
superirredundancy could be applied is formula revision or
update~\cite{pepp-08,kats-mend-91}: these transformations are known to
potentially increase the size of the changed formula~\cite{cado-etal-99};
minimizing it~\cite{libe-05} is a problem where superirredundancy could be
applied. In general, every mechanism that transforms a formula in whichever way
(update, summarize, expand, etc.) is subject to minimizing, and
superirredundancy applies. Finally, given that some sufficient conditions to
superirredundancy are computationally easy (like replacing variables with
values and checking the resulting formula for separation of variables), they
may also be used as a simple preliminary test when performing formula
minimization.

Superredundancy is a derivation property. As such, it depends on the syntax of
the formula. Therefore, it is not the same as any semantical property like
implication, prime implication, redundancy in the set of prime implicates or
essentiality. It depends on the syntax because it is based on resolution, and
resolution is a restricted form of entailment: it does not allow adding
arbitrary literals to clauses. In the other way around, entailment is
resolution plus expansion. The large corpus of research on automated
reasoning~\cite{fitt-12,harr-09} offers numerous alternative forms of
derivation that work even when formulae are not in clausal form, like natural
deduction and Frege systems. Some variant of superredundancy may be defined for
them.

\appendix

\section*{Complete proof}
\label{section-complete}

The following is the complete proof of correctness of the reduction presented
in Section~\ref{section-example}.

\begin{theorem}

The problem of establishing the existence of a formula $B$ such that $||B||
\leq k$ and $B \equiv A$ is \np-hard.

\end{theorem}

\proof Proof is by reduction from propositional satisfiability. An arbitrary
CNF formula $F$ is shown satisfiable if and only if a Horn formula $A$ is
equivalent to one of size bounded by $k$.

Let the CNF formula be $F = \{f_1,\ldots,f_m\}$ and $X=\{x_1,\ldots,x_n\}$ its
variables. The formula $A$ is built over an extended alphabet comprising the
variables $X$ and the additional variables
{} $E = \{e_1,\ldots,e_n\}$,
{} $T = \{t_1,\ldots,t_n\}$,
{} $C = \{c_1,\ldots,c_m\}$,
{} $R = \{r_1,\ldots,r_n\}$ and
{} $S = \{s_1,\ldots,s_n\}$,
{} $q$.
The formula $A$ and the integer $k$ are as follows.

\begin{eqnarray*}
A &=& A_F \cup A_T \cup A_C \cup A_B					\\
A_F &=&
\{x_i \vee \neg q \mid x_i \in X\} \cup
\{e_i \vee \neg q \mid x_i \in X\}					\\
A_T &=&
\{\neg x_i \vee t_i, \neg e_i \vee t_i \mid x_i \in X\}			\\
A_C &=&
\{\neg x_i \vee c_j \mid      x_i \in f_j ,~ f_j \in F\} \cup
\{\neg e_i \vee c_j \mid \neg x_i \in f_j ,~ f_j \in F\}		\\
A_B' &=&
\{\neg t_1 \vee \cdots \vee \neg t_n \vee
  \neg c_1 \vee \cdots \vee \neg c_m \vee
  x_i \vee \neg r_i, r_i \vee \neg q \mid x_i \in X\} \cup		\\
&&
\{\neg t_1 \vee \cdots \vee \neg t_n \vee
  \neg c_1 \vee \cdots \vee \neg c_m \vee
  e_i \vee \neg s_i, s_i \vee \neg q \mid x_i \in X\}			\\
\end{eqnarray*}

Before formally proving that the reduction works, a short summary of why it
works is given. All clauses of $A$ but $A_F$ are superirredundant: all minimal
equivalent formulae contain them. The bound $k$ allows only one clause of $A_F$
for each $i$. Combined with the clauses of $A_T$ they entail $t_i \vee \neg q$.
If $F$ is satisfiable, they also combine with the clauses $A_C$ to imply all
clauses $c_j \vee \neg q$. Resolving these clauses with $A_B$ produces all
clauses $x_i \vee \neg q$ and $e_i \vee \neg q$, including the ones not in the
selection. This way, a formula that contains one clause of $A_F$ for each index
$i$ implies all of $A_F$, but only if $F$ is satisfiable.

The following figure shows how $e_1 \vee q$ is derived from $x_1 \vee q$ and
$e_2 \vee q$, when the formula is $F = \{f_1,f_2\}$ where $f_1 = x_1 \vee x_2$
and $f_2 = \neg x_1 \vee \neg x_2$. These clauses translate into
{} $A_C = \{
{}	\neg x_1 \vee c_1,
{}	\neg x_2 \vee c_1,
{}	\neg e_1 \vee c_2,
{}	\neg e_2 \vee c_2
{} \}$.

\begin{center}
\setlength{\unitlength}{5000sp}%
\begingroup\makeatletter\ifx\SetFigFont\undefined%
\gdef\SetFigFont#1#2#3#4#5{%
  \reset@font\fontsize{#1}{#2pt}%
  \fontfamily{#3}\fontseries{#4}\fontshape{#5}%
  \selectfont}%
\fi\endgroup%
\begin{picture}(3504,3675)(6589,-7726)
\thinlines
{\color[rgb]{0,0,0}\put(7501,-4486){\line( 0,-1){225}}
}%
{\color[rgb]{0,0,0}\put(7501,-5011){\line( 0,-1){300}}
}%
{\color[rgb]{0,0,0}\put(7501,-5986){\line( 0,-1){300}}
}%
{\color[rgb]{0,0,0}\put(6601,-4936){\line( 1,-1){300}}
}%
{\color[rgb]{0,0,0}\put(6601,-5236){\line( 1, 1){300}}
}%
{\color[rgb]{0,0,0}\put(7501,-6586){\line( 0,-1){225}}
}%
{\color[rgb]{0,0,0}\put(8551,-5311){\line( 1, 0){1500}}
}%
{\color[rgb]{0,0,0}\put(8551,-6061){\line( 1, 0){1500}}
}%
{\color[rgb]{0,0,0}\put(8551,-6586){\line( 1, 0){1500}}
}%
{\color[rgb]{0,0,0}\put(8551,-4711){\line( 1, 0){1500}}
}%
{\color[rgb]{0,0,0}\put(6601,-5986){\line( 1,-1){300}}
}%
{\color[rgb]{0,0,0}\put(6601,-6286){\line( 1, 1){300}}
}%
{\color[rgb]{0,0,0}\put(10051,-4411){\vector( 0,-1){2475}}
}%
{\color[rgb]{0,0,0}\put(7051,-4711){\vector( 1, 0){900}}
}%
{\color[rgb]{0,0,0}\put(7051,-4786){\vector( 2,-1){900}}
}%
{\color[rgb]{0,0,0}\put(7051,-6511){\vector( 2, 1){900}}
}%
{\color[rgb]{0,0,0}\put(7051,-6586){\vector( 1, 0){900}}
}%
\put(7501,-4411){\makebox(0,0)[b]{\smash{{\SetFigFont{12}{24.0}
{\rmdefault}{\mddefault}{\updefault}{\color[rgb]{0,0,0}$\neg x_1 \vee t_1$}%
}}}}
\put(7501,-6961){\makebox(0,0)[b]{\smash{{\SetFigFont{12}{24.0}
{\rmdefault}{\mddefault}{\updefault}{\color[rgb]{0,0,0}$\neg e_2 \vee t_2$}%
}}}}
\put(6751,-4711){\makebox(0,0)[b]{\smash{{\SetFigFont{12}{24.0}
{\rmdefault}{\mddefault}{\updefault}{\color[rgb]{0,0,0}$x_1 \vee \neg q$}%
}}}}
\put(6751,-5116){\makebox(0,0)[b]{\smash{{\SetFigFont{12}{24.0}
{\rmdefault}{\mddefault}{\updefault}{\color[rgb]{0,0,0}$e_1 \vee \neg q$}%
}}}}
\put(6751,-6211){\makebox(0,0)[b]{\smash{{\SetFigFont{12}{24.0}
{\rmdefault}{\mddefault}{\updefault}{\color[rgb]{0,0,0}$x_2 \vee \neg q$}%
}}}}
\put(6751,-6586){\makebox(0,0)[b]{\smash{{\SetFigFont{12}{24.0}
{\rmdefault}{\mddefault}{\updefault}{\color[rgb]{0,0,0}$e_2 \vee \neg q$}%
}}}}
\put(8251,-4711){\makebox(0,0)[b]{\smash{{\SetFigFont{12}{24.0}
{\rmdefault}{\mddefault}{\updefault}{\color[rgb]{0,0,0}$\neg q \vee t_1$}%
}}}}
\put(8251,-6586){\makebox(0,0)[b]{\smash{{\SetFigFont{12}{24.0}
{\rmdefault}{\mddefault}{\updefault}{\color[rgb]{0,0,0}$\neg q \vee t_2$}%
}}}}
\put(7501,-5461){\makebox(0,0)[b]{\smash{{\SetFigFont{12}{24.0}
{\rmdefault}{\mddefault}{\updefault}{\color[rgb]{0,0,0}$\neg x_1 \vee c_1$}%
}}}}
\put(7501,-7711){\makebox(0,0)[b]{\smash{{\SetFigFont{12}{24.0}
{\rmdefault}{\mddefault}{\updefault}{\color[rgb]{0,0,0}$\neg e_1 \vee c_2$}%
}}}}
\put(7501,-7486){\makebox(0,0)[b]{\smash{{\SetFigFont{12}{24.0}
{\rmdefault}{\mddefault}{\updefault}{\color[rgb]{0,0,0}$\neg x_2 \vee c_1$}%
}}}}
\put(8251,-5311){\makebox(0,0)[b]{\smash{{\SetFigFont{12}{24.0}
{\rmdefault}{\mddefault}{\updefault}{\color[rgb]{0,0,0}$\neg q \vee c_1$}%
}}}}
\put(7501,-5911){\makebox(0,0)[b]{\smash{{\SetFigFont{12}{24.0}
{\rmdefault}{\mddefault}{\updefault}{\color[rgb]{0,0,0}$\neg e_2 \vee c_2$}%
}}}}
\put(8251,-6061){\makebox(0,0)[b]{\smash{{\SetFigFont{12}{24.0}
{\rmdefault}{\mddefault}{\updefault}{\color[rgb]{0,0,0}$\neg q \vee c_2$}%
}}}}
\put(9976,-4261){\makebox(0,0)[b]{\smash{{\SetFigFont{12}{24.0}
{\rmdefault}{\mddefault}{\updefault}{\color[rgb]{0,0,0}$\neg t_1 \vee \neg t_2 \vee \neg c_1 \vee \neg c_2 \vee e_1$}%
}}}}
\put(10051,-7111){\makebox(0,0)[b]{\smash{{\SetFigFont{12}{24.0}
{\rmdefault}{\mddefault}{\updefault}{\color[rgb]{0,0,0}$e_1 \vee \neg q$}%
}}}}
\end{picture}%
\nop{
      -x1 v t1           -t1 v -t2 v -c1 v -c2 v n1 v -r1
          |                    |
       ---+---> -q v t1 -------+
x1 v -q                        |
       ---+---> -q v c2 -------+
          |                    |
      -x1 v c2                 |
                               |
(n1 v -q)                      |
                               |
                               |
(x2 v -q)                      |
                               |
      -n2 v c1                 |
          |                    |
      ----+---> -q v c1 -------+
n2 v -q                        |
      ----+---> -q v t2 -------+
          |                    |
      -n2 v t2                 V
                         -q v n1 v -r1 ----+----> n1 v -q
                                           |
      -x2 v c1                          r1 v -q
      -n1 v c2
}
\end{center}

For each index $i$, at least one among $x_i \vee \neg q$ and $e_i \vee \neg q$
is necessary for deriving $\neg q \vee t_i$, which is entailed by $A$.
Alternatively, $\neg q \vee t_i$ itself is necessary for the formula to be
equivalent. Either way, for each index $i$ at least a two-literal clause is
necessary.

The claim is formally proved in four steps: first, the superirredundant clauses
are identified; second, an equivalent formula of size $k$ is built if $F$ is
satisfiable; third, the necessary clauses in every equivalent formula are
identified; fourth, if $F$ is unsatisfiable every equivalent formula is proved
to have size greater than $k$.

\

{\bf Superirredundancy.}

The claim requires $A$ to be minimal, which follows from all its clauses being
superirredundant by Lemma~\ref{minimal}. Most of them survive forgetting; the
reduction is based on these being superirredundant. Instead of proving
superirredundancy in two different but similar formulae, it is proved in their
union.

In particular, the clauses
{} $A_T \cup A_C \cup A_B'$
are shown superirredundant in
{} $A_F \cup A_T \cup A_C \cup A_B'$.

Superirredundancy is proved via Lemma~\ref{set-value}: a substitution simplify
{} $A_F \cup A_T \cup A_C \cup A_B'$
enough to prove superirredundancy easily, for example because its clauses do
not resolve and Lemma~\ref{no-resolution} applies.

\begin{itemize}

\item

Replacing all variables $x_i$, $e_i$, $t_i$ and $c_j$ with true removes from
{} $A_F \cup A_T \cup A_C \cup A_B'$
all clauses of $A_F$, $A_T$, $A_C$ and all clauses of $A_B'$ but $r_i \vee \neg
q$ and $s_i \vee \neg q$. The remaining clauses contain only the literals
$r_i$, $s_i$ and $\neg q$. Therefore, they do not resolve. Since none is
contained in another, they are all superirredundant by
Lemma~\ref{no-resolution}. This proves the superirredundancy of all clauses
$r_i \vee \neg q$ and $s_i \vee \neg q$.

\item

Replacing all variables $q$, $r_i$ and $s_i$ with false removes from
{} $A_F \cup A_T \cup A_C \cup A_B'$
all clauses but $A_T \cup A_C$. These clauses contain only the literals $\neg
x_i$, $\neg e_i$, $t_i$ and $c_j$. Therefore, they do not resolve. Since they
are not contained in each other, Lemma~\ref{no-resolution} proves them
superirredundant.

\item

Replacing all variables with false except for all variables $t_i$ and $c_j$ and
the single variable $x_h$ removes all clauses from
{} $A_F \cup A_T \cup A_C \cup A_B'$
but
{} $\neg x_h \vee t_h$,
{} $\neg t_1 \vee \cdots \vee \neg t_n \vee
{}  \neg c_1 \vee \cdots \vee \neg c_m \vee
{}  x_h \vee \neg r_h$
and all clauses
{} $\neg x_h \vee c_j$ with $x_h \in f_j$.
They only resolve in tautologies. Therefore, their resolution closure only
contains them. Removing
{} $\neg t_1 \vee \cdots \vee \neg t_n \vee
{}  \neg c_1 \vee \cdots \vee \neg c_m \vee
{}  x_h \vee \neg r_h$
from the resolution closure leaves only
{} $\neg x_h \vee t_h$
and all clauses
{} $\neg x_h \vee c_j$ with $x_h \in f_j$.
They do not resolve since they do not contain opposite literals. Since
{} $\neg t_1 \vee \cdots \vee \neg t_n \vee
{}  \neg c_1 \vee \cdots \vee \neg c_m \vee
{}  x_h \vee \neg r_h$
is not contained in them, it is not entailed by them. This proves it
superirredundant. A similar replacement proves the superirredundancy of each
{} $\neg t_1 \vee \cdots \vee \neg t_n \vee
{}  \neg c_1 \vee \cdots \vee \neg c_m \vee
{}  e_h \vee \neg s_h$.

\end{itemize}

These points prove that the clauses in
{} $A_T \cup A_C \cup A_B'$
are superirredundant in $A$. The only clauses that may be superredundant are
$A_F$.

\

{\bf Formula $F$ is satisfiable.}

Let $M$ be a model satisfying $F$. The set $A_R'$ is defined as comprising the
clauses $x_i \vee \neg q$ such that $M \models x_i$ and the clauses $e_i \vee
\neg q$ such that $M \models \neg x_i$. The Horn formula $A_R' \cup A_C \cup
A_T \cup A_B'$ has size $k$. It is equivalent to $A_R \cup A_T \cup A_C \cup
A_B$. This is proved by showing that it entails every clause in $A_R$,
including the only clauses of $A$ it does not contain.

Since $M$ satisfies every clause $f_j \in F$, it satisfies at least a literal
of $f_j$: for some $x_i$, either $x_i \in f_j$ and $M \models x_i$ or $\neg x_i
\in f_j$ and $M \models \neg x_i$. By construction, $x_i \in f_j$ implies $\neg
x_i \vee c_j \in A_C$ and $\neg x_i \in f_j$ implies $\neg e_i \vee c_j \in
A_C$. Again by construction, $M \models x_i$ implies $x_i \vee \neg q \in A_R'$
and $M \models \neg x_i$ implies $e_i \vee \neg q \in A_R'$. As a result,
either
{} $x_i \vee \neg q \in A_R'$ and $\neg x_i \vee c_j \in A_C$
or 
{} $e_i \vee \neg q \in A_R'$ and $\neg e_i \vee c_j \in A_C$.
In both cases, the two clause resolve in $c_j \vee q$.

Since $M$ satisfies either $x_i$ or $\neg x_i$, either $x_i \vee \neg q \in
A_R'$ or $e_i \vee \neg q \in A_R'$. The first clause resolve with $\neg x_i
\vee t_i$ and the second with $\neg e_i \vee t_i$. The result is $t_i \vee \neg
q$ in both cases.

Resolving all these clauses $t_i \vee \neg q$ and $c_j \vee q$ with
{} $\neg t_1 \vee \cdots \vee \neg t_n \vee
{}  \neg c_1 \vee \cdots \vee \neg c_m \vee
{}  x_i \vee \neg r_i$
and then with $r_i \vee \neg q$, the result is $x_i \vee \neg q$. In the same
way, resolving these clauses with
{} $\neg t_1 \vee \cdots \vee \neg t_n \vee
{}  \neg c_1 \vee \cdots \vee \neg c_m \vee
{}  e_i \vee \neg s_i$
and $s_i \vee \neg q$ produces $e_i \vee \neg q$. This proves that all clauses
of $A_R$ are entailed.

\

{\bf Necessary clauses}

All CNF formulae that are equivalent to $A_F \cup A_T \cup A_C \cup A_B'$ and
have minimal size contain
{} $A_T \cup A_C \cup A_B'$
because these clauses are superirredundant. Therefore, these formulae are
{} $A_N \cup A_T \cup A_C \cup A_B'$
for some set of clauses $A_N$. This set $A_N$ is now proved to contain either
{} $x_h \vee \neg q$,
{} $x_h \vee \neg r_i$,
{} $e_h \vee \neg q$,
{} $e_h \vee \neg s_i$ or
{} $t_h \vee \neg q$
for each index $h$. Let $M$ and $M'$ be the following models.

\begin{eqnarray*}
M &=&
\{x_i = e_i = t_i = \true \mid i \not= h\} \cup
\{x_h = e_h = t_h = \false\} \cup \\
&&
\{c_j = \true\} \cup
\{q = \true\} \cup
\{r_i = \true, s_i = \true\}
\\
M' &=&
\{x_i = e_i = t_i = \true \mid i \not= h\} \cup
\{x_h = e_h = t_h = \true \} \cup \\
&&
\{c_j = \true\} \cup
\{q = \true\} \cup
\{r_i = \true, s_i = \true\}
\end{eqnarray*}

The three clauses are falsified by $M$. Since the two of them $xh \vee \neg q$
and $e_h \vee \neg q$ are in $A_F$, this set is also falsified by $M$. As a
result, $M$ is not a model of
{} $A_F \cup A_T \cup A_C \cup A_B'$.
This formula is equivalent to
{} $A_N \cup A_T \cup A_C \cup A_B'$,
which is therefore falsified by $M$. In formulae,
{} $M \not\models A_N \cup A_T \cup A_C \cup A_B'$.

The formula $A_N \cup A_T \cup A_C \cup A_B'$ contains a clause falsified by
$M$. Since $M \models A_T \cup A_C \cup A_B'$, this clause is in $A_N$ but not
in $A_T \cup A_C \cup A_B'$. In formulae, $M \not\models c$ for some $c \in
A_N$ and $c \not\in A_T \cup A_C \cup A_B'$. This clause is entailed by
{} $A_F \cup A_T \cup A_C \cup A_B'$
because this formula entails all of
{} $A_N \cup A_T \cup A_C \cup A_B'$,
and $c$ is in $A_N$. In formulae,
{} $A_F \cup A_T \cup A_C \cup A_B' \models c$.

This clause $c$ contains either
{} $x_h$, $e_h$ or $t_h$.
This is proved by deriving a contradiction from the assumption that $c$ does
not contain any of these three literals. Since $M \not\models c$, the clause
$c$ contains only literals that are falsified by $M$. Not all of them: it does
not contain
{} $x_h$, $e_h$ and $t_h$
by assumption. It does not contain $\neg x_h$, $\neg e_h$ and $\neg t_h$ either
because it would otherwise be satisfied by $M$. As a result, $c$ is also
falsified by $M'$, which is the same as $M$ but for the values of
{} $x_h$, $e_h$ and $t_h$.
At the same time, $M'$ satisfies
{} $A_F \cup A_T \cup A_C \cup A_B'$,
contradicting
{} $A_F \cup A_T \cup A_C \cup A_B' \models c$.
This contradiction proves that $c$ contains either
{} $x_h$, $e_h$ or $t_h$.

From the fact that $c$ contains either $x_h$, $e_h$ or $t_h$, that is a
consequence of $A_F \cup A_T \cup A_C \cup A_B'$, and that is in a minimal-size
formula, it is now possible to prove that $c$ contains either
{} $x_h \vee \neg q$,
{} $x_h \vee \neg r_i$,
{} $e_h \vee \neg q$,
{} $e_h \vee \neg s_i$ or
{} $t_h \vee \neg q$.

Since $c$ is entailed by
{} $A_F \cup A_T \cup A_C \cup A_B'$,
a subset of $c$ follows from resolution from it:
{} $A_F \cup A_T \cup A_C \cup A_B' \vdash c'$ with $c' \subseteq c$.
This implies
{} $A_N \cup A_T \cup A_C \cup A_B' \models c'$
by equivalence. If $c' \subset c$, then
{} $A_N \cup A_T \cup A_C \cup A_B'$
would not be minimal because it contained a non-minimal clause $c \in A_N$.
Therefore,
{} $A_F \cup A_T \cup A_C \cup A_B' \vdash c$.

The only two clauses of
{} $A_F \cup A_T \cup A_C \cup A_B'$
that contain $x_h$ are $x_h \vee \neg q$ and
{} $\neg t_1 \vee \cdots \vee \neg t_n \vee
{}   \neg c_1 \vee \cdots \vee \neg c_m \vee
{}   x_h \vee \neg r_h$.
They contain either $\neg q$ or $\neg r_h$. These literals are only resolved
out by clauses containing their negations $q$ and $r_h$. No clause contains $q$
and the only clause that contains $r_h$ is $r_h \vee \neg q$, which contains
$\neg q$. If a result of resolution contains $x_h$, it also contains either
$\neg q$ or $\neg r_h$. This applies to $c$ because it is a result of
resolution.

The same applies if $c$ contains $e_h$: it also contains either $\neg q$ or
$\neg s_i$.

The case of $t_h \in c$ is a bit different. The only two clauses of
{} $A_F \cup A_T \cup A_C \cup A_B'$
that contain $t_h$ are $\neg x_h \vee t_h$ and $\neg e_h \vee t_h$. Since both
are in $A_T$ and $c \not\in A_T$, they are not $c$. The first clause $\neg x_h
\vee t_h$ only resolves with $x_i \vee \neg q$ or
{} $\neg t_1 \vee \cdots \vee \neg t_n \vee
{}   \neg c_1 \vee \cdots \vee \neg c_m \vee
{}   x_h \vee \neg r_h$,
but resolving with the latter generates a tautology. The result of resolving
$\neg x_h \vee t_h$ with $x_i \vee \neg q$ is $t_h \vee \neg q$; no clause
contains $q$. Therefore, $c$ can only be $t_h \vee \neg q$. The second clause
$\neg e_h \vee t_h$ leads to the same conclusion.

In summary, $c$ contains either
{} $x_h \vee \neg q$,
{} $x_h \vee \neg r_i$,
{} $e_h \vee \neg q$,
{} $e_h \vee \neg s_i$ or
{} $t_h \vee \neg q$.
In all these cases it contains at least two literals. This is the case for
every index $h$; therefore, $A_N$ contains at least $n$ clauses of two
literals. Every minimal CNF formula equivalent to
{} $A_R \cup A_T \cup A_C \cup A_B$
has size at least $2 \times n$ plus the size of $A_T \cup A_C \cup A_B$. This
sum is exactly $k$. This proves that every minimal CNF formula expressing
forgetting contains at least $k$ literal occurrences. Worded differently, every
CNF formula expressing forgetting has size at least $k$.

\

{\bf Formula $F$ is unsatisfiable}

The claim is that no CNF formula of size $k$ expresses forgetting if $F$ is
unsatisfiable. This is proved by deriving a contradiction from the assumption
that such a formula exists.

It has been proved that the minimal CNF formulae equivalent to
{} $A_F \cup A_T \cup A_C \cup A_B'$
are
{} $A_N \cup A_T \cup A_C \cup A_B'$
for some set $A_N$ that contains clauses that include either
{} $x_h \vee \neg q$,
{} $x_h \vee \neg r_i$,
{} $e_h \vee \neg q$,
{} $e_h \vee \neg s_i$ or
{} $t_h \vee \neg q$
for each index $h$.

If $A_N$ contains other clauses, or more than one clause for each $h$, or these
clauses contain other literals, the size of
{} $A_N \cup A_T \cup A_C \cup A_B'$
is larger than
{} $k = 2 \times n + ||A_T|| + ||A_C|| + ||A_B'||$,
contradicting the assumption. This proves that every formula of size $k$ that
is equivalent to
{} $A_F \cup A_T \cup A_C \cup A_B'$
is equal to
{} $A_N \cup A_T \cup A_C \cup A_B'$
where $A_N$ contains exactly one clause among
{} $x_h \vee \neg q$,
{} $x_h \vee \neg r_i$,
{} $e_h \vee \neg q$,
{} $e_h \vee \neg s_i$ or
{} $t_h \vee \neg q$
for each index $h$.

The case
{} $x_h \vee \neg r_h \in A_N$
is excluded. It would imply
{} $A_F \cup A_T \cup A_C \cup A_B' \models x_h \vee \neg r_h$,
which implies the redundancy of
{} $\neg t_1 \vee \cdots \vee \neg t_n \vee
{}   \neg c_1 \vee \cdots \vee \neg c_m \vee
{}   x_h \vee \neg r_h \in A_B$
contrary to its previously proved superirredundancy. A similar argument proves
{} $e_h \vee \neg s_h \not\in A_N$.

The conclusion is that every formula of size $k$ that is equivalent to
{} $A_F \cup A_T \cup A_C \cup A_B'$
is equal to
{} $A_N \cup A_T \cup A_C \cup A_B'$
where $A_N$ contains exactly one clause among
{} $x_h \vee \neg q$,
{} $e_h \vee \neg q$,
{} $t_h \vee \neg q$
for each index $h$.

If $F$ is unsatisfiable, all such formulae are proved to be satisfied by a
model that falsifies
{} $A_F \cup A_T \cup A_C \cup A_B'$,
contrary to the assumed equivalence.

Let $M$ be the model that assigns $q = \true$ and $t_i=\true$, and assigns
$x_i=\true$ and $e_i=\false$ if $x_i \vee \neg q \in A_N$ and $x_i=\false$ and
$e_i=\true$ if $e_i \vee \neg q \in A_N$ or $t_i \vee \neg q \in A_N$. All
clauses of $A_N$ and $A_T$ are satisfied by $M$.

This model $M$ can be extended to satisfy all clauses of $A_C \cup A_B'$. Since
$F$ is unsatisfiable, $M$ falsifies at least a clause $f_j \in F$. Let $M'$ be
the model obtained by extending $M$ with the assignments of $c_j$ to false, all
other variables in $C$ to true and all variables $r_i$ and $s_i$ to true. This
extension satisfies all clauses of $A_B$ either because it sets $c_j$ to false
or because it sets $r_i$ and $s_i$ to true. It also satisfies all clauses of
$A_C$ that do not contain $c_j$ because it sets all variables of $C$ but $c_j$
to true.

The only clauses that remain to be proved satisfied are the clauses of $A_C$
that contain $c_j$. They are
{} $\neg x_i \vee c_j$ for all $x_i \in f_j$
and
{} $\neg e_i \vee c_j$ for all $\neg x_i \in f_j$.
Since $M'$ falsifies $f_j$, it falsifies every $x_i \in f_j$; therefore, it
satisfies $\neg x_i \vee c_j$. Since $M'$ falsifies $f_j$, it falsifies every
$\neg x_i \in f_j$; since by construction it assigns $e_i$ opposite to $x_i$,
it falsifies $e_i$ and therefore satisfies $\neg e_i \vee c_j$.

This proves that $M'$ satisfies
{} $A_N \cup A_T \cup A_C \cup A_B'$.
It does not satisfy
{} $A_F \cup A_T \cup A_C \cup A_B'$.
If $x_1 \vee \neg q \in A_N$, then $M'$ sets $x_1$ to true and $e_1$ to false;
therefore, it does not satisfy $e_1 \vee \neg q \in A_R$. Otherwise, $M'$ sets
$x_1$ to false and $e_1$ to true; therefore, it does not satisfy $x_1 \vee \neg
q$.

This contradicts the assumption that
{} $A_N \cup A_T \cup A_C \cup A_B'$
is equivalent to
{} $A_F \cup A_T \cup A_C \cup A_B'$.
The assumption that it has size $k$ is therefore false.~\qed


\let\c=\cedilla
\bibliographystyle{plain}

\end{document}